\documentclass[12pt,preprint]{aastex}

\def\arcsec {$^{\prime \prime}$}

\def\etal   {{et~al.\/}}

\def\mo     {{$M_{\odot}$}}
\def\lo     {{$L_{\odot}$}}

\bibpunct[]{(}{)}{,}{a}{}{,}

\begin{document}

\shorttitle{GLIMPSE Mid-IR Excesses}
\shortauthors{Uzpen, B. \etal}

\title{A GLIMPSE into the Nature of Galactic Mid-IR Excesses}

\author{B. Uzpen \altaffilmark{1}, H.~A. Kobulnicky \altaffilmark{1}, 
D.~R. Semler \altaffilmark{2}, T. Bensby \altaffilmark{3} \altaffilmark{4}, 
\and C. Thom \altaffilmark{5} }


\altaffiltext{1}{University of Wyoming, Dept. of Physics \&
Astronomy, Dept. 3905, Laramie, WY 82071}

\altaffiltext{2}{Columbia Astrophysics Laboratory, Columbia University, 550 West 120th Street, New York, NY 10027}

\altaffiltext{3}{University of Michigan, Department of Astronomy
500 Church St, Ann Arbor, MI 48109-1042}

\altaffiltext{4}{European Southern Observatory, Alonso de Cordova 3107,
Vitacura, Casilla 19001, Santiago Chile}

\altaffiltext{5}{The University of Chicago, Department of Astronomy \& Astrophysics
5640 S. Ellis Ave, Chicago, IL 60637}

\begin{abstract}
We investigate the nature of the mid-IR excess for 31
intermediate-mass stars that exhibit an 8 $\mu$m excess in either the
Galactic Legacy Infrared Mid-Plane Survey Extraordinaire or the
Mid-Course Space Experiment using high resolution optical spectra to
identify stars surrounded by warm circumstellar dust. From these data
we determine projected stellar rotational velocities and estimate
stellar effective temperatures for the sample.  We estimate stellar
ages from these temperatures, parallactic distances, and evolutionary
models.  Using MIPS [24] measurements and stellar parameters we
determine the nature of the infrared excess for 19 GLIMPSE stars. We
find that 15 stars exhibit H$\alpha$ emission and four exhibit
H$\alpha$ absorption.  Assuming that the mid-IR excesses arise in
circumstellar disks, we use the H$\alpha$ fluxes to model and estimate
the relative contributions of dust and free-free emission.  Six stars
exhibit H$\alpha$ fluxes that imply free-free emission can plausibly
explain the infrared excess at [24]. These stars are candidate
classical Be stars.  Nine stars exhibit H$\alpha$ emission, but their
H$\alpha$ fluxes are insufficient to explain the infrared excesses at
[24], suggesting the presence of a circumstellar dust component.
After the removal of the free-free component in these sources, we
determine probable disk dust temperatures of
T$_{disk}\simeq$300--800~K and fractional infrared luminosities of
$L_{IR}/L_{*}$ $\simeq$10$^{-3}$. These nine stars may be
pre-main-sequence stars with transitional disks undergoing disk
clearing.  Three of the four sources showing H$\alpha$ absorption
exhibit circumstellar disk temperatures $\simeq$300--400~K,
$L_{IR}/L_{*}$ $\simeq$10$^{-3}$, IR colors $K$-[24]$<$ 3.3, and are
warm debris disk candidates. One of the four H$\alpha$ absorption
sources has $K$-[24]$>$ 3.3 implying an optically thick outer disk and
is a transition disk candidate.
\end{abstract}


\section{Introduction}

With its launch in 1983, the \textit{Infrared Astronomical Satellite}
($IRAS$) opened up a new wavelength window and enabled identification
of many classes of stellar sources with circumstellar material,
providing new insights into star formation and evolution.  In one of
the most exciting discoveries enabled by $IRAS$, \citet{Aumann:1984}
reported the first detection of an infrared excess owing to dust in
the Vega system, and this excess was later explained as a debris
disk. To date, many debris disks have been found with \textit{IRAS},
the \textit{Infrared Space Observatory} \textit{(ISO)}, and the
\textit{Spitzer Space Telescope} (\textit{Spitzer})
(\citealt{Backman:1993}; \citealt{Spangler:2001};
\citealt{Meyer:2004}). Most known debris disks show evidence of cool
dust at wavelengths $>$ 20 $\mu$m, and the majority of these surround
intermediate-mass stars (\citealt{Backman:1993}; \citealt{Chen:2006};
\citealt{Hillenbrand:2008}).  Only two $IRAS$ Point Source Catalog
debris disk sources, the intermediate-mass A-type stars $\beta$
Pictoris and $\zeta$ Leporis, exhibited a mid-IR ($\leq$ 12 $\mu$m)
excess detectable by $IRAS$ \citep{Aumann:1991}.  $Spitzer$
observations of $IRAS$-discovered debris disks indicate that most are
devoid of warm dust and small particles (sizes $<$ 10 $\mu$m)
\citep{Chen:2006}.  The rarity of mid-IR excess sources may imply that
dust particles emitting at terrestrial temperatures are uncommon
around main-sequence stars.

Advances in infrared technologies coupled with large-aperture
ground-based telescopes and $Spitzer's$ mid-IR sensitivity with IRAC
enable new searches for warm/hot circumstellar dust that $IRAS$ did
not detect.  Even with such improvements in sensitivity over $IRAS$,
only five known intermediate-mass stars (HD 3003, HD 38678, HD 75416,
HD 105234, and HD 181296) are surrounded by debris disks with
temperatures analogous to our asteroid belt ($>$ 200 K)
(\citealt{Chen:2001}; \citealt{Rhee:2007} and references therein).
$Spitzer$ observations have also revealed a number of mid-IR excess
objects that are primordial disks undergoing disk clearing processes
similar to the solar-mass star TW Hya (\citealt{Calvet:2005};
\citealt{Furlan:2007}).  ``Transition disks'' may be evolved
protoplanetary disks with inner holes caused by clearing of the
initial gas-dominated disks from the inside out
\citep{Strom:1989}. These disks would be optically thin at near- and
mid-IR wavelengths and optically thick in the far-IR.  These sources
would show weak to no accretion signatures
\citep{Sicilia-Aguilar:2006b}. Intermediate-mass candidate transition
disks have only recently been discovered (\citealt{Hernandez:2006};
\citealt{Sicilia-Aguilar:2006a}; \citealt{Hernandez:2007}).  Such
transition disks are especially interesting as possible evolutionary
links between the optically thick Herbig AeBe disk systems and the
older optically thin debris disk systems.

In \citet{Uzpen:2005,Uzpen:2007} we exploited this new mid-IR
capability and used $Spitzer$ to identify stars with 8 $\mu$m infrared
excesses. In \citet{Uzpen:2007}, we identified 230 stars that
exhibited 8 $\mu$m mid-IR excesses out of 4,015 stars investigated in
the \textit{Spitzer} Galactic Legacy Infrared MidPlane Survey
Extraordinaire (GLIMPSE; \citealt{Benjamin:2003}) and MidCourse Space
Experiment (\textit{MSX}; \citealt{Egan:2003}) catalogs. Out of the
230 stars with 8 $\mu$m excesses, 183 are of spectral types O or early
B. The excess for these stars is plausibly explained by free-free
emission. Out of the 47 remaining stars of spectral types B8 or later,
five stars were shown to be false excess sources, and three sources
were not yet confirmed at 24 $\mu$m \citep{Uzpen:2007}. In systems
with mid-IR excesses we modeled the 8 and 24 $\mu$m excesses as simple
blackbodies to characterize the color temperature and fractional
infrared luminosity of the circumstellar disk. The results were
consistent with massive debris disks and/or transitional disks with
disk temperatures ranging between $\approx$200--800~K and fractional
infrared luminosities ($L_{IR}/L_{*}$)
$\approx$10$^{-4}$--10$^{-2}$. However, infrared photometry alone is
not sufficient to unambiguously determine the nature of the
extraphotospheric emission in these sources.  Because of their
scarcity, identifying new transition disks and warm debris disks is an
important step toward understanding the evolutionary path of
circumstellar disks.

The nature of the mid-IR excesses emanating from circumstellar disks
around intermediate-mass stars (2--8\ \mo) varies greatly and
partially depends on age \citep{Zuckerman:2001}. Young stars form
circumstellar disks through the collapse of the initial molecular
cloud as a method to conserve angular momentum
\citep{Shu:1987}. Intermediate-mass pre-main-sequence Herbig AeBe
stars accrete gaseous material via their circumstellar disks as they
evolve towards main-sequence stars (\citealt{Herbig:1960};
\citealt{Waters:1998}).  Circumstellar dust disks, presumably due to
planetesimal collisions, surround some main-sequence stars
(\citealt{Backman:1993}; \citealt{Lagrange:2000}). Circumstellar
gaseous disks that emit excesses owing to free-free emission surround
other main-sequence stars during late main-sequence or
post-main-sequence evolution and are known as classical Be stars
(\citealt{Struve:1931}; \citealt{Porter:2003};
\citealt{McSwain:2005}). Circumstellar disks can surround evolved
giants as well (e.g., \citealt{Jura:1999}; \citealt{Verhoelst:2007};
\citealt{Deroo:2007}). By understanding the nature of the
circumstellar disks, we can better understand how these disks evolve.
 
The two most common stellar contaminants in surveys for stars with
debris disks are Herbig AeBe stars and classical Be stars.  Herbig
AeBe stars are young stars accreting circumstellar material via their
disks (\citealt{Strom:1972}; \citealt{Hillenbrand:1992};
\citealt{Hartmann:1993}). Emission lines and near-infrared excesses
evidence this accretion \citep{Hillenbrand:1992}. \citet{Waters:1998}
suggest the following as criteria for Herbig AeBe stars: (1) an
emission-line A or B star, (2) an infrared excess owing to
circumstellar dust, either hot or cool, and (3) luminosity class
III--V.  As the primordial disk dissipates the emission lines weaken
and the infrared excess decreases, appearing only at longer
wavelengths \citep[e.g.,][]{The:1994}.  Classical Be stars are
typically rapidly rotating B stars with H$\alpha$ emission and an IR
excess from free-free emission (\citealt{Gehrz:1974};
\citealt{Balona:2000}). These stars are variable in both emission-line
strength and luminosity \citep{Dachs:1988}. The circumstellar disks
for classical Be stars are thought to be mechanisms for mass loss,
unlike Herbig AeBe circumstellar disks which facilitate accretion
\citep{Balona:2000}. Classical Be stars are similar to Herbig AeBe
stars in that they both exhibit emission lines due to circumstellar
gas (\citealt{Struve:1931}; \citealt{Herbig:1960}). The greatest
distinction between Herbig AeBe stars and classical Be stars is the
nature of the infrared excess---dust in the former and thermal
bremsstrahlung in the latter.

In this paper we present results of optical spectroscopic analysis for
30 mid-IR excess sources identified in the \textit{Spitzer} GLIMPSE
and $MSX$ catalogs to determine the nature of infrared excess
\citep{Uzpen:2007}. By investigating the nature of the infrared excess
through optical spectral features, we can distinguish between various
types of circumstellar disks that persist around intermediate-mass
stars.  Our goal is to remove classical Be and Herbig AeBe stars from
our sample to identify transition and warm debris disk candidates. In
\S 2 we describe how we obtained new optical echelle spectroscopy.  We
discuss the results of our rotational velocity analysis in \S 3.  In
\S 4 we present our methods for determining effective temperatures. In
\S 5 we discuss the spectra including the H$\alpha$, \ion{Fe}{2}
$\lambda$ 5317 \AA, Paschen, \ion{O}{1}, and \ion{Ca}{2} IR triplet
line profiles and any implications on the nature of the disk from the
lines.  We describe how we estimate stellar ages through main-sequence
fitting utilizing our determined fundamental parameters in \S 6. In \S
7 we discuss how all of the derived parameters provide insight into
the nature of the infrared excess, how we use these parameters to
estimate the contribution of free-free emission associated with our
sources, and how we draw conclusions regarding the evolutionary state
of the circumstellar disk. In \S 8 we describe the individual sources
and classify the infrared excesses. Finally we use spectral classification
schemes for stars with infrared excesses, such as classical
Be stars, with our stars to determine if the classifications
are consistent in \S 9. We also apply our classification schemes
to well-known stellar samples to determine their classification
in \S 9.

\section{Target Selection and Observations}

We observed 30 of 42 stellar sources that exhibited an 8 $\mu$m mid-IR
excess in \citet{Uzpen:2007}, 25 of which have an excess at 8 and 20+
$\mu$m. The purpose of these observations was to identify
circumstellar disk characteristics (e.g., Balmer emission, metallic
emission, forbidden-line emission), and stellar classification
features (e.g, \ion{He}{1}, and \ion{Mg}{2} absorption) based on
optical spectral features to determine stellar and circumstellar
properties. We chose stars with both known and unknown circumstellar
evolutionary state.

We conducted high-resolution, high-signal-to-noise ratio spectroscopic
observations at the Magellan Telescope with the Magellan Inamori
Kyocera Echelle (MIKE; \citealt{Bernstein:2003}) double echelle
spectrograph on the nights of 26--28 September 2006 and 7--8 April
2007.  The spectral coverage was $\simeq$3400--9500 \AA \ with a
resolving power of R=37,000 at 4000 \AA \ and R=29,000 at 6500 \AA \,
measured from arc spectra.  The typical S/N was $\simeq$100:1 in the
stellar continuum.  We conducted standard data reductions in IRAF
including flat fielding with a continuum lamp and wavelength
calibration using a ThAr arc lamp, but no flux calibration was
applied.

We also observed 13 stars to use as rotational velocity standards
listed below the solid horizontal line in Table~\ref{trot}. These
stars were used to validate our technique and compare our determined
values to those in the literature. Seven of these stars had projected
rotational velocities determined in \citet{Royer:2002a}.  Ten of these
stars had projected rotational velocities determined in
\citet{Slettebak:1975}. Four stars had their rotational velocities
determined in both papers and provide a direct comparison between the
two techniques. We will discuss these stars further in \S 3.

\section{Rotational Velocities}

The most common methods for measuring the projected rotational
velocities are through the computation of the first zero of a Fourier
Transform of a given line profile (\citealt{Carroll:1933};
\citealt{Gray:1992}; \citealt{Royer:2005}) or through the measurement
of a line full width half maximum (FWHM) \citep{Slettebak:1975}. The
FWHM of \ion{He}{1} $\lambda$ 4471 \AA, \ion{Mg}{2} $\lambda$ 4481
\AA, and \ion{Fe}{1} $\lambda$ 4476 \AA \ are shown to correlate with
\textit{v} sin \textit{i} in \citet{Slettebak:1975} and is the method
we adopt to measure projected rotational velocity. However, this
method is less reliable at lower rotational velocities because of the
non-linearity of the FWHM/projected rotational velocity relationship,
and the calibration of \citet{Slettebak:1975} has been shown to be
10--12$\%$ lower than measurements made using the Fourier Transform
\citep{Royer:2002a}. In order to improve upon existing linear
relations between \ion{He}{1} $\lambda$ 4471 \AA, \ion{Mg}{2}
$\lambda$ 4481 \AA, and \ion{Fe}{1} $\lambda$ 4476 \AA \ we measure
the FWHM of the rotationally broadened stellar models of
\citet{Munari:2005} of solar metallicity at a resolving power of
R=20,000 from 21,000--7,500~K.

\begin{figure}
\hbox{
    \includegraphics[width=2.0in,angle=90]{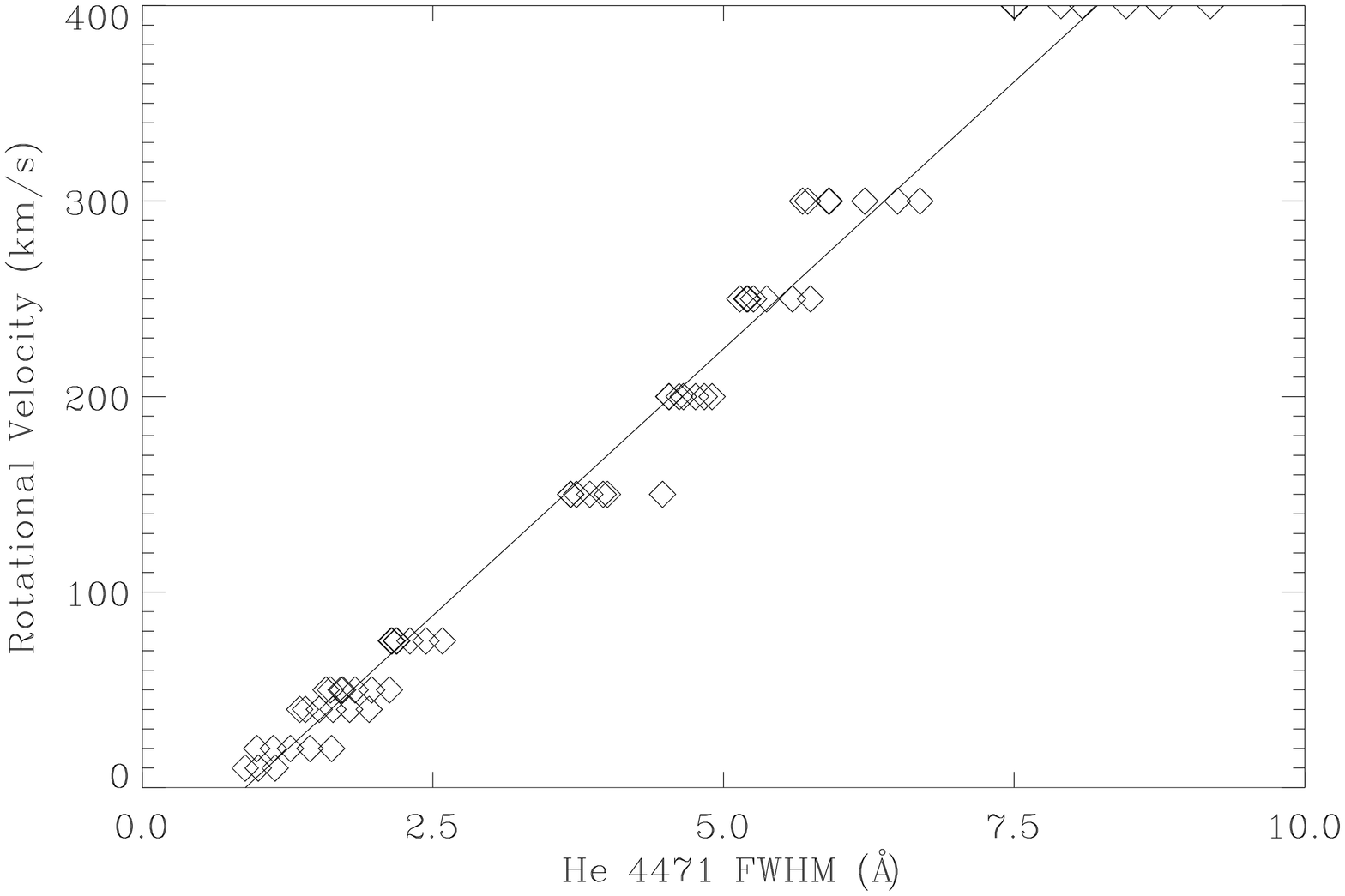}
    \includegraphics[width=2.0in,angle=90]{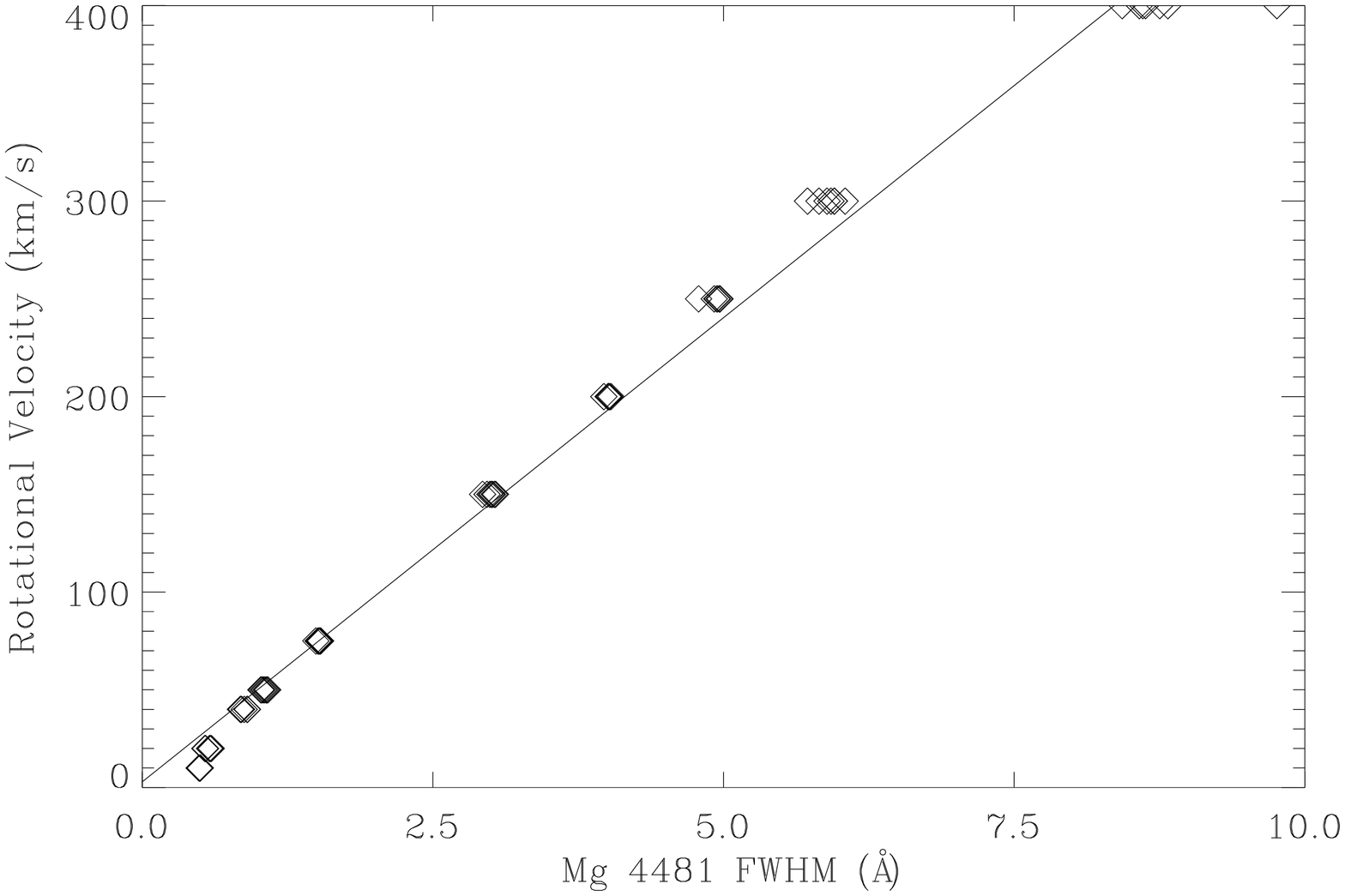}
}
\hbox{
    \includegraphics[width=2.0in,angle=90]{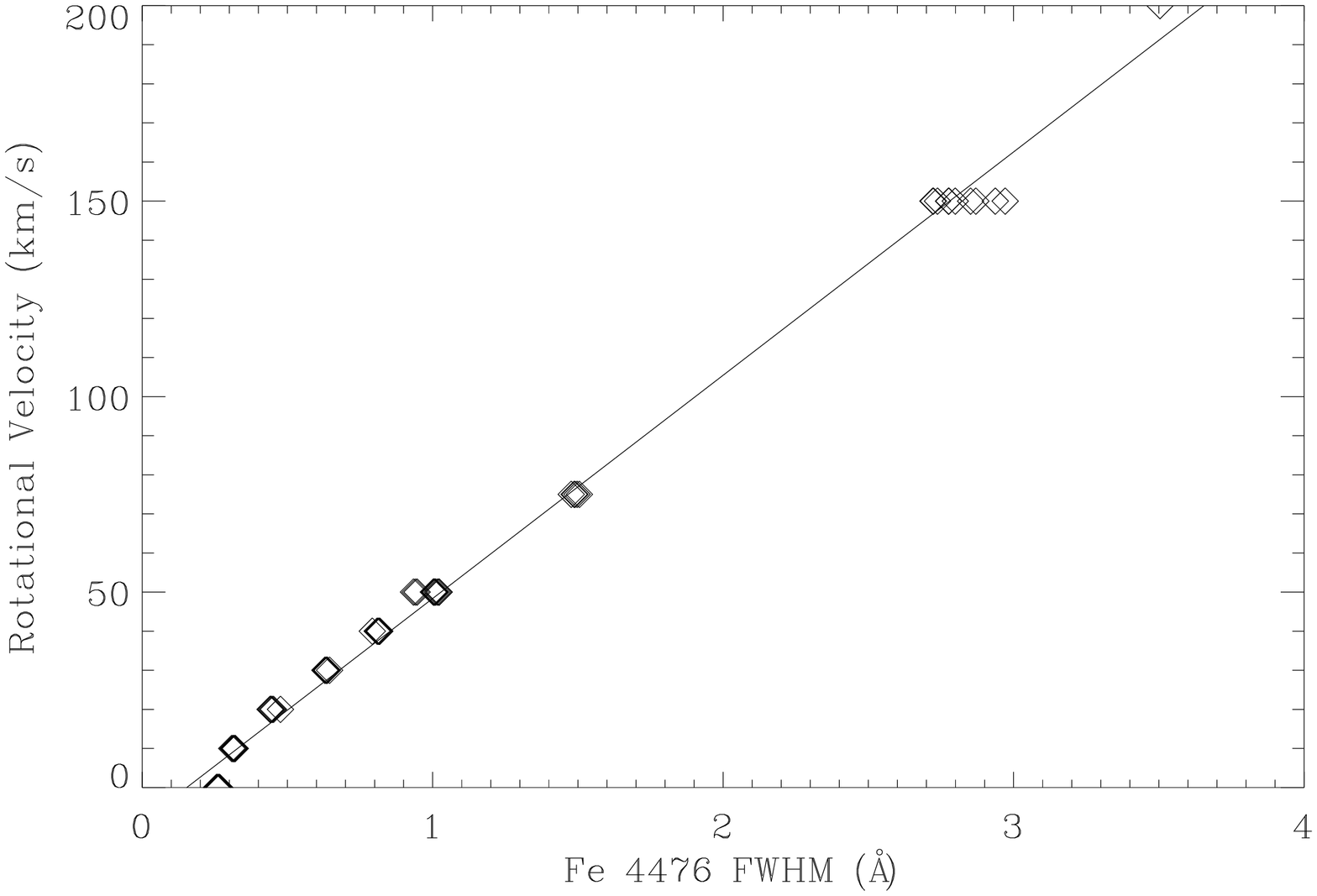}
    \includegraphics[width=2.0in,angle=90]{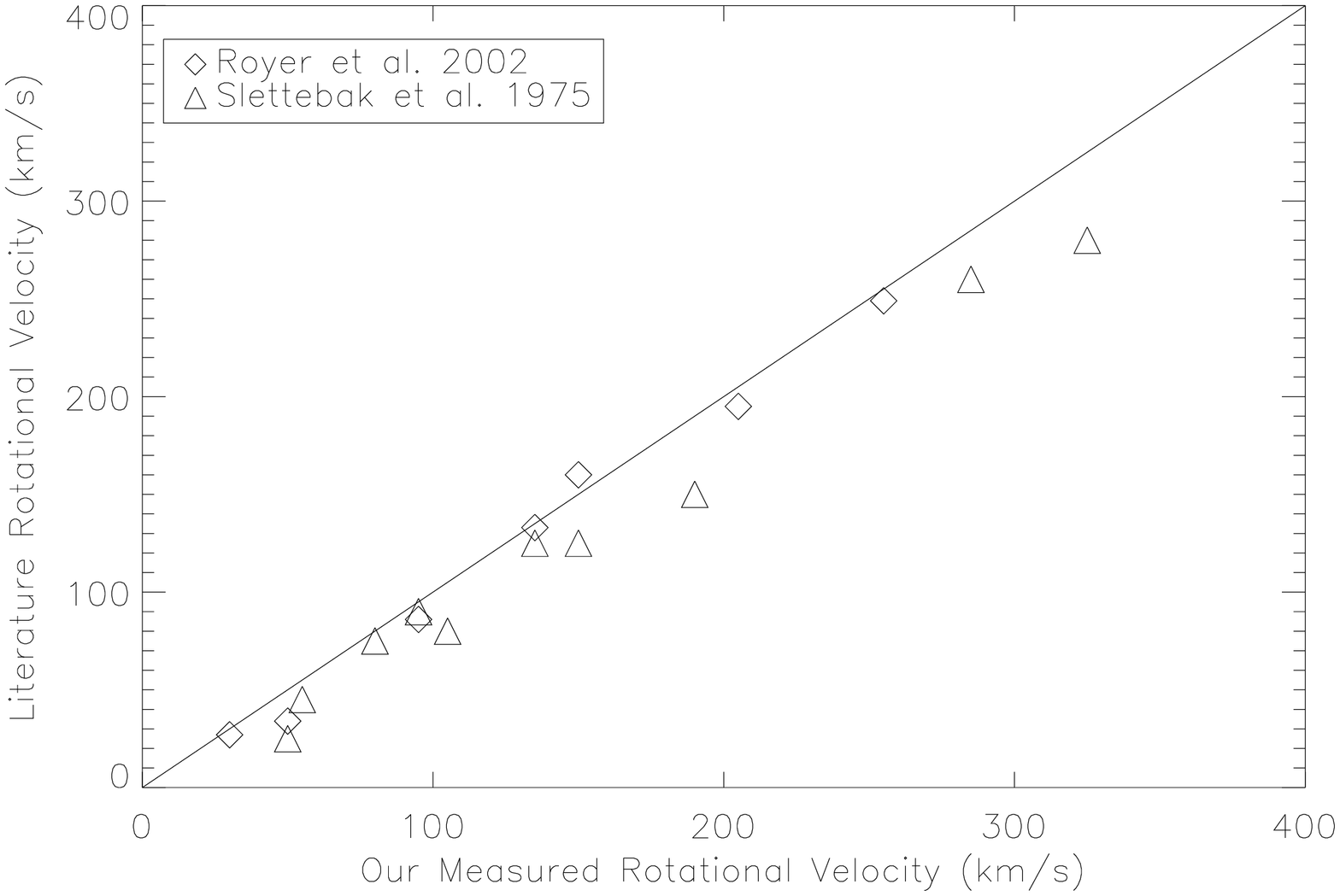}
}
 \caption{(upper left) The relation between projected rotational
 velocity and \ion{He}{1} $\lambda$ 4471 \AA\ FWHM of \citet{Munari:2005} 
 stellar models, where the multiple points at each velocity reflect 
 measurements over a range of effective temperatures. The \ion{He}{1}
 $\lambda$ 4471 \AA\ line contains Stark broadening that results in a
 greater dispersion in the FWHM measurement. The solid line shows the
 linear least squares fit.  (upper right) The relation
 between projected rotational velocity and \ion{Mg}{2} $\lambda$ 4481
 \AA\ FWHM.  (lower left) The relation between projected rotational
 velocity and \ion{Fe}{1} $\lambda$ 4476 \AA\ FWHM.  (lower right) A
 comparison of projected rotational velocities of \citet{Royer:2002a}
 and \citet{Slettebak:1975} versus our rotational velocities. The solid
 line shows the 1:1 relationship in this panel. Our measurements are
 consistent with those of \citet{Royer:2002a} but $\simeq$10$\%$
 larger than those of \citet{Slettebak:1975}, as explained in the
 text.}
    \label{FWHM}
\end{figure}

Figure~\ref{FWHM} shows the relationships between projected rotational
velocity and the FWHM measured over a range of stellar temperatures by
the author B.U. of \ion{He}{1} $\lambda$ 4471 \AA \ (upper left),
\ion{Mg}{2} $\lambda$ 4481 \AA \ (upper right), and \ion{Fe}{1}
$\lambda$ 4476 \AA \ (lower left). The \ion{He}{1} $\lambda$ 4471 \AA
\ line profile includes a Stark broadening component. Since Stark
broadening has a non-linear dependence on temperature this component
degraded the repeatability of consistent measurements that caused a
greater dispersion in the FWHM measurement. We derived the following
equations for a linear least squares fit between the measured stellar
absorption line FWHM and the projected rotational velocity:
\begin{equation}
\textit{v} \ sin \ \textit{i}\ (km\ s^{-1})=54.25 \times He(FWHM)-46.58\ (km\ s^{-1}),
\end{equation}
\begin{equation}
\textit{v} \ sin \ \textit{i}\ (km\ s^{-1})=47.64 \times Mg(FWHM)+2.01\ (km\ s^{-1}),
\end{equation}
\begin{equation}
\textit{v} \ sin \ \textit{i}\ (km\ s^{-1})=57.04 \times Fe(FWHM)-8.73\ (km\ s^{-1}).
\end{equation}

\noindent 
The Mg FWHM relation is not linear below $\simeq$40 km s$^{-1}$ due to
resolution of the line multiplets or above 300 km s$^{-1}$ due to
blends with the \ion{He}{1} $\lambda$ 4471 \AA \ line, while the Fe
FWHM relation is linear at velocities $\geq$10 km s$^{-1}$.

Table~\ref{trot} lists the \textit{v} sin \textit{i} measurements of
our target stars and rotational velocity standards and compares them
with values listed in the literature, when
available. Figure~\ref{FWHM}(lower right) shows a comparison between
our measurements and literature rotational values from
\citet{Slettebak:1975} and \citet{Royer:2002a,Royer:2002b}.  Our
values are consistent with those derived from
\citet{Royer:2002a,Royer:2002b} and are 10--15\% higher than those of
\citet{Slettebak:1975}.  This offset compared to
\citet{Slettebak:1975} is consistent with results from other projected
rotational velocity studies. We estimate our uncertainties in
projected rotational velocity at 10$\%$, given in Table~\ref{trot},
using the FWHM method based on both the dispersion of the models from
the best-fit line and measurement uncertainties in the FWHM.  When two
rotational velocity indicators are present, the reported rotational
velocity is the weighted mean of both measurements, and the
uncertainty is the uncertainty of the mean.

\section{Surface Temperatures}

We determined effective temperatures by comparing model equivalent
width (EW) measurements of \ion{He}{1} $\lambda$ 4009 \AA, \ion{He}{1}
$\lambda$ 4471 \AA, \ion{Mg}{2} $\lambda$ 4481 \AA, \ion{Fe}{2}
$\lambda$ 4233 \AA, and \ion{Ca}{1} $\lambda$ 4227 \AA \ of
\citet{Munari:2005} R=20,000 solar metallicity models to our spectral
measurements. For our spectra we used the models in the range 7,000~K
$\leq$ T$_{eff}$ $\leq$ 21,000~K, with log \textit{g}=4.5 for
T$_{eff}$ $<$ 10,000~K, and log \textit{g}=4.0 for T$_{eff}$ $\geq$
10,000~K. We used models with rotational broadening between 0 km
s$^{-1}$ and 400 km s$^{-1}$. The effective temperature step size
varies from $\Delta$T$_{eff}$=1000~K for 12,000~K $<$ T$_{eff}$ $\leq$
21,000~K, $\Delta$T$_{eff}$=500~K for 10,000~K $\leq$ T$_{eff}$ $<$
12,000~K and $\Delta$T$_{eff}$=250~K for 7,000~K $\leq$ T$_{eff}$ $<$
10,000~K.

\begin{figure}
\hbox{
    \includegraphics[width=2.0in,angle=90]{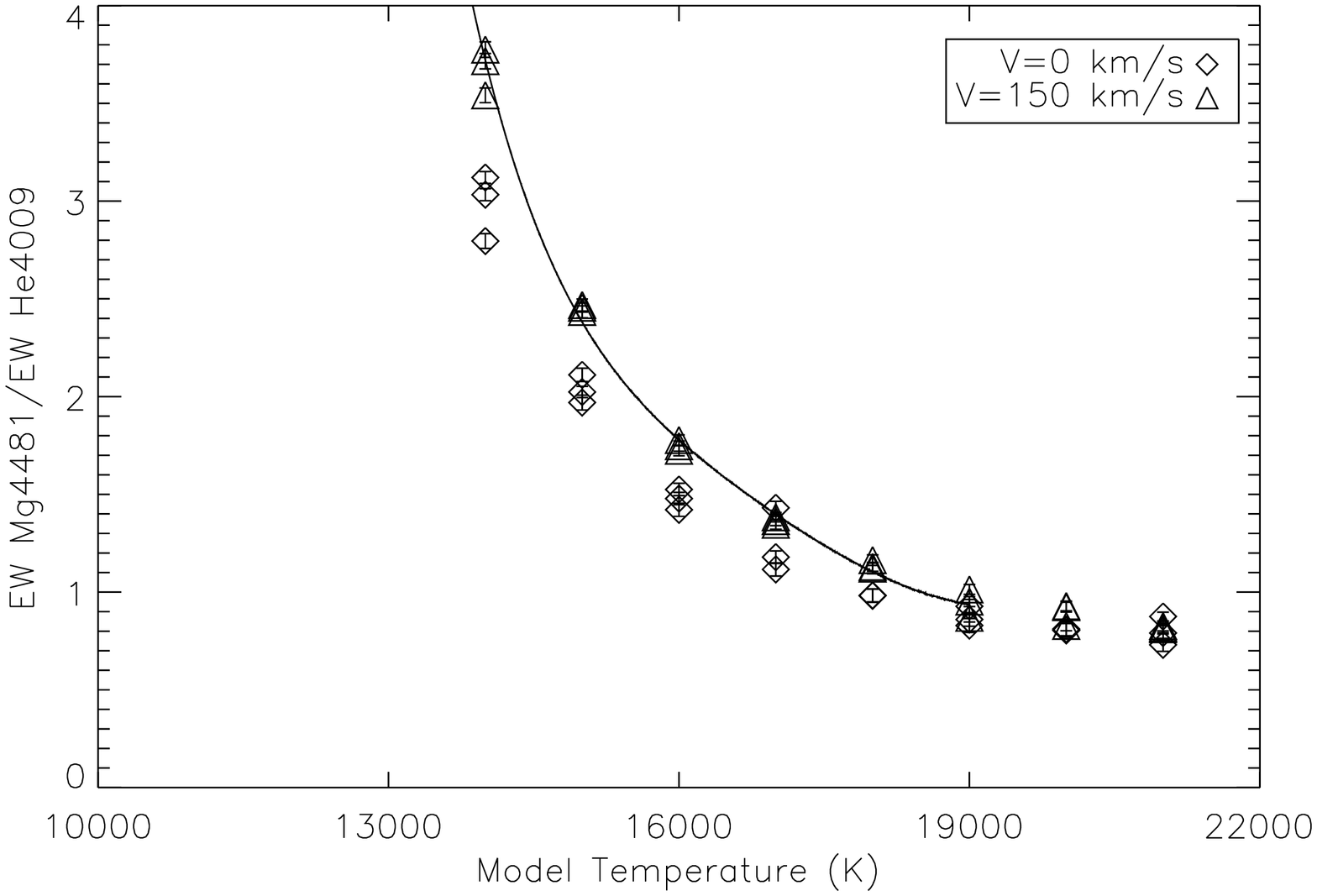}
    \includegraphics[width=2.0in,angle=90]{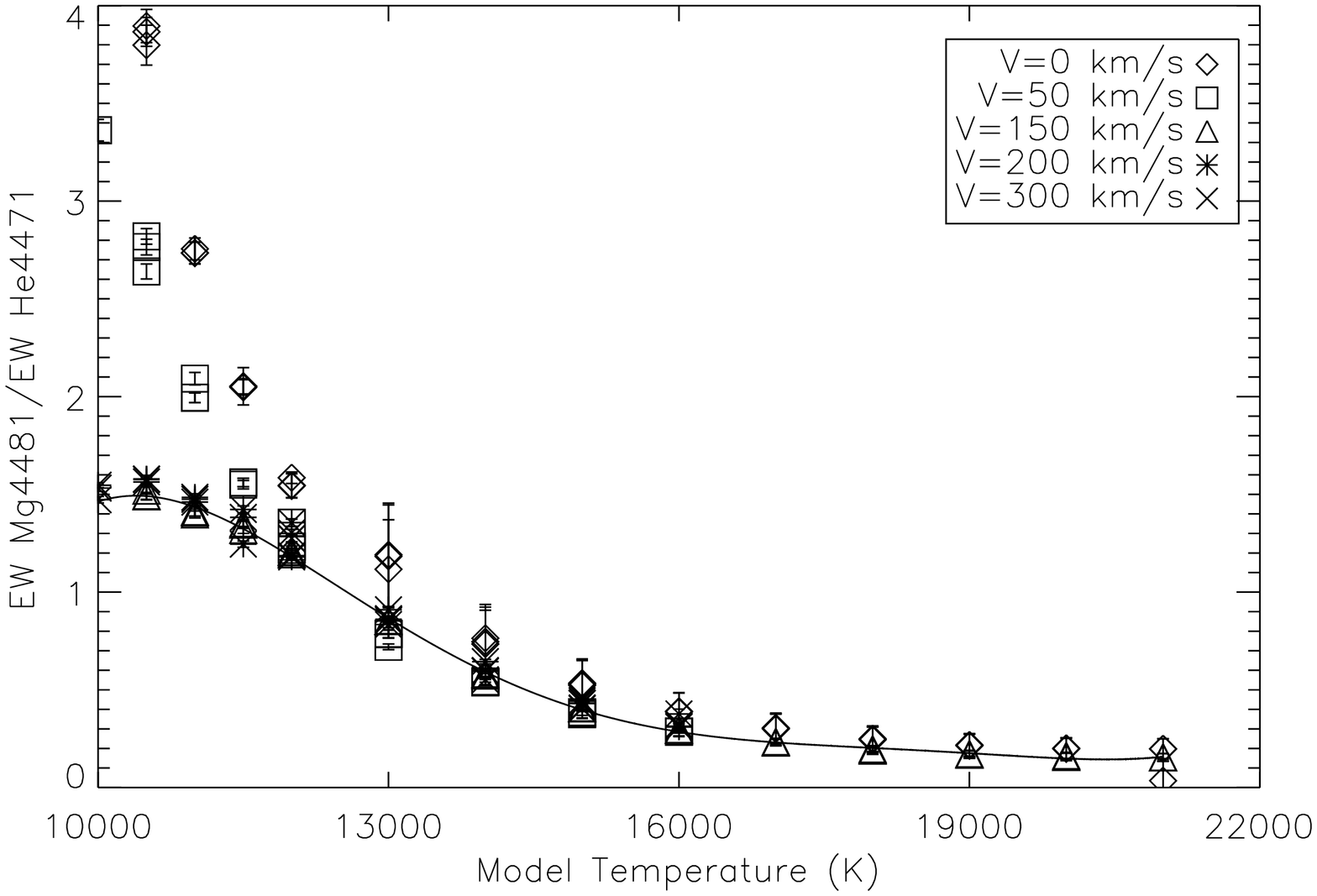}
}
\hbox{
    \includegraphics[width=2.0in,angle=90]{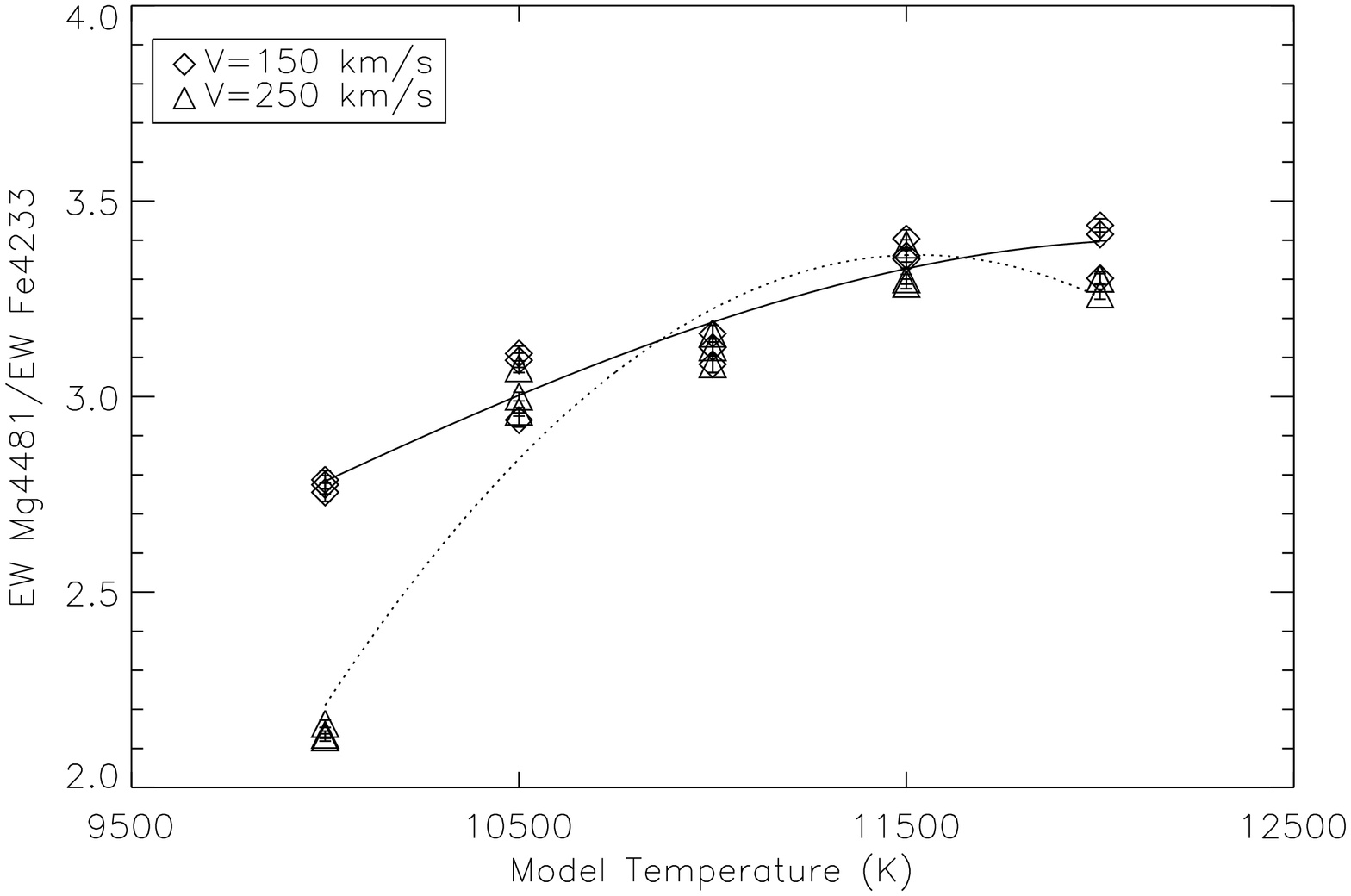}
    \includegraphics[width=2.0in,angle=90]{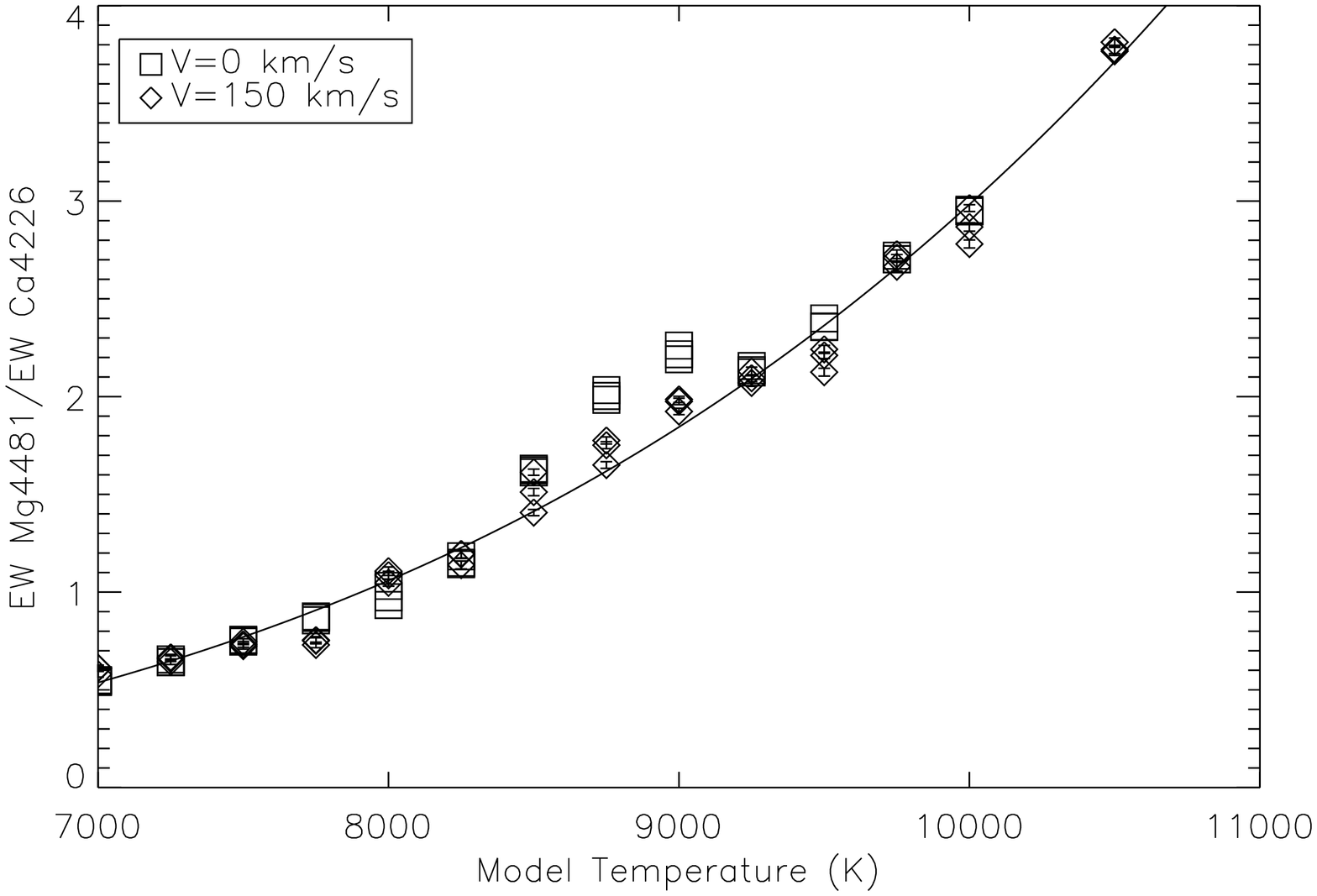}
}
 \caption{(upper left) The relation between model ratios of EW
 \ion{Mg}{2} $\lambda$ 4481 \AA/EW \ion{He}{1} $\lambda$ 4009 \AA\ and
 effective temperature. The multiple data points at each
 abscissa/ordinate are due to multiple measurements of the same model
 and shows the systematic errors associated with this
 relationship. The solid curve shows the best fit relation of EW ratio
 to effective temperature for the models with 150 km s$^{-1}$
 projected rotational velocity.  (upper right) The relation between EW
 \ion{Mg}{2} $\lambda$ 4481 \AA/EW \ion{He}{1} $\lambda$ 4471 \AA\ and
 effective temperature.  (lower left) The relation between EW
 \ion{Mg}{2} $\lambda$ 4481 \AA/EW \ion{Fe}{1} $\lambda$ 4233 \AA\ and
 effective temperature. This relationship depends upon rotational
 velocity, and the dotted curve shows the best fit relation of EW
 ratio to effective temperature at 250 km s$^{-1}$. (lower right) The
 relation between EW \ion{Mg}{2} $\lambda$ 4481 \AA/EW \ion{Ca}{1}
 $\lambda$ 4227 \AA\ and effective temperature.}
    \label{temp}
\end{figure}

The equivalent width ratios of \ion{Mg}{2} $\lambda$ 4481
\AA/\ion{He}{1} $\lambda$ 4009 \AA, \ion{Mg}{2} $\lambda$ 4481
\AA/\ion{He}{1} $\lambda$ 4471 \AA, \ion{Mg}{2} $\lambda$ 4481
\AA/\ion{Fe}{2} $\lambda$ 4233 \AA, and \ion{Mg}{2} $\lambda$ 4481
\AA/\ion{Ca}{1} $\lambda$ 4227 \AA \ form monotonic relations
dependent on both rotational velocity and temperature spanning
T=21,000--7,000~K, as shown in Figure~\ref{temp}. The solid curves in
Figure~\ref{temp} are polynomials representing the best-fit relation
between the EW ratio and stellar effective temperature for the models
with projected rotational velocity of 150 km s$^{-1}$.  The multiple
data points at each abscissa/ordinate are due to multiple measurements
of the same model and show the magnitude of the random uncertainties
associated with measuring the EW.  The dotted curve in
Figure~\ref{temp} (lower left panel) represents the best fit
polynomial between the EW ratio of \ion{Mg}{2} $\lambda$ 4481
\AA/\ion{Fe}{2} $\lambda$ 4233 \AA\ for models with rotational
velocity of 250 km s$^{-1}$ (triangles). We chose Mg and He because
these lines are prominent in B and early A stars, and He strength
decreases rapidly in late B stars. Fe lines start to appear in early A
stars and increase in strength toward later spectral types. Ca lines
increase in strength and number starting at late A stars and
continuing through F stars. At rotational velocities $>$ 150 km
s$^{-1}$ the \ion{Mg}{2} $\lambda$ 4481 \AA/\ion{He}{1} $\lambda$ 4009
\AA\ and \ion{Mg}{2} $\lambda$ 4481 \AA/\ion{He}{1} $\lambda$ 4471
\AA\ ratios are independent of rotational velocity and thus provide
the best estimates of T$_{eff}$. At rotational velocities $<$ 150 km
s$^{-1}$ the \ion{Mg}{2} $\lambda$ 4481 \AA/\ion{Ca}{1} $\lambda$ 4227
\AA\ ratios are independent of rotational velocity and provide the
best estimates of T$_{eff}$. The \ion{Mg}{2} $\lambda$ 4481
\AA/\ion{Fe}{2} $\lambda$ 4233 \AA\ relationship depends upon
rotational velocity over all velocity ranges, unlike the other EW
ratios.  Therefore, this ratio is not generally as sensitive to
effective temperature, but we use it over a limited temperature range,
10,000--12,000~K, where other lines ratios are not useful.

We determined the effective temperature utilizing one or more
equivalent width ratios in Figure~\ref{temp} for our target stars and
present them in Table~\ref{ttemp}, along with spectral types from the
literature. The uncertainties on effective temperatures were
determined by adding in quadrature the uncertainty derived from
equivalent width fitting with one-half the model step size. The former
uncertainty component approximately accounts for the random
uncertainty in measuring the equivalent width, which is dominated by
the uncertainty in the definition of the stellar continuum. In
general, most derived effective temperature measurements are
consistent with the stellar spectral classification, within
uncertainties.

\begin{figure}
    \includegraphics[width=4.0in,angle=90]{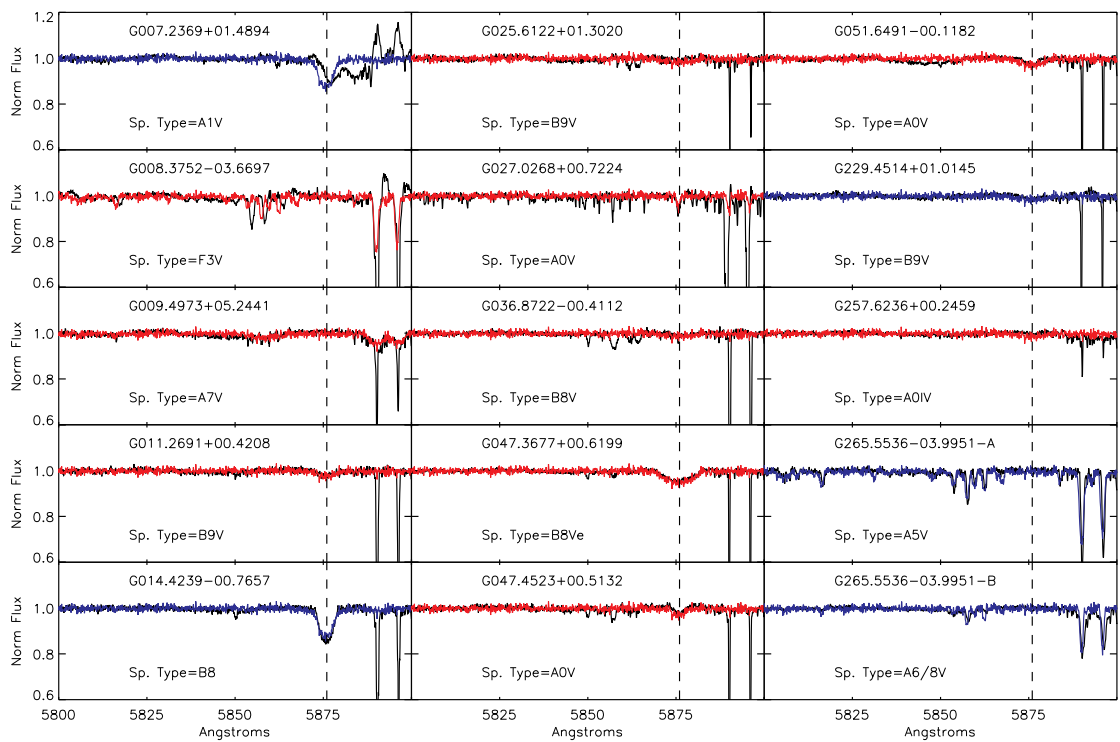}
    \caption{\ion{He}{1} $\lambda$ 5876 \AA\ profiles (denoted by the
dashed vertical line) of mid-infrared excess stars overplotted with
best fit \citet{Munari:2005} stellar models.  Red curves show the
stellar models for stars in which the measured stellar temperature
falls in between the two nearest stellar models. Blue curves show the
stellar models for stars in which the measured temperature does not
fall in between the nearest stellar models. G014.4239-00.7657,
G229.4514+01.0145, G265.5536-03.9951-A and B are overplotted by blue
curves indicating the discrepancy between the \ion{He}{1} line
and measured temperature.}
    \label{he1}
\end{figure}

\begin{figure}
    \includegraphics[width=4.0in,angle=90]{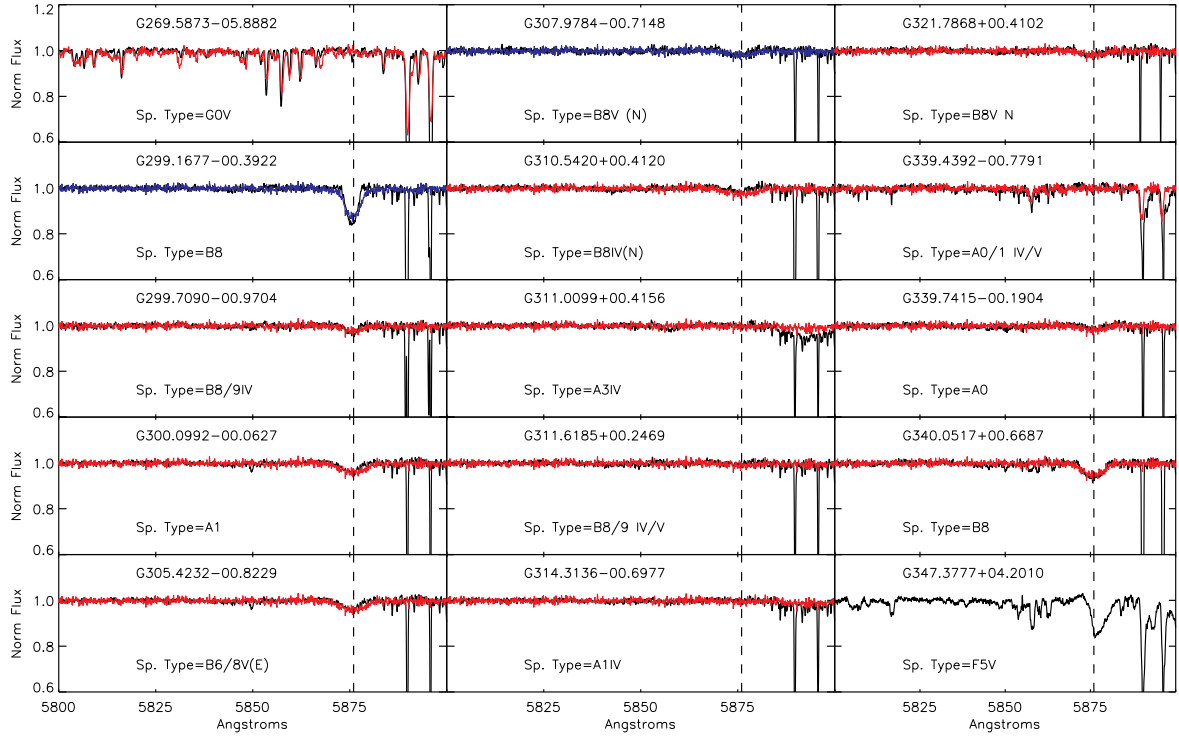}
    \caption{\ion{He}{1} 5876 \AA\ profiles of mid-infrared excess
      stars overplotted with best fit \citet{Munari:2005} stellar
      models using the same symbolism as
      Figure~\ref{he1}. G299.1677-00.3922 and G307.9784-00.7148 are
      overplotted by blue curves indicating the discrepancy
      between the \ion{He}{1} line and measured temperature.}
    \label{he2}
\end{figure}

We found several late B stars to have effective temperatures hotter
than their spectral classifications would indicate. Such an effect
might result if, for example, some stars were deficient in heavy
metals, resulting in smaller metal-to-helium EW ratios in
Figure~\ref{temp} and systematically higher indicated effective
temperatures.  In order validate the derived temperatures and test
whether this approach is sensitive to our assumption of solar
metallicity, we compare in Figures~\ref{he1} and \ref{he2} our
spectra with the \citet{Munari:2005} stellar models, degraded to
similar signal-to-noise ratios, in the region near \ion{He}{1}
$\lambda$ 5876 \AA.  Figures~\ref{he1} and \ref{he2} show the
spectral region near \ion{He}{1} $\lambda$ 5876 \AA\ (denoted by the
dashed vertical line) with the best fitting stellar model over-plotted
in either red or blue. The narrow strong lines to the right of
\ion{He}{1} $\lambda$ 5876 \AA\ are the \ion{Na}{1} $\lambda$ 5890/5896
\AA\ doublet, generally of interstellar origin except in the later
type stars which have an atmospheric component.  Red curves (24
out of 30 stars) show the best fit stellar model and indicate that the
temperature measured from EW ratios fell within one model temperature
step size from the best fit model. Blue curves (6 out of 30
stars) show the best fit stellar temperature model and indicate that
the model deviated by more than one model temperature step size from
the temperature determined from EW ratios.  For G014.4239-00.7657 the
best fit stellar model was T$_{eff}$=18,000~K as it is for
G299.1677-00.3922 and G007.2369+01.4894. G014.4239-00.7657 and
G299.1677-00.3922 have the hottest measured T$_{eff}$ and our
relationship for \ion{Mg}{2} $\lambda$ 4481 \AA/\ion{He}{1} $\lambda$
4471 \AA\ is not very sensitive in this temperature regime. For both
stars the \ion{Mg}{2} $\lambda$ 4481 \AA/\ion{He}{1} $\lambda$ 4471
\AA\ derived temperature is $\sim$600 K lower than \ion{Mg}{2}
$\lambda$ 4481 \AA/\ion{He}{1} $\lambda$ 4009 \AA\ ratio pushing the
weighted mean to a lower temperature. If we ignore the \ion{Mg}{2}
$\lambda$ 4481 \AA/\ion{He}{1} $\lambda$ 4471 \AA\ temperature, the
\ion{Mg}{2} $\lambda$ 4481 \AA/\ion{He}{1} $\lambda$ 4009 \AA\
temperature, with uncertainties, falls between the 17,000 and 18,000~K
models, as is found from the best fit models. For the remainder of
this paper we will use T$_{eff}$=16,900$\pm$800 K for
G014.4239-00.7657 and T$_{eff}$=16,450$\pm$1030 K for
G299.1677-00.0392. G007.2369+01.4894 is a known Herbig AeBe star, see
\S 8 for more details. The \ion{Na}{1} D lines in emission and strong
broad \ion{He}{1} $\lambda$ 5876 \AA\ absorption are indicators of
circumstellar gas. This star exhibits line profile variations
consistent with circumstellar activity (\citealt{Catala:1989};
\citealt{Pogodin:1994}). The variable lines and circumstellar gas
explain the discrepancy between temperatures measured from the
\ion{He}{1} $\lambda$ 5876 \AA\ line fitting and EW ratios.
G229.4514+01.0145 exhibits stronger \ion{He}{1} $\lambda$ 5876 \AA\
absorption than the \ion{Mg}{2} $\lambda$ 4481 \AA/\ion{He}{1}
$\lambda$ 4471 \AA\ ratio temperature of T$_{eff}$=10,220$\pm$390~K
implies.  This is a suspected Herbig AeBe star, see \S 8, and, like
G007.2369+01.4894, it is better fit with a hotter model,
T$_{eff}$=12,000~K.  G265.5536-03.9951-A and B are binaries and their
interactions may effect their spectra. For G265.5536-03.9951-A we
found a better fit with a cooler model of 6,500~K rather than the
measured T$_{eff}$ of 7,040$\pm$170~K, while G265.5536-03.9951-B is
better fit with a hotter model of 8,250~K rather than the measured
T$_{eff}$ of 7,660$\pm$140~K.  G307.9784-00.7148 was also better fit
with a cooler stellar model, T$_{eff}$=12,000~K rather than the EW
ratio measured T$_{eff}$=13,220$\pm$520~K. With the exception of the
probable pre-main-sequence stars, the best fit models all lie within
two stellar models of the measured temperatures based on EW
ratios. The sources with discrepant model fit temperatures and EW
ratio measured temperatures may show indication of circumstellar gas,
see \S 5.  In summary, the six stars that show small discrepancies
between temperatures derived from He $\lambda$ 5876 \AA\ fits and EW
ratios are explainable in most cases by circumstellar material and
activity rather than metallicity enhancements or deficiencies.

The discrepancy between literature spectral types and derived
temperature for some objects in Table~\ref{ttemp} may be in part due
to differences between observational parameters and theoretical
stellar models.  For example, under the MK system, the definition of a
B8 main-sequence star requires the \ion{He}{1} $\lambda$ 4471 \AA\ EW
to be equal to the \ion{Mg}{2} $\lambda$ 4481 \AA\ EW.  Using this MK
system definition of a B8 main-sequence star and \citet{Munari:2005}
models we derive a temperature of $\simeq$12,600~K through
interpolation of the model grids.  However, the observational
effective temperature for a B8 main-sequence star is 11,400~K
\citep{Cox:2000}.  We do not present revised spectral types for the
late B stars in Table~\ref{ttemp} due to the small systematic
temperature difference between optical classification criteria and
theoretical models.

\section{Spectral Features}

Metallic and hydrogen emission lines in the spectra of both classical
Be and Herbig AeBe stars have been well studied (e.g.,
\citealt{Finkenzeller:1984}; \citealt{Andrillat:1990}).  For example,
the lines of \ion{He}{1} $\lambda$ 5876 \AA\ , \ion{Na}{1} D,
H$\alpha$, \ion{Fe}{2}, \ion{O}{1}, and \ion{Ca}{2} IR triplet have
been studied as indicators of circumstellar material around Herbig
AeBe stars (\citealt{Finkenzeller:1984}; \citealt{Hamann:1992};
\citealt{Bohm:1995}; \citealt{Hernandez:2004}).  The spectra of
classical Be stars exhibit, in addition to hydrogen emission,
\ion{Ca}{2} IR triplet, \ion{O}{1} $\lambda$ 8446 and \ion{O}{1}
$\lambda$ 7772 \AA, and \ion{Fe}{2} in emission
(\citealt{Andrillat:1990}; \citealt{Jaschek:1993};
\citealt{Hanuschik:1996}). By comparing line strengths with stellar
models it is possible to identify emission and absorption lines
arising in circumstellar material and identify the type of
circumstellar disk.

\subsection{Hydrogen Lines}

\subsubsection{Balmer Lines}

\begin{figure}
    \includegraphics[width=4.0in,angle=90]{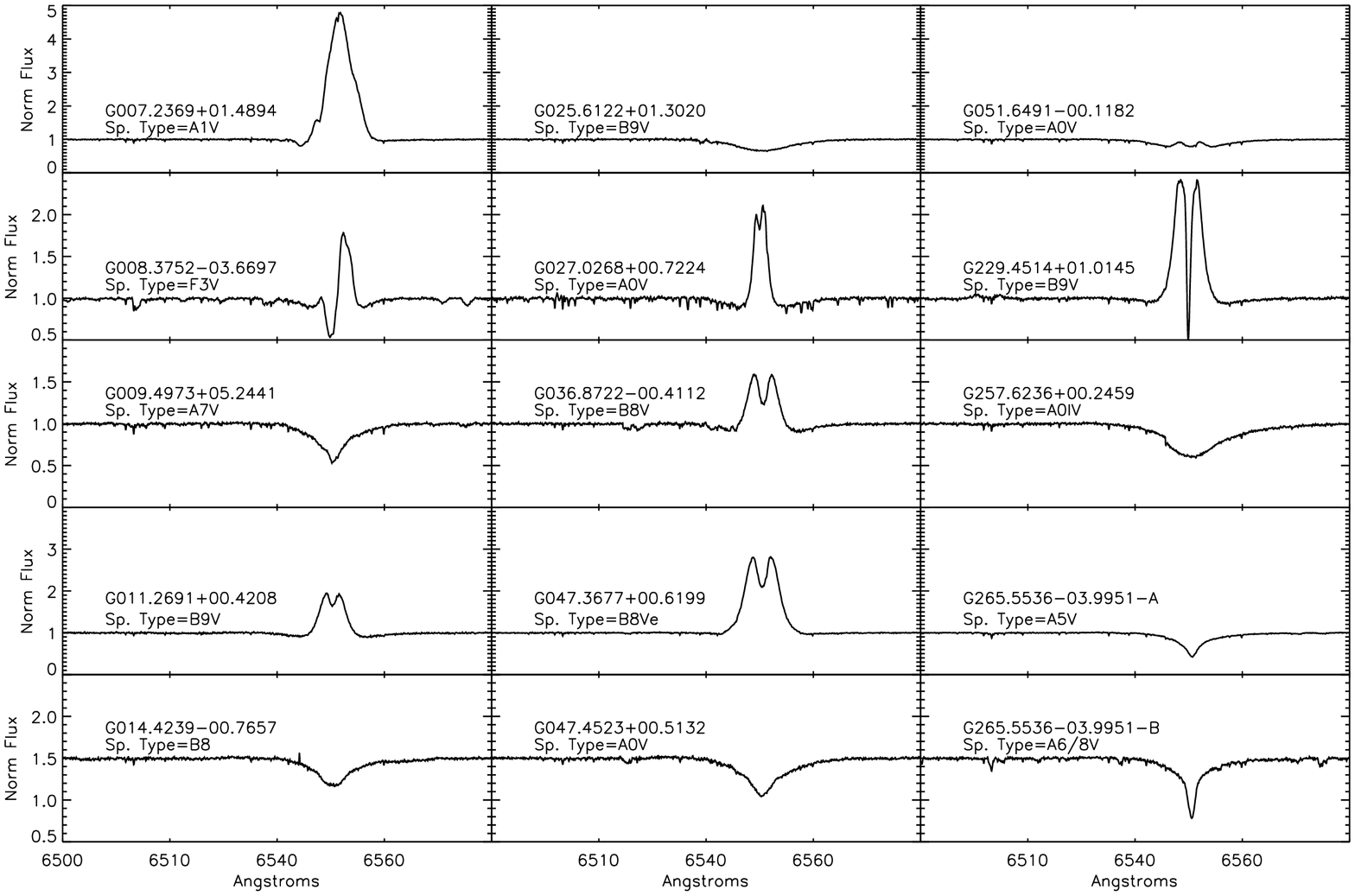}
    \caption{H$\alpha$ profiles of mid-infrared excess stars. The
    majority of our sources exhibit a double-peaked emission profile.}
    \label{ha1}
\end{figure}

\begin{figure}
    \includegraphics[width=4.0in,angle=90]{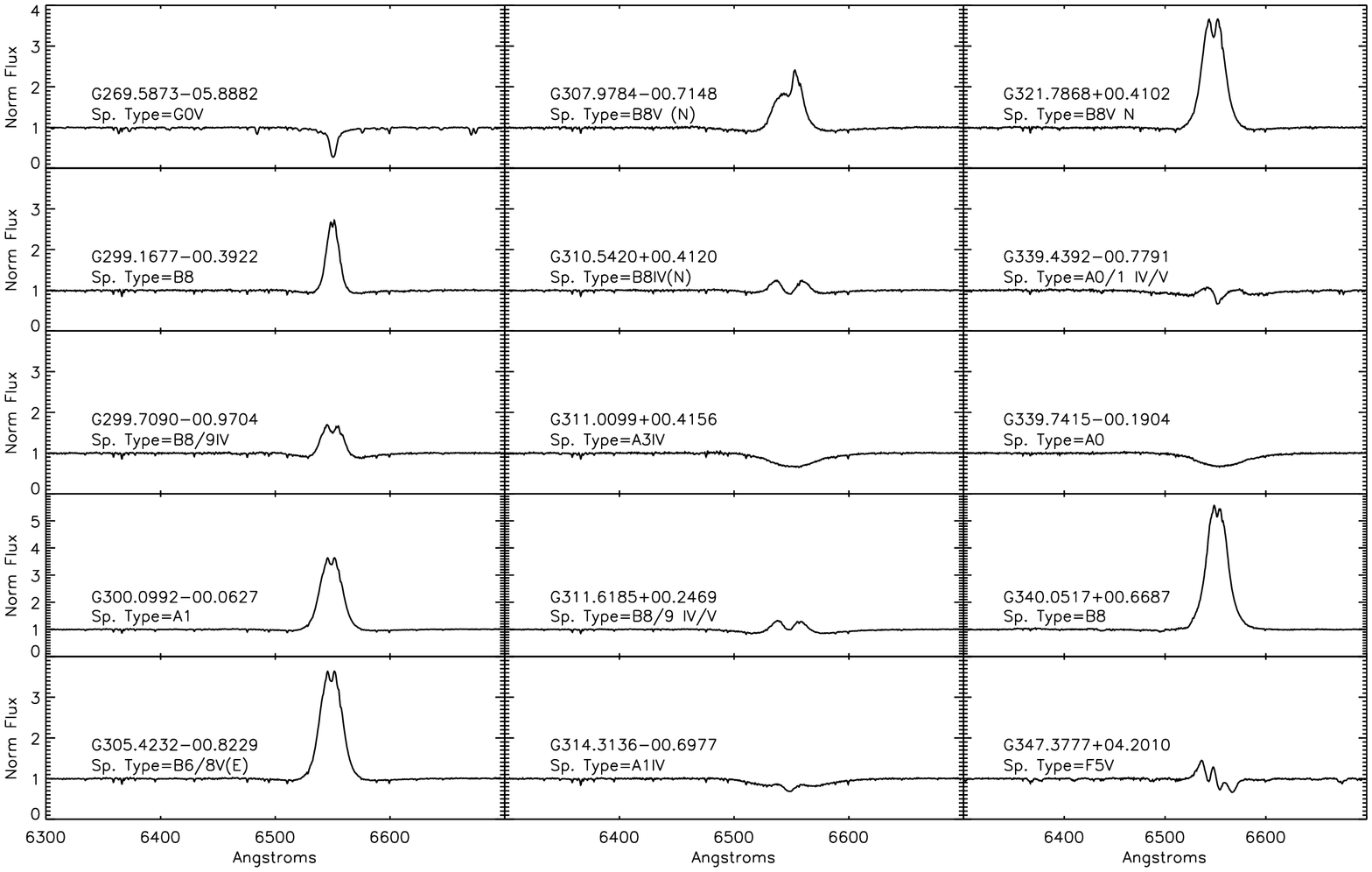}
    \caption{H$\alpha$ profiles of mid-infrared excess stars. The
    majority of our sources exhibit a double-peaked emission profile.}
    \label{ha2}
\end{figure}

The nature of an infrared excess can, in part, be determined through
H$\alpha$ spectroscopy.  By definition, classical Be stars and Herbig
AeBe stars exhibit Balmer emission lines owing to their circumstellar
gaseous disk. In contrast, debris disks do not exhibit any Balmer
emission lines and are devoid of significant amounts of circumstellar
gas \citep{Lagrange:2000}. We measured the H$\alpha$ equivalent width
for our targets and noted the five H$\alpha$ profile types in
Table~\ref{tha}. These profile types include single-peaked emission
profiles (S), double-peaked emission profiles (D), absorption profiles
(A), P-Cygni emission profiles (P), and inverse P-Cygni emission
profiles (IP). For complex profiles, double-peaked emission, or
P-Cygni features, we fit multiple Gaussian profiles to the H$\alpha$
feature and report the sum of the EWs in Table~\ref{tha}.  We
performed the fitting process manually five times, and we adopt the
rms dispersion of the summed EWs as our estimate of the 1$\sigma$
uncertainty on EW(H$\alpha$). We then utilized \citet{Munari:2005}
stellar model H$\alpha$ absorption measurements to correct the
H$\alpha$ equivalent width for underlying absorption in order to
estimate the true EW of the emission. This measure,
EW(H$\alpha$$_{corr}$), is then used for the remainder of this paper.
Normalized H$\alpha$ spectra are shown in Figures~\ref{ha1} \&
\ref{ha2}. One-third of the sample, ten stars, exhibit H$\alpha$
absorption profiles.  Of the remaining twenty stars that exhibit
H$\alpha$ emission, only one star exhibits single-peaked emission, one
star exhibits inverse P-Cygni, and one star exhibits P-Cygni profiles.
The majority of the stars that exhibit Balmer emission show a
double-peaked emission profile (17 stars).  This result is not
surprising given that the majority of both classical Be and Herbig
AeBe stars exhibit double-peaked H$\alpha$ profiles
(\citealt{Porter:2003}; \citealt{Finkenzeller:1984}).

\subsubsection{Paschen Lines}

\begin{figure}
    \includegraphics[width=4.0in,angle=90]{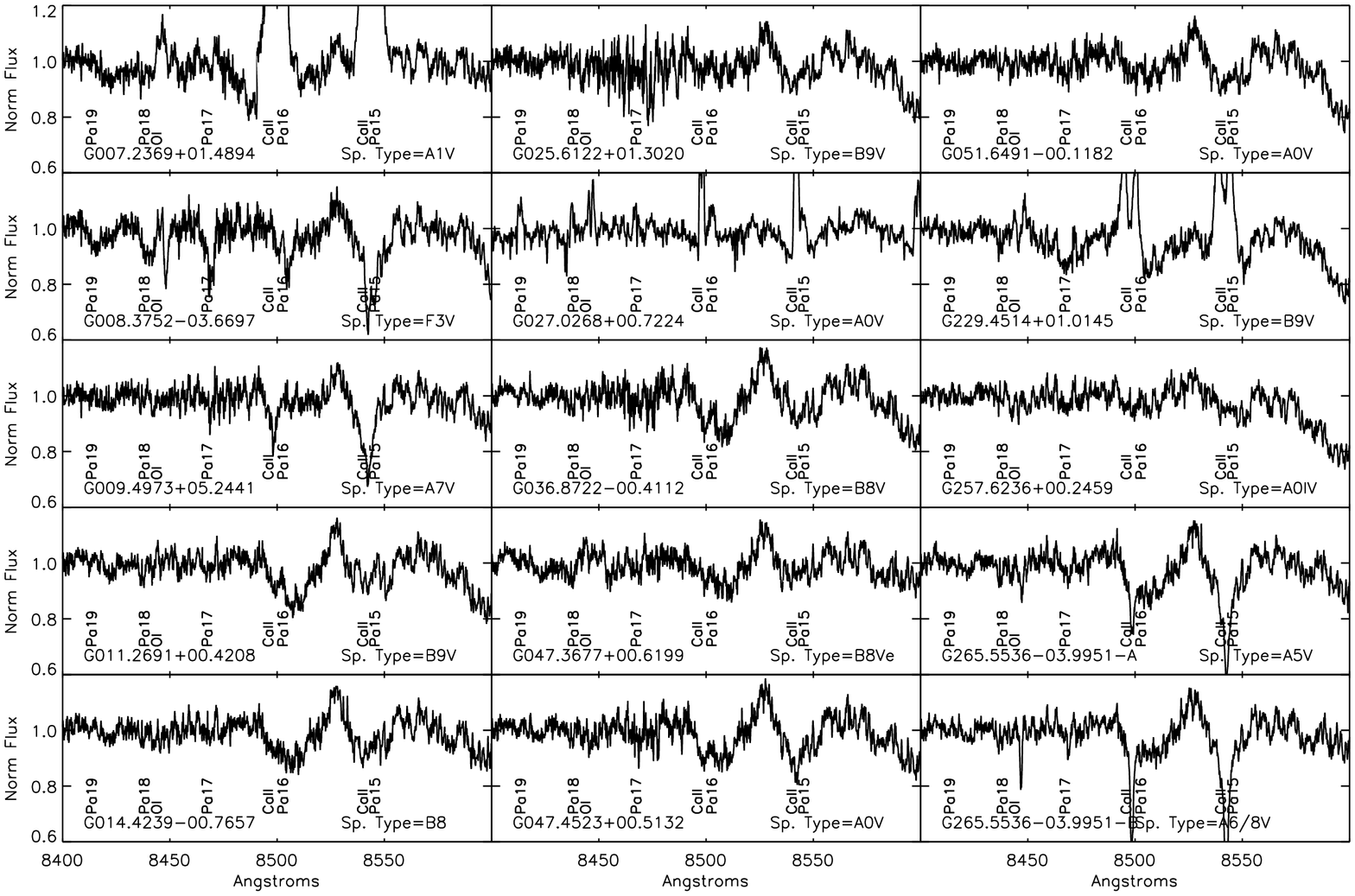}
    \caption{The spectral region 8400--8600 \AA\ showing several
    Paschen lines, Pa15-19, \ion{Ca}{2} $\lambda$ 8498 and 8542 \AA\,
    and \ion{O}{1} $\lambda$ 8446 \AA.}
    \label{pa1}
\end{figure}

\begin{figure}
    \includegraphics[width=4.0in,angle=90]{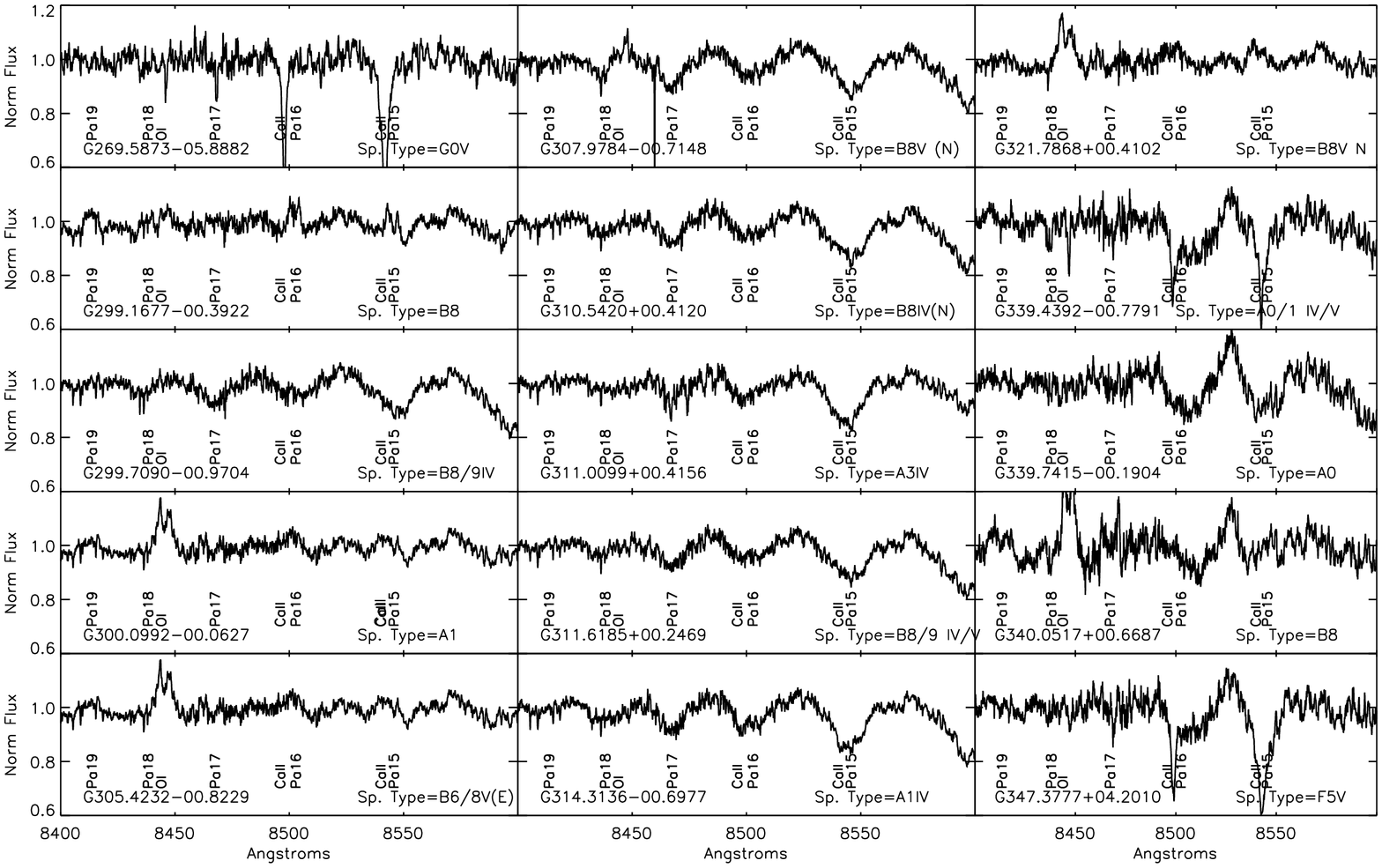}
    \caption{The spectral region 8400--8600 \AA\ showing several
    Paschen lines, Pa15-19, \ion{Ca}{2} $\lambda$ 8498 and 8542 \AA\,
    and \ion{O}{1} $\lambda$ 8446 \AA.}
    \label{pa2}
\end{figure}

Both early- and late-type classical Be stars exhibit Paschen lines in
emission with the later type only exhibiting weak emission not
reaching the continuum \citep{Andrillat:1990}.  Herbig AeBe stars have
been found to exhibit Paschen emission, although the Paschen lines are
significantly weaker than the metallic Fe and Ca emission lines in the
red spectral region \citep{Catala:1986}. Figures~\ref{pa1} and
\ref{pa2} shows spectral regions 8400--8600 \AA\ and Paschen lines
Pa15--19.  Lower order Paschen lines either lie off our spectral
coverage or fall in regions of poor atmospheric transparency. Very few
stars exhibit Paschen emission: G027.0268+00.7224, G299.1677-00.3992,
G321.7686+00.4102, and G340.0517+00.6687. Table~\ref{tspec} lists the
spectral features used for confirmation of circumstellar gas and
determination of the nature of the circumstellar disk. Paschen
emission is listed as column 3 in Table~\ref{tspec}.  For classical Be
stars, emission from metallic lines of \ion{Fe}{2}, \ion{O}{1}
$\lambda$ 8446, \ion{O}{1} $\lambda$ 7772 \AA\, and \ion{Ca}{2} is
found only when the Paschen lines are in emission
(\citealt{Andrillat:1990}; \citealt{Jaschek:1993}).  Herbig AeBe stars
may exhibit \ion{O}{1}, \ion{Fe}{2}, and \ion{Ca}{2} emission with or
without Paschen emission \citep{Catala:1986}.

\subsection{Metallic Lines}

\subsubsection{\ion{Ca}{2} IR Triplet}

\begin{figure}
    \includegraphics[width=4.0in,angle=90]{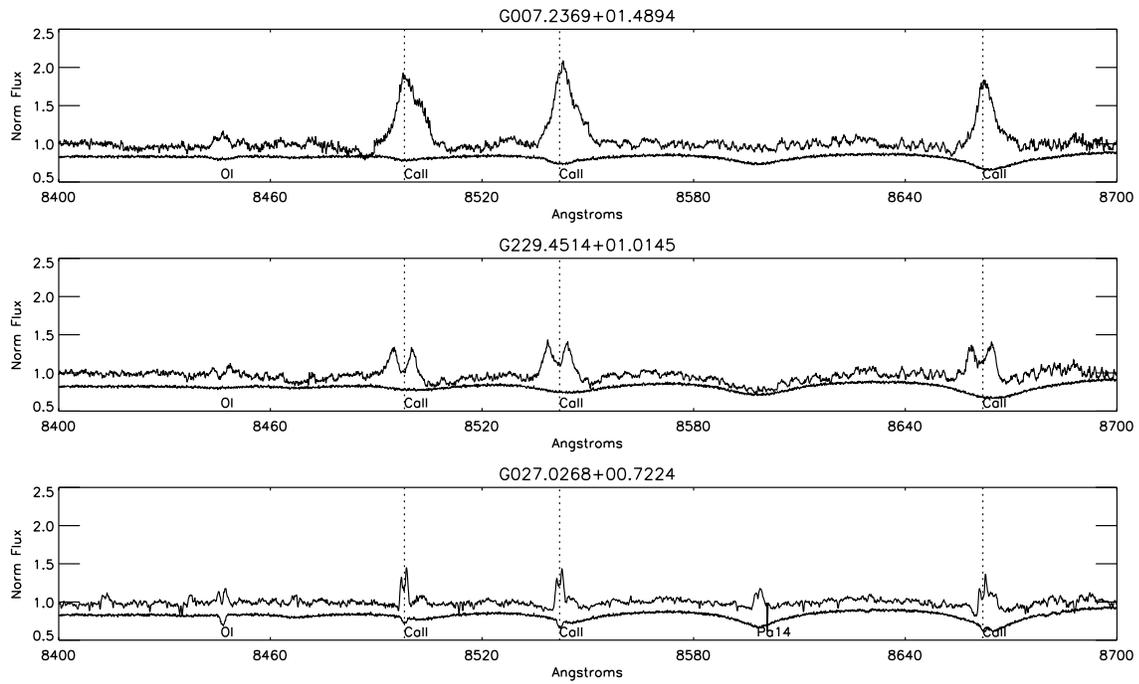}
    \caption{The spectral region 8400--8700 \AA\ showing 
    the three stars within our sample that exhibit \ion{Ca}{2}
    emission. \citet{Munari:2005} stellar models of nearest
    measured temperature are shown below. All three
    sources also exhibit \ion{O}{1} emission, and G027.0268+00.7224
    exhibits Paschen emission.}
    \label{ca2}
\end{figure}

For classical Be stars, \citet{Andrillat:1990} found that \ion{Ca}{2}
is only seen in emission if Paschen lines were seen in
emission. \citet{Andrillat:1988} found that \ion{Ca}{2} emission was
observed in those stars that exhibited a large $IRAS$ excess, $>$0.6
magnitudes at 12 $\mu$m.  For Herbig AeBe stars, the \ion{Ca}{2} IR
triplet is shown to correlate with the H$\alpha$ emission strength
\citep{Catala:1986}. Only three of our sources exhibited \ion{Ca}{2}
IR triplet emission, G007.2369+01.4894, G027.0268+00.7224, and
G229.4514+01.0145, and their spectra are shown in Figure~\ref{ca2}.
The best fitting EW ratio determined stellar model is shown beneath
the stellar spectrum. G007.2368+01.4894 was also shown to exhibit
\ion{Ca}{2} emission in \citet{Hamann:1992}, and their profile appears
similar to our spectrum. G229.4514+01.0145 and G027.0268+00.7224 show
double peaked \ion{Ca}{2} emission.  These two stars also exhibit
double peaked H$\alpha$ emission.  We also find that \ion{Ca}{2}
emission strength is correlated with H$\alpha$ emission strength, but
our sample is not large, containing only three sources.  Therefore, we
do not further investigate this correlation.

\subsubsection{\ion{O}{1} $\lambda$ 8446 \AA\ Emission and \ion{O}{1} $\lambda$ 7772 \AA\ Absorption}

\ion{O}{1} $\lambda$ 8446 \AA\ emission is seen in both
Herbig AeBe stars and classical Be stars. The \ion{O}{1} excitation is
due to \ion{O}{1} $\lambda$ 1025 \AA\ fluorescence with Lyman $\beta$
that subsequently excites the related \ion{O}{1} $\lambda$ 8446 \AA\
\citep{Bowen:1947}. \citet{Andrillat:1990} found that classical Be
stars only exhibit \ion{O}{1} $\lambda$ 8446 \AA\ emission if Paschen
emission is also present. \citet{Jaschek:1993} found a correlation
between \ion{O}{1} $\lambda$ 8446 \AA\ emission and \ion{O}{1}
$\lambda$ 7772 \AA\ emission with the former about four times stronger
than the later. \citet{Jaschek:1993} also found that Be stars which
exhibit both \ion{O}{1} $\lambda$ 7772 \AA\ in emission also exhibit
\ion{Fe}{2} $\lambda$ 7712 \AA\ in emission. \citet{Hamann:1992} found
that some Herbig AeBe stars exhibit anomalously large \ion{O}{1}
$\lambda$ 7772 \AA\ absorption and attribute it to a large
circumstellar envelope. \ion{O}{1} $\lambda$ 7772 \AA\ has also been
seen in emission for some Herbig AeBe stars (e.g.,
\citealt{Hamann:1992}).  Column five of Table~\ref{tspec} lists
whether \ion{O}{1} $\lambda$ 8446~\AA\ is found in emission. Column
six of Table~\ref{tspec} indicates whether \ion{O}{1} $\lambda$ 7772
\AA\ is found in absorption stronger than stellar models predict (s)
or in emission (e).

\subsubsection{\ion{Fe}{2} Emission}

\begin{figure}
    \includegraphics[width=4.0in,angle=90]{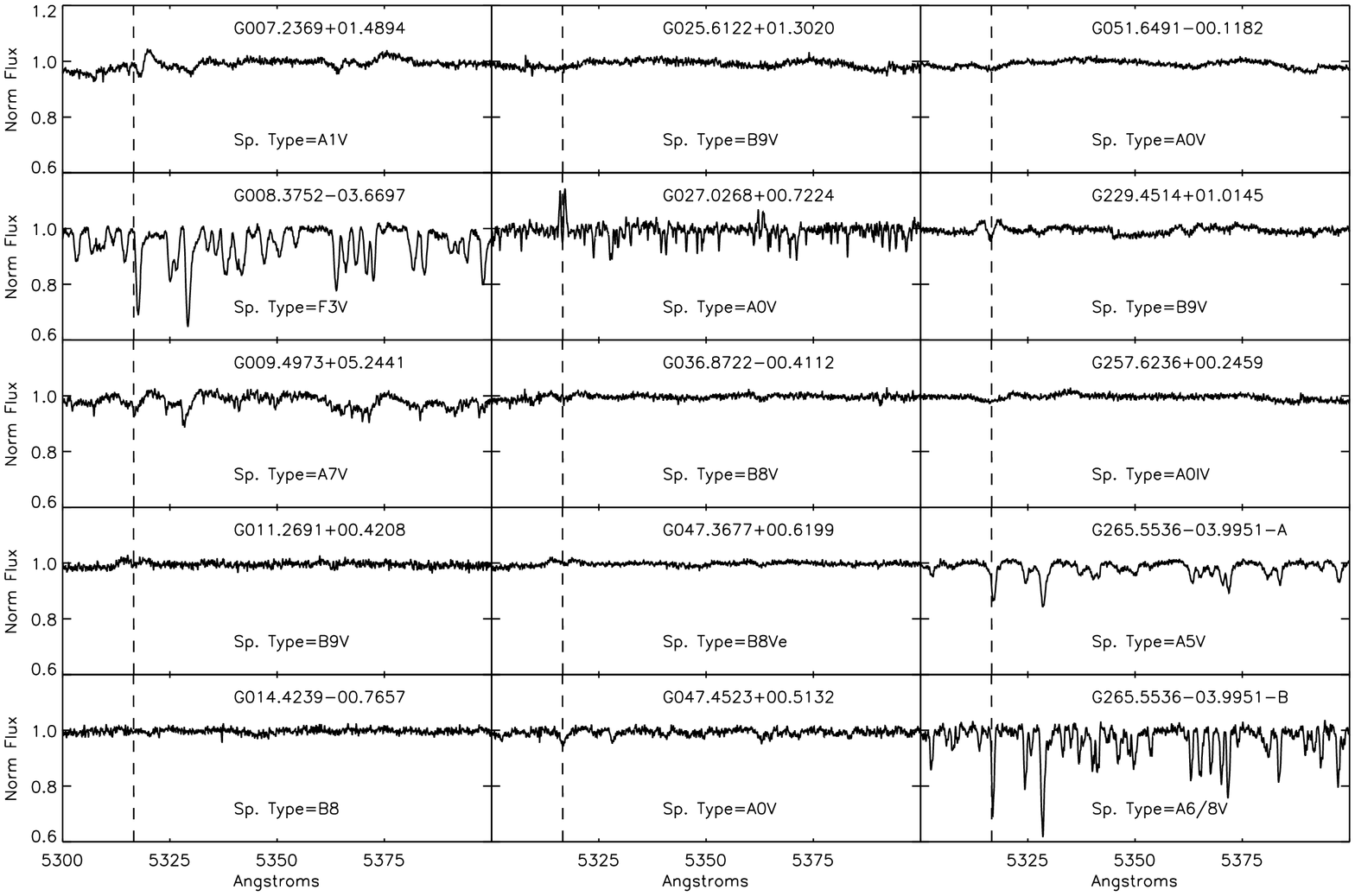}
    \caption{\ion{Fe}{2} $\lambda$ 5317 \AA\ profiles of mid-infrared
    excess stars. \ion{Fe}{2} is indicated by the dashed vertical
    line. \ion{Fe}{2} emission is not very common among our sources.}
    \label{fe1}
\end{figure}

\begin{figure}
    \includegraphics[width=4.0in,angle=90]{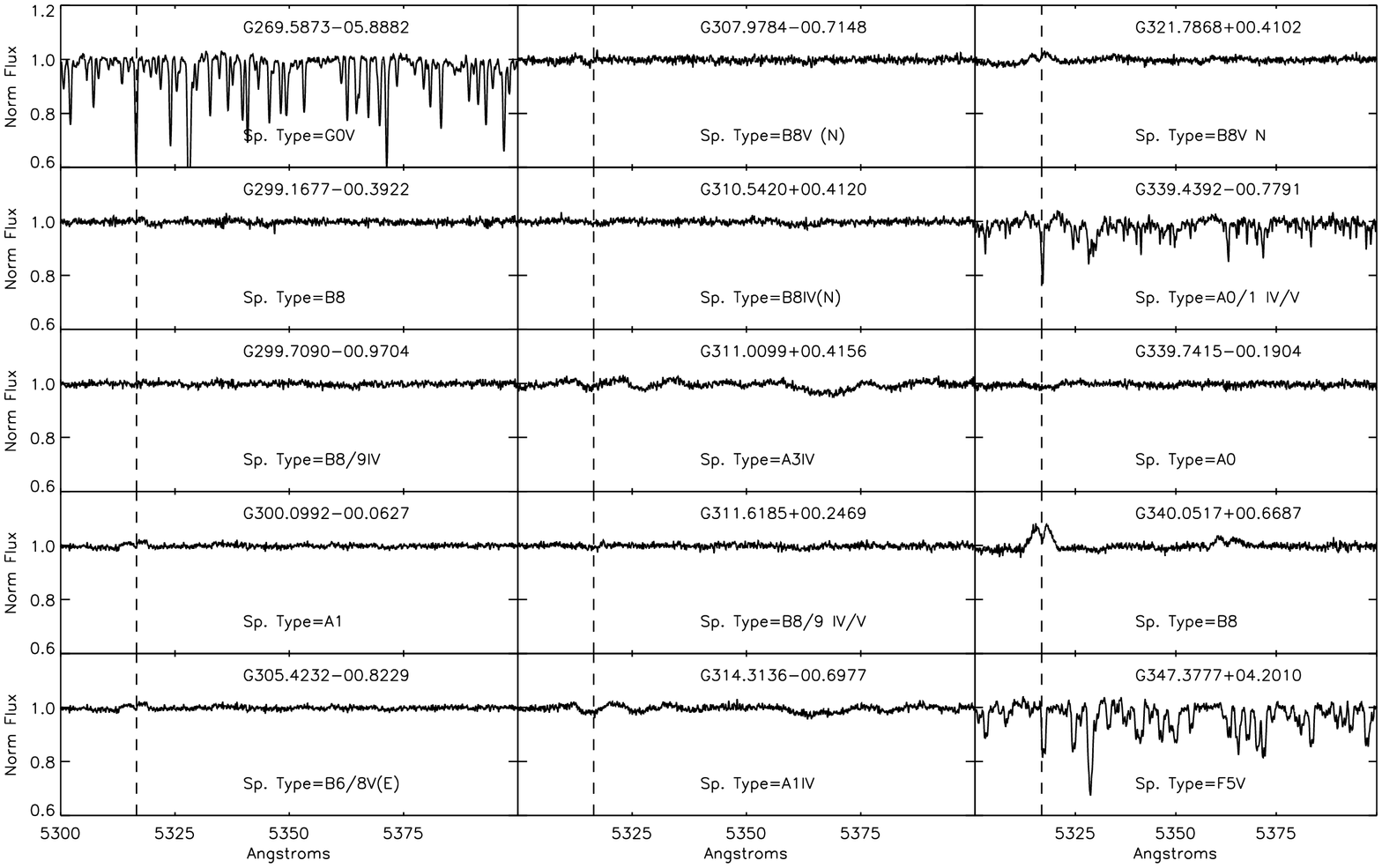}
    \caption{\ion{Fe}{2} $\lambda$ 5317 \AA\ profiles of mid-infrared
    excess stars.  \ion{Fe}{2} is indicated by the dashed vertical
    line. \ion{Fe}{2} emission is not very common among our sources.}
    \label{fe2}
\end{figure}

Classical Be stars exhibit metallic lines in emission, especially
\ion{Fe}{2} (\citealt{Jaschek:1980}; \citealt{Hanuschik:1996}).
Presence or absence of \ion{Fe}{2} is one criterion for classification
of classical Be stars into subgroups \citep{Jaschek:1980}. We show
continuum-normalized spectra near \ion{Fe}{2} $\lambda$ 5317 \AA\ in
Figures~\ref{fe1} and \ref{fe2}. Only four sources exhibit \ion{Fe}{2}
in emission: G027.0268+00.7224 (A0V), G229.4514+01.0145 (B9V),
G321.7868+00.4102 (B8VN), and G340.0517+00.6682 (B8).
\citet{Jaschek:1980} Group I sources, stars with \ion{Fe}{2} emission,
are of spectral type B0--B6. These four stars may be an extension to
later spectral types for Group I classical Be stars. Herbig AeBe stars
have also been shown to exhibit \ion{Fe}{1} and \ion{Fe}{2} emission,
but the EW of H$\alpha$ for such sources show strong emission,
EW(H$\alpha$) $<$ -25 \AA\ \citep{Hernandez:2004}. G340.0517+00.6682
and G321.7868+00.4102 are the only two of the four sources to exhibit
strong H$\alpha$ emission, EW(H$\alpha_{corr}$)=-39.38 and -26.81 \AA\
and \ion{Fe}{2} emission.

\subsubsection{Broad Absorption}

Herbig AeBe stars have been shown to exhibit absorption lines such as
\ion{He}{1} $\lambda$ 5876 \AA\ and \ion{O}{1} $\lambda$ 7772 \AA\
that are stronger and broader than expected from the stellar
photosphere \citep{Catala:1986}.  G007.2369+01.4894 exhibits strong
\ion{He}{1} $\lambda$ 5876 \AA\ as noted in \S 3 and seen in the top
left panel of Figure~\ref{he1}.  G229.4514+01.0145 may also exhibit
strong \ion{He}{1} $\lambda$ 5876 \AA\ circumstellar absorption as
indicated by the higher temperature stellar model needed to fit the
line than measured by \ion{He}{1} and \ion{Mg}{2} EW ratios (see \S
3). Several stars exhibit strong broad absorption features at
$\lambda$ $\sim$4400, $\sim$4418, and $\sim$4455 \AA. These lines do
not appear to correspond with known diffuse interstellar bands (DIBs)
or expected photospheric atomic absorption transitions
\citep{Jenniskens:1994}. The stars that exhibit broad non-stellar
absorption lines at these wavelengths are G007.2369+01.4894,
G025.6122+01.3020, G051.6491-00.1182, and G229.4514+01.0145.  These
absorption lines may imply that G025.6122+01.3020 and
G051.6491-00.1182 are surrounded by circumstellar gaseous material
similar to the material surrounding Herbig AeBe stars. We will discuss
the implications of the stellar absorption/emission lines in context
of the nature of the infrared excess in \S 8.

\section{Age Determination}

When stars are not members of known stellar clusters with
well-determined ages, stellar ages are generally estimated by fitting
theoretical isochrones to effective temperatures and bolometric
luminosities, provided that some distance estimate is available
\citep{Song:2001}.  Using our derived T$_{eff}$ and trigonometric
parallaxes from $Hipparcos$/Tycho, we estimated stellar ages by
fitting stellar parameters to \citet{Siess:2000} evolutionary model
tracks.  $Hipparcos$ parallaxes were the preferred distance measures
and used when available (11 stars). In only one instance we used a Tycho
parallax.  If neither parallax was available, we adopted absolute
visual magnitudes derived in the literature (4 stars) using methods
such as H$\beta$ luminosity and intrinsic Stromgren colors of
luminosity classes (see \citealt{Seidensticker:1989};
\citealt{Westin:1985}; \citealt{Kozok:1985}). We list parallactic
distances in Table~\ref{tage} except in the cases where noted. We
then interpolated a bolometric correction using our derived effective
temperatures in conjunction with \citet{Cox:2000} effective
temperatures and bolometric corrections.  We applied these bolometric
corrections to the Tycho-2 visual magnitudes and correct for distance
and extinction to derive the absolute bolometric magnitudes.

\begin{figure}
\includegraphics[width=4.0in,angle=90]{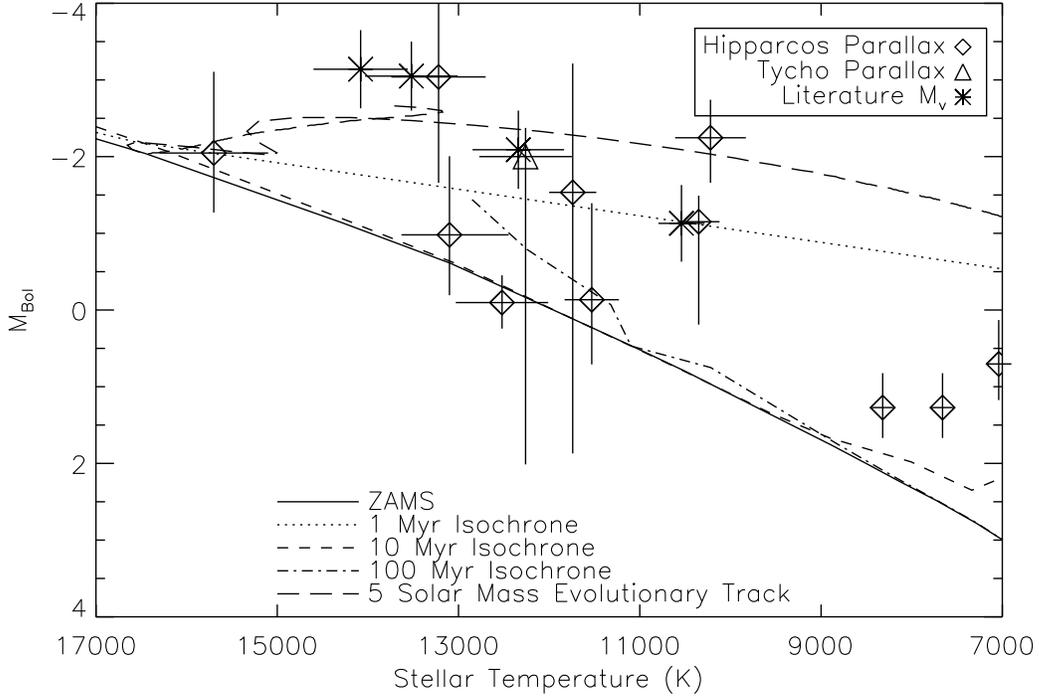}
 \caption{HR diagram for sixteen of our target stars overlaid with
\citet{Siess:2000} stellar evolutionary models for 7,000 $<$ T$_{eff}$
$<$ 17,000~K. The symbols show the various methods used to determine
the absolute magnitude, diamonds for $Hipparcos$ parallax, triangle
for Tycho parallax, and asterisks for stars with absolute magnitudes
taken from the literature. The zero-age main-sequence is shown by the
solid line. Isochrones are shown by the dotted, dashed, and dash-dot
relations. The majority of our stars lie above the main-sequence and
are consistent with 3--5\ \mo\ stars.  The three coolest stars are
above the main-sequence and consistent with pre-main-sequence objects.}
    \label{ages}
\end{figure}

Figure~\ref{ages} shows our derived absolute bolometric magnitudes and
effective temperatures (i.e., the HR diagram) with isochrones derived
from \citet{Siess:2000} Z=0.02 stellar evolutionary models
over-plotted.  Figure~\ref{ages} shows the range of stellar
temperatures consistent with $\sim$2--5\ \mo\ stars.  Stars with
absolute bolometric magnitudes determined from $Hipparcos$ parallaxes
are denoted by the diamonds, those with bolometric magnitudes
determined from Tycho parallaxes are shown by triangles, and those for
which literature absolute magnitudes were used are shown by
asterisks. The zero-age main-sequence is shown by the solid
line. Isochrones for 1, 10, and 100 Myr are shown by the dotted,
dashed, and dash-dot relations. The majority of our stars are
consistent with 3--5\ \mo \ stellar models.  Most of our stars lie
above the main-sequence indicating either a late pre-main-sequence
(few Myr) or late main-sequence stellar age ($\simeq$ 100 Myr).  We
estimate a range of possible stellar ages by taking the full range of
\citet{Siess:2000} model ages consistent with the derived temperatures
and luminosities, taking into account the measurement uncertainties.
Table~\ref{tage} lists the possible age ranges for the stars with
sufficient measurements and compares them to other results in the
literature. The age ranges we determined for the three stars having
published age estimates were consistent with those measurements.  We
were able to determine the age ranges of two stars identified in
literature as pre-main-sequence stars and found that the ages were
consistent with such a state.  Most published measurements of stellar
ages have uncertainties of order a few $\%$ of the stellar lifetime
because they are either nearby (small uncertainties in luminosity) or
derived from main-sequence cluster fitting. Our age measurements
typically have errors of 10$\%$ or greater, so the derived values
should be regarded as general age ranges.

\section{Implication of Fundamental Parameters on the Nature of the 
Mid-IR Excesses}

Stellar parameters, such as rapid rotation and H$\alpha$ emission can
provide insight into the nature of the infrared excess.  On average,
classical Be stars rotate more rapidly than Herbig AeBe stars.
H$\alpha$ emission can be used to determine the amount of free-free
emission expected from the circumstellar disk. H$\alpha$ emission is
also found to decrease as pre-main-sequence stars evolve toward the
main-sequence \citep{Manoj:2006}. Broadband infrared colors are
commonly used to discriminate between various circumstellar disks
(i.e., \citealt{Hartmann:2005}). We use the fundamental parameters
derived above to determine the nature of the circumstellar
excess in the following subsections.

\subsection{Large Rotational Velocities}

Intermediate-mass stars rotate at a wide variety of velocities
depending on how they formed and their evolutionary
state. \citet{Finkenzeller:1985} found that Herbig AeBe stars rotate
at values of 100 $<$ \textit{v} sin \textit{i} $<$ 225 km
s$^{-1}$. \citet{Bohm:1995} predict that higher mass Herbig AeBe stars
($>$ 4\ \mo) should gain angular momentum as they approach the
zero-age main-sequence. \citet{Bohm:1995} also predict that zero-age
main-sequence stars in the mass range 2.6\ \mo\ $<$ M$_{*}$ $<$ 4.0 \
\mo\ should rotate at 205$\pm$55 km s$^{-1}$ and stars with M$_{*}$
$>$ 4.0\ \mo\ should rotate at 180$\pm$90 km s$^{-1}$ based on
pre-main-sequence projected rotational velocities.  Classical Be stars
are typically rapid rotators having \textit{v} sin \textit{i} $>$
200 km $^{-1}$.  In Figure~\ref{comp} we compare the rotational
velocities for our sample of 20 GLIMPSE late-B and early A stars
(dashed histogram), to B8--B9.5 stars from \citet{Abt:2002} (solid
histogram) and classical Be stars from \citet{Hanuschik:1996}
(dash-dot histogram).  Our 20 GLIMPSE stars rotate between 25 km
s$^{-1}$ $<$ \textit{v} sin \textit{i} $<$ 310 km s$^{-1}$ with a mean
and dispersion of 209$\pm$76 km s$^{-1}$.  \citet{Abt:2002} found that
the rotational velocity distribution for B8--9.5 stars is bimodal with
peaks at both 50 km s$^{-1}$ and 250 km s$^{-1}$. The mean projected
rotational velocity and dispersion of the \citet{Abt:2002} late B star
sample is 148$\pm$84 km s$^{-1}$. The mean projected rotational
velocity and dispersion of \citet{Hanuschik:1996} classical Be stars
is 217$\pm$17 km s$^{-1}$.  Thus, our sample has comparatively high
rotational velocities, more consistent with the HAeBe stars and Be
stars than the field B stars of \citet{Abt:2002}.

We compared our sample rotational velocities to both the classical Be
sample and the B8--B9.5 sample using a K-S test to determine if our
sample could be drawn from either of these other distributions.  There
is no large sample of Herbig AeBe stars with projected rotational
velocity measurements as they approach the zero-age main-sequence, so
we cannot make a comparison to these objects.  The probability that
our sample could be derived from the \citet{Abt:2002} sample is $<$
0.1\%. The probability that our sample may be derived from the
\citet{Hanuschik:1996} sample is 8\%.  Although the mean projected
rotational velocity of our GLIMPSE stars is consistent with those
of the \citet{Hanuschik:1996} classical Be stars, it is unlikely that
our sample consists of solely Be stars.  This result may imply that
our sample is a mixed group of sources including both normal B stars,
classical Be stars, and even Herbig AeBe stars.

\subsection{Infrared Colors as a Discriminant for Terrestrial 
Temperature Dust}

Using the working definition of \citet{Waters:1998} for Herbig AeBe
stars, the only differing criterion between classical Be and Herbig
AeBe stars is the nature of the infrared excess. While Herbig AeBe
stars have an infrared excess due to circumstellar dust and gas,
classical Be stars owe their infrared excess to free-free emission
in ionized gas. 

Color-color plots are often used to distinguish various types of
infrared sources (e.g., \citealt{Walker:1989};
\citealt{Zhang:2004}). Using a color-color plot in both the near- and
mid-infrared we can distinguish between sources that have an excess
owing to free-free emission (and free-bound emission) and sources that
have dust emission. In order to model the expected contribution of
free-free emission to the infrared continuum we utilize the following
equation from \citet{Tucker:1975} to estimate the free-free flux.
The free-free emissivity at a given frequency, $\nu$, is

\begin{equation}
j_{ff}(\nu)=5.44\times10^{-39}
\frac{g_{ff}Z^{2}_{i}N_{e}N_{p}}{\sqrt{T}}e^{-h\nu/kT}
(erg\ cm^{-3}\  s^{-1}\ sr^{-1}\ Hz^{-1})
\end{equation}

\noindent where N$_{e}$ and N$_{p}$ are the electron and proton
densities, T is the electron temperature, and g$_{ff}$ is the Gaunt
factor. We assumed an electron temperature of 10,000~K, a pure
hydrogen gas, Z=1, and N$_{e}$=N$_{p}$=3$\times$10$^{11}$
cm$^{-3}$. The values of electron densities for classical Be stars
vary and we use the mean value from \citet{Gehrz:1974}
($n_e=3\times10^{11}$ cm$^{-3}$). Even though classical Be stars exhibit
such large electron densities within their circumstellar disks, it has
been shown that such disks are still optically thin at optical wavelengths
\citep{McDavid:2001}. We used Gaunt factors from Figure~5 of
\citet{Karzas:1961} by fitting a polynomial to parametrize the Gaunt
factors as a function of wavelength for T$\simeq$10,000~K. Equation~5
gives an expression for the Gaunt factor for 10,000~K where $\lambda$
is in $\mu$m

\begin{equation}
g_{ff}=33.513-4.041 \times Log(\lambda)+0.106 \times Log(\lambda)^{2}+0.001 \times Log(\lambda)^{3}.
\end{equation}

In the near-infrared, the contribution of free-bound emission is not
negligible.  \citet{Ashok:1984} found the contributions from free-free
and free-bound to the total nebular emission at 2.2 $\mu$m are 63\%
and 37\%, respectively, for the physical conditions of \citet{Gehrz:1974}. 
We therefore include free-bound emission in our
calculation of nebular emission.  The free-bound emissivity 
as a function of frequency is

\begin{equation}
j_{\lambda,n}^{fb}=1.4\times10^{-33}n^{-3}N_{e}N_{p}T^{-3/2}Z^{4}e^{(\chi_{n}-h\nu)/kT}
(erg\ cm^{-3}\ s^{-1}\ sr^{-1}\ Hz^{-1}),
\end{equation}

\noindent where $n$ is the excitation level of hydrogen
\citep{Tucker:1975}.  We adopted the same values for $N_e$, $N_p$, and
$T$ as for the free-free emissivity above.

We used the above equations for free-free and free-bound emission to
model the infrared colors of a two-component spectral energy
distribution consisting of a stellar atmosphere and a thermal
bremsstrahlung excess.  We simulated the free-free contribution to the
overall SED by adding increasing amounts of free-free (and free-bound)
emission (normalized at 8 $\mu$m) to a B8V stellar model from
\citet{Kurucz:1993}.  We varied the free-free contribution from 0.01\%
of the stellar blackbody at 8 $\mu$m to 100 times the stellar
blackbody at 8 $\mu$m.  Varying the electron temperature of the
circumstellar component between 5,000~K and 15,000~K has a negligible
effect on the broadband IR colors.  We also modeled the colors
expected for a two-component spectral energy distribution consisting
of a stellar blackbody and a cooler blackbody representative of
circumstellar dust at temperatures 100--1100~K.  We then used these
colors to determine if we can distinguish between free-free emission
and dust emission in color space.

Figure~\ref{color} shows a plot of 8-24 $\mu$m vs. $K$-8 $\mu$m colors
for the 19 GLIMPSE stars with [24] excesses.  Open circles represent
GLIMPSE sources having H$\alpha$ in emission and open stars represent
GLIMPSE sources having H$\alpha$ in absorption.  Classical Be stars
($asterisks$) are those from \citet{Zhang:2004}.  We utilized $MSX$ A
band and $IRAS$ 25 $\mu$m as 8 $\mu$m and 24 $\mu$m measurements for
the sources from \citet{Zhang:2004,Zhang:2005}
respectively. Thirty-five of the \citet{Zhang:2004,Zhang:2005} stars
have $MSX$ A band and $IRAS$ 25 $\mu$m measurements. Both Be catalogs,
\citet{Zhang:2004,Zhang:2005}, draw on Be stars from
\citet{Jaschek:1982}, however at least 7 of the 35
\citet{Jaschek:1982} sources have some indication in the literature
that they may be Herbig Be stars rather than classical Be stars. The
arrow shows the reddening vector for A$_{V}$=5.0. The solid curve
denotes a B8V main-sequence star with A$_{V}$=1.0 and increasing
contribution from a free-free component at 8 $\mu$m. The solid symbols
denote where the circumstellar contribution at 8 $\mu$m constitutes
5\% ($diamond$), 33\% ($hexagon$) and 50\% ($square$) of the total 8
$\mu$m flux.  The dashed curves denote a B8V main-sequence star with
A$_{V}$=1.0 and an 800~K or 300~K blackbody circumstellar
component. We added the blackbody contribution in a similar manner as
the free-free contribution, and the solid symbols denote 5\%, 33\%,
and 50\% fractional circumstellar contributions.

Figure~\ref{color} shows that this color-space can distinguish between
free-free and cool dust origins of the IR excesses. Points to the left
of the solid curve may be explained by either hot dust or free-free
emission. Thus, hot dust is indistinguishable from free-free emission
in this color space. Points to the right of the solid curve can only
be explained by a cool (terrestrial or cooler) dust component.  The
majority (13 of 15) of sources with H$\alpha$ emission ({\it open
circles}) lie to the left of the solid line, consistent with them
being classical Be stars or stars with hot circumstellar dust.  The
majority of the sources with H$\alpha$ absorption ({\it open stars})
lie to the right of the solid line, consistent with a cool
circumstellar dust component.  The majority of the classical Be stars
({\it asterisks}) lie to the left of the solid line, as expected for
free-free excesses and consistent with their classification.  There
are, however, 14 \citet{Jaschek:1982} sources that fall to the right
of the solid curve. They are HD 6343, HD 37318, HD 251726, HD 52721,
HD 53367, HD 90177, HD 141926, BD-11 4667, HD 174571, HD 204722, HD
240010, HD 37967, HD 259431, and HD 50083. This sub-sample of 14 stars
contains five of the seven possible Herbig Be stars within the 35 star
\cite{Jaschek:1982} sample. The two remaining possible Herbig Be stars
are HD 50138 and HD 101412.  These two sources show the greatest $K$-8
color and fall off the top of the plotted region and to the left of
the solid curve. HD 50138 is also listed as a possible B[e] star
\citep[e.g.,][]{Bjorkman:1998}. Therefore, it is possible that some of
the stars classified as classical Be stars may actually be Herbig Be
stars.  The two H$\alpha$ emission GLIMPSE sources that fall to the
right of the solid curve are G036.8722+00.6199 and G311.6185+00.2469.
H$\alpha$ absorption sources that fall to the right of the solid curve
are G014.4239-00.7657, G311.0099+00.4156, and G339.7415-00.1904.
These five sources are candidates for objects with terrestrial
temperature circumstellar dust and are discussed in detail below.

\subsection{Individual Free-free Contributions}

A quantitative analysis of the H$\alpha$ and mid-IR excess fluxes can
help distinguish between circumstellar dust and ionized gas
better than IR colors alone. Specifically, we want to determine if the
free-free excess in the mid-IR predicted by the H$\alpha$ flux can
account for the observed IR excess or if a dust component is
required. We use the H$\alpha$ flux of each star to predict the
gaseous circumstellar contribution to the 8 $\mu$m and 24 $\mu$m
continuum, under the assumption that the H$\alpha$, free-free, and
free-bound emitting volumes are coincident.  Using this method one can
then verify that the color-color plot in Figure~\ref{color}
discriminates between infrared excesses with free-free emission and
those with an excess due to cool dust. We adopt the emissivity at
H$\beta$ from \citet{Brocklehurst:1971} for 10,000~K,

\begin{equation}
4 \pi j_{H\beta} = 1.24 \times 10^{-25} N_{e} N_{p}\ 
(ergs\ cm^{-3} \ s^{-1}),
\end{equation}

\noindent where N$_{e}$ and N$_{p}$ are the number densities of
electrons and protons in cm$^{-3}$, respectively. We used the
ratio of H$\alpha$ to H$\beta$ emission coefficients from
\citet{Brocklehurst:1971} for Case B,

\begin{equation}
\frac{j_{H\alpha}}{j_{H\beta}}=2.87.
\end{equation}

\noindent Finally the flux due to free-free emission can be derived utilizing
the flux at H$\alpha$ and the emissivity at H$\alpha$,

\begin{equation}
F_{ff}(\nu)=F_{H\alpha}\frac{j_{ff}({\nu})}{j_{H\alpha}},
\end{equation}

\noindent where \textit{j}$_{ff}{(\nu)}$ is the thermal bremsstrahlung
emissivity at a given frequency, and \textit{j}$_{H\alpha}$ is the
emissivity of H$\alpha$. We used an electron temperature of 10,000~K
but the exact choice of N$_{e}$, N$_{p}$, and $T$ is not critical
to the ratio of $j$$_{ff}$/$j$$_{H\alpha}$. The ratio is independent
of gas density and only slowly varies with $T$.

We calculate the H$\alpha$ flux by multiplying the
absorption-corrected equivalent width of the H$\alpha$ line,
EW(H$\alpha_{corr}$), by the reddening-corrected flux at R
band. Dereddened R fluxes are listed in Table~\ref{tha} computed from
R-band magnitudes of \citet{Monet:1998} \& \citet{Monet:2003}. We
adopt extinction estimates from \citet{Uzpen:2007}, which may include
circumstellar absorption and may be larger than true interstellar
extinction. The resultant H$\alpha$ flux thus predicts a maximum
possible free-free flux, ignoring possible variability and optical
depth effects.

As a means of verifying our calculations, we used the observed
H$\alpha$ fluxes of \ion{H}{2} regions from \citet{Viallefond:1983},
\citet{Viallefond:1986a}, and \citet{Viallefond:1986b} to predict the
free-free radio fluxes. We compared our predicted radio fluxes to the
measured values and found that the predicted 6 or 21 cm fluxes are a
factor of $\simeq$2 deficient relative to measured
values. \citet{Israel:1980} found typical extinctions of 1--2
magnitudes at visual wavelengths for \ion{H}{2} regions. The H$\alpha$
fluxes from \citet{Viallefond:1986a} do not take into account
extinction, and this may be part of the reason for the deficit.
Non-thermal processes in the radio may also contribute to the radio
flux.

Figure~\ref{G36} shows a comparison of the data and models for one star,
G036.8722-00.4112. The modeled free-free component plus stellar
blackbody (dash-dot-dot-dot curve) is not sufficient to explain the
measured excess at 24 $\mu$m for this source. A stellar blackbody
with additional blackbody due to dust (thin solid curve) fits all the
measured data and is the likely explanation for the excess.

We applied our free-free model to all stars with H$\alpha$ emission
and measured [24] flux. Table~\ref{tff} gives the measured [24] flux
(from \citealt{Uzpen:2007}) for all 20 GLIMPSE stars and derived flux
contribution at [24] from free-free emission for the 15 stars with
emission and [24] excess. The excesses from six sources are
attributable to circumstellar gas since the removal of the free-free
component results in $<$ 3 $\sigma$ excesses. We designate such
objects as Type I sources.  The remaining nine stars with emission and
[24] measurements have excesses $>$ 3 $\sigma$ after the free-free
components were removed. We designate these as Type II sources. We
note that in none of our sources does the removal of the free-free
component reduce the 8 $\mu$m or 24 $\mu$m flux below photospheric
levels. In all cases, the correction for circumstellar free-free
contributions either reduces the 8 and 24 $\mu$m measurements to the
photospheric levels or leaves a significant excess at these bands.

\subsection{Variability}

Both classical Be and Herbig AeBe stars exhibit variability at
H$\alpha$ (\citealt{Dachs:1988}; \citealt{Waters:1998}). Our sources
may also be variable at H$\alpha$, and that may explain some of the
Type II sources that exhibit an excess $>$ 3.0 $\sigma$ at [8.0]
and/or [24] when the free-free component is removed. In order to
assess H$\alpha$ variability among this sample, we investigated
H$\alpha$ profile variations for a sub-sample of four northern Type I
or II stars and a Balmer absorption star, G011.2691+00.4208,
G036.8722-00.4112 and G047.3677+00.6199, G047.4523+00.5132, and
G051.6491-00.1182.  Ideally all Type II sources would be investigated
for variability but five of the Type II stars are in the southern
hemisphere and not observable from the Wyoming Infrared Observatory
(WIRO) (G300.0992-00.0627, G305.4232-00.8229, G307.9784-00.7148,
G314.3136-00.6977, and G340.0517+00.6687).

We obtained optical spectroscopy of all stars with the WIRO-Spec
instrument on WIRO on the nights of 2005 August 14 and 15 using the
1435 $l$ mm$^{-1}$ volume-phase holographic grating in first
order. This instrument is an integral field unit spectrograph
consisting of 293 densely packed fibers in a 19 $\times$ 20
configuration with each fiber projecting to 1\arcsec \ on the sky.
This grating provides a dispersion of 0.7 \AA\ pixel$^{-1}$, a spectral
resolution of 1.4 \AA,\ and wavelength coverage of $\simeq$$1400$ \AA
\ from $\simeq$5750--7150 \AA. Standard reduction techniques were
implemented, including flat fielding using a continuum lamp and
wavelength calibration using a CuAr arc lamp, but no flux calibration
was performed.  We achieved a typical signal-to-noise ratio of 100:1.

We also obtained optical spectroscopy using the WIRO-Longslit
spectrograph on the nights of 2006 September 13 and 14.  The
WIRO-Longslit is a low-resolution spectrograph with a dispersion of
$1.1$ \AA\ pixel$^{-1}$, a resolution of $2.6$ \AA \ using a 2\arcsec \
slit width, and covers $\simeq$$4400-6700$ \AA\ when using the 600
$l$ mm$^{-1}$ grating in second order.  We used a GG455 blocking
filter to remove third order contamination. We performed the same
reduction procedures for the longslit data as WIRO-Spec data, and the
signal-to-noise ratios were similar.

G011.2691+00.4208, G036.8722-00.4112, G051.6491-00.1182 displayed a
double-peaked emission profile all three nights.  G047.4523+00.5132
exhibited absorption all three nights.  G047.3677+00.6199 displayed a
double-peaked emission profile on the nights of 2005 August 15 and
2006 September 27.  However, on the night of 2006 September 13 the
star exhibited a single-peaked emission profile. The
EW(H$\alpha_{corr}$) changed $\simeq$25$\%$ and may be due to
H$\alpha$ variability. A variation in H$\alpha$ may result in a
lower-than-expected calculation of the free-free emission.  We
estimated at what level the H$\alpha$ EW must vary to account for the
excess at [24]. We find that the EW(H$\alpha_{corr}$) must double for
the free-free emission to explain the excess at [24]. It is
interesting to note that even if the EW(H$\alpha_{corr}$) were a
factor of two larger, the excess is still significant at [8.0] even
after removal of the free-free component.

For the six Type II stars in the southern hemisphere, we are unable
to verify any H$\alpha$ variability because we have only one
observation of these stars. The H$\alpha$ variability
combined with the non-contemporaneous nature of the H$\alpha$ and IR
data, may lead us to over- or underestimate the circumstellar
free-free contribution to the [8.0] and [24] bands. If the
circumstellar free-free contribution is underestimated, we expect to
see a residual IR excess. The nine Type II sources may be examples of
this effect. However, we find no instances of the converse, i.e.,
sources where the H$\alpha$ flux over predicts of the circumstellar
free-free contribution, thereby creating a deficit at [8.0] or
[24]. We find that the six southern Type II sources need to exhibit
H$\alpha$ equivalent width variation at two or three times the
measured levels to explain the excesses as free-free emission. Thus,
it is unlikely that the Type II sources are to be explained
as objects owing their excess to free-free emission.

Optical depth effects at H$\alpha$ may also lead to underestimates of
the free-free emission. The H$\alpha$ optical depth would have to
exceed $\tau$$\approx$1 in order to explain the excesses of Type II
sources by free-free emission. As mentioned above,
\citet{McDavid:2001} shows that a Be stellar disk with an electron
density of order 10$^{12}$ cm$^{-3}$ is optically thin, with the
maximum $\tau$ $<$ 1, at optical wavelengths.

\subsection{Physical Nature of IR Excess Sources}

Table~\ref{tsum} lists the derived circumstellar parameters for the
nine Type II stars with the free-free emission component
removed. Circumstellar disk temperature and fractional infrared
luminosity were determined in the same manner as
\citet{Uzpen:2007}. Briefly, we fit the SED as a single temperature
blackbody to model the IR excess. Table~\ref{tsum} also lists the
derived circumstellar parameters for the four non-emission line stars,
hereafter referred to as Type III. We list the circumstellar disk
parameters from \citet{Uzpen:2007} for the five $MSX$ stars with
$\lambda$ $>$ 20 $\mu$m observations.

We searched for correlations between derived circumstellar parameters
with stellar parameters to provide insight into the nature of the
excess. Figure~\ref{diskt} shows circumstellar disk temperature (upper
panel) and stellar effective temperature (lower panel) versus
EW(H$\alpha_{corr}$). We found a significant inverse-correlation
between EW(H$\alpha_{corr}$) and circumstellar disk temperature for
Type II stars. The correlation coefficient of -0.81 for the nine Type
II sources means there is a $<$ 1 $\%$ probability of obtaining such a
high degree of correlation by chance.  Such a correlation is expected
if the H$\alpha$ emission is arising within a gaseous and dusty
circumstellar disk undergoing a clearing of the inner disk. That is,
the strength of H$\alpha$ decreases and the dust temperature drops as
the inner disk becomes anemic. If the excess were due to free-free
emission, we would expect the strength of H$\alpha$ equivalent width
to be correlated with stellar effective temperature. The hotter stars
generate more ionizing photons when compared to cooler stars and
exhibit a larger excess \citep{Gehrz:1974}. We find no such
correlation between stellar temperature and H$\alpha$ equivalent width
for Type I, Type II or Type III stars, see Figure~\ref{diskt}.

Using our age estimates we investigated whether there is a correlation
between EW(H$\alpha_{corr}$), age, and $L_{IR}/L_{*}$ in
Figure~\ref{3in1}. Omitting the four pre-main-sequence stars from
$MSX$, we used the nine Type II stars, but only six of the nine
have age estimates. We plotted the younger age for Type II stars, when
available. Using the six Type II sources and the one Type III source
with an age, we find that age is strongly correlated with H$\alpha$
equivalent width with a correlation coefficient of 0.97 indicating $<$
1\% probability of obtaining such a high probability by chance.  Using
the older age when available, there is no correlation between
EW(H$\alpha_{corr}$) and age.  \citet{Manoj:2006} found that
H$\alpha$ line strength decreases as Herbig AeBe stars age.  If those
stars are part of the evolved tail to Herbig AeBe stars we would
expect to see such a correlation between EW(H$\alpha_{corr}$) and age.

Using the above fundamental parameters, above correlations, and
spectral features we can determine the circumstellar disk
classification for our entire sample of 30 mid-IR excess sources. We
indicate our circumstellar disk classification in
Table~\ref{tsum}. Pre-main-sequence Herbig AeBe stars and T-Tauri
stars are indicated by H and TT respectively.  Debris disks are
denoted by D in Table~\ref{tsum}. Stars with an excess owing to
free-free emission are classical Be stars and listed as B. Finally,
stars with a transitional disk system are labeled with T. Five of the
$MSX$ stars do not have a confirmed excess at $\lambda$ $>$ 20 $\mu$m
and thus, we do not classify their possible circumstellar disk. We
find that utilizing Figure~\ref{color} as a first order discriminate
is useful but may not be the only parameter necessary to determine the
nature of the excess. Each object is discussed in the following
section.

\section{Individual Objects}

In the subsections below, we discuss individual sources. When
available, we list the HD catalog number, the GLIMPSE or $MSX$ ID
number, and the literature spectral type. In some cases, the spectral
type found in the literature is discrepant from our results.
Figure~\ref{flow} is a flowchart illustrating the criteria we use to
assign an evolutionary state to each source.  In summary, we require
that our candidate debris disks not exhibit any emission lines and
exhibit $K$-[24] colors $<$ 3.3 magnitudes (e.g.,
\citealt{Hernandez:2007}).  Transition disk systems may not exhibit
Paschen emission lines, may exhibit \ion{O}{1} $\lambda$ 8446 \AA\
emission, may exhibit \ion{O}{1} $\lambda$ 7772 \AA\ in emission or
absorption, may not exhibit \ion{Fe}{2} emission, and must have an
excess not entirely explainable by free-free emission.  Classical Be
stars must have Balmer emission, may exhibit Paschen emission, may
exhibit \ion{Fe}{2} emission, may exhibit \ion{O}{1} emission, and
must have an excess explainable entirely by free-free emission.
Herbig AeBe stars may have essentially the same optical spectral
characteristics as classical Be stars, and are therefore difficult to
distinguish based on optical spectra alone.  The infrared excess in
HAeBe stars is dominated by dust rather than free-free emission,
therefore knowledge of the SED at wavelengths longer than $\sim$10
microns is essential for unambiguous classification.  However, for our
sample, HAeBe stars either exhibit \ion{O}{1} $\lambda$ 7772 \AA\
absorption and \ion{Ca}{2} IR emission or exhibit \ion{O}{1} emission,
Paschen emission, and an excess not entirely explainable by free-free
emission.

\subsection{Candidate Debris Disk Systems}

\textbf{HD 231033 (G047.4523+00.5132)-A0V}- As one of the few stars in
our sample that lacks H$\alpha$ emission, the derived disk temperature
(355$\pm$8~K) and $L_{IR}/L_{*}$ (1.2$\times$10$^{-3}$) make
this one of the best candidates for being a warm debris disk system.
This source falls just to the left of the solid line in
Figure~\ref{color}, in the region where excesses due to either
free-free emission or hot dust are permissible.  The disk temperature
and fractional infrared luminosity are similar to, but slightly higher
than, typical debris disks. Since this star does not exhibit any
optical emission lines, we classify the circumstellar disk as a
massive debris disk. This star exhibits stronger \ion{O}{1} $\lambda$
7772 \AA\ absorption than the stellar temperature would imply.  This
may indicate the presence of a small amount of circumstellar gas. This
star was observed multiple times at H$\alpha$ with both WIRO and
Magellan and was not observed to exhibit H$\alpha$ variability. The
age derived for this star (log age(yr)=6.11--6.37 or 7.58--8.19) is
consistent with a main-sequence or late-stage main-sequence star,
which agrees with its circumstellar disk classification. The large
uncertainty in age is due to the large uncertainties associated with
the Tycho parallactic measurement (d=909$^{+91}_{-755}$~pc).

\textbf{HD 121808 (G311.0099+00.4156)-A3IV}- This star lacks optical
emission lines of any kind, making it another strong debris disk
candidate.  Like G047.4523+00.5132, it exhibits stronger \ion{O}{1}
$\lambda$ 7772 \AA\ absorption than the stellar temperature would
indicate, suggesting that it may also be surrounded by a small amount
of circumstellar gas. Circumstellar disk parameters
(T$_{disk}$=315$^{+4}_{-3}$~K and
$L_{IR}/L_{*}$=2.7$\times$10$^{-4}$) are consistent with a
warm, massive debris disk.  There is insufficient data in the
literature to make a distance estimate, and therefore, an age
estimate, for this star.

\textbf{HD 151017 (G339.7415-00.1904)-A0}- This star does not show any
spectral features indicative of the presence of circumstellar
gas. Derived circumstellar disk parameters (T$_{disk}$=311$\pm$6~K and
$L_{IR}/L_{*}$=1.0$\times$10$^{-4}$) are consistent with a
warm, massive debris disk. There is insufficient data in the
literature to make a distance estimate, and therefore, an age
estimate, for this star.

\subsection{Candidate Transition Disk Systems}

Since transition disks around intermediate-mass stars have only
recently been identified, we discuss some general properties of our
eight candidate transition disk systems as a whole. The four
transition disk sources with determined disk temperatures above 500~K
exhibit H$\alpha$ emission and \ion{O}{1} $\lambda$ 8446 \AA\
emission. Two of these four sources exhibit \ion{O}{1} $\lambda$ 7772
\AA\ in emission and two exhibit \ion{O}{1} $\lambda$ 7772 \AA\
absorption consistent with stellar temperatures. Three of the four
transition disk candidates with disk temperatures below 500~K exhibit
H$\alpha$ emission. Two of these four exhibit \ion{O}{1} $\lambda$
7772 \AA\ absorption stronger than the best-matched stellar model, and
two exhibit \ion{O}{1} $\lambda$ 7772 \AA\ absorption consistent with
stellar temperatures. Three of the four stars with disk temperatures
below 500~K are also in the region of color-color space in
Figure~\ref{color} not explainable by free-free emission. The change
in the \ion{O}{1} lines from emission in systems with hot
circumstellar disks to absorption in systems with cooler circumstellar
disks may be an evolutionary effect as the disk clears. Hotter
circumstellar dust would be located nearer the star, and the lack of
gaseous emission as the inner disk radius becomes larger may be direct
evidence of disk clearing. More transitional disk systems need to be
identified and studied to provide statistical significance to this
trend.

\textbf{HD 168246 (G014.4239-00.7657)-B8}- This star does not exhibit
H$\alpha$ emission and has an intermediate value of projected
rotational velocity, 165$\pm$ 10 km s$^{-1}$. $L_{IR}/L_{*}$
for this star, 1.3$\times$10$^{-2}$, is higher than typical for debris
disks.  The derived disk temperature of T$_{disk}$=191~K is the
coolest in our sample and suggests that the circumstellar material is
located at a relatively large radius of $\sim$75 AU.  We find that this
star exhibits anomalously strong \ion{O}{1} $\lambda$ 7772 \AA\
absorption compared to the adopted stellar model, suggesting the
presence of circumstellar gas. Given the lower temperature derived for
this disk and the large $K$-[24] color (5.15 mag), this disk may be
optically thick at [24]. This system may contain a transitional disk
at the end of its clearing phase. We also find that this star lies
near a strong 70 $\mu$m extended source, raising the possibility that
the infrared emission at [24] is unrelated to the star.  We find that
the difference between the position of the strong [24] point source
and the [8.0] point source is 1.1\arcsec, i.e., small compared to the
6 \arcsec\ resolution at [24]. In order to determine the chance of
a random coincidence between the two sources we used the [24] source
density in this field, 54 stars over an area of 0.01 deg$^{2}$
(0.1$\times$0.1 deg. region), to determine that there are
4.2$\times$10$^{-4}$ sources per arcsecond$^{2}$.  The probability of
a [24] source randomly falling within 1.1\arcsec\ of any given [8.0]
source is 0.2\%.  Thus, these two sources are probably
related. Although unlikely, there is a chance that the excess for this
source may be due to background cirrus.

\textbf{(G036.8722-00.4112)-B8V}- This star exhibits double-peaked
H$\alpha$ emission but no other optical emission lines.  Disk
temperature (T$_{disk}$=427~K) and $L_{IR}/L_{*}$
(7.2$\times$10$^{-4}$) are similar to, but slightly higher than,
debris disks. Given that this star falls in the region of 8-24 $\mu$m
vs. $K$-8 $\mu$m color space not explainable by free-free emission, we
classify this star as having a transitional disk. The possible ages
for this star are widely variant, similar to either a late-stage
pre-main-sequence star or late-stage main-sequence star. Given the
nature of the excess and the large difference between the possible
ages, we adopt the younger age range of log age(yr)=6.11--6.32.

\textbf{HD 180398 (G047.3677+00.6199)-B8Ve}- This star has been
studied as a classical Be star \citep{Jaschek:1992}. Our Magellan
H$\alpha$ profile shows the same shape recorded by
\citet{Jaschek:1992}, but we also observe profile variability
consistent with the report of \citet{Hamdy:1991}.  This star exhibits
\ion{O}{1} $\lambda$ 8446 \AA\ emission but not Paschen emission. We
classify this source as a transition disk. The derived disk
temperature of 784$^{+19}_{-24}$~K is one of the highest in our
sample, as is its H$\alpha$ equivalent width
(EW(H$\alpha$$_{corr}$)=-22.47). The determined $L_{IR}/L_{*}$
of 2.8$\times$10$^{-3}$ is consistent with a massive debris disk or
transition disk.  The age of this star is somewhere between a
late-stage pre-main-sequence star and an early-stage main-sequence
star (log age(yr)=6.87--7.72), consistent with a transitional
disk. This star lies in a region of color-color space explainable
either by free-free emission or a hot dust excess.

\textbf{(G300.0992-00.0627)-A1}- This star was tentatively associated
with an X-ray source detected by the \textit{Uhuru} Satellite
\citep{Gursky:1972}.  However, the \textit{Rosat} Faint Source Catalog
position for the X-ray source is 580$\pm$31 \arcsec\ from the GLIMPSE
IR position, effectively ruling out an association with the IR excess
source \citep{Voges:2000}.  This star exhibits strong double-peaked
H$\alpha$ emission, \ion{O}{1} $\lambda$ 8446 and 7772 \AA\ emission,
without \ion{Fe}{2} or Paschen emission. We characterize the infrared
excess as due to a transition disk.  Circumstellar disk parameters of
T$_{disk}$=526$^{+41}_{-32}$~K and
$L_{IR}/L_{*}$=2.1$\times$10$^{-3}$ indicate a disk between
those of Herbig AeBe stars and those of debris disks.  This may be an
early form of transitional disk.

\textbf{HD 114757 (G305.4232-00.8209)-B6/8V(E)}-
\citet{Seidensticker:1989} derived M$_{V}$= -1.89$\pm$0.44,
A$_{V}$=1.11$\pm$0.13 mag, and distance modulus of
10.03$\pm$0.46 mag. H$\alpha$ exhibits a double-peaked emission-line
profile. This star exhibits a warm circumstellar disk temperature
(T$_{disk}$=675~K) and moderate $L_{IR}/L_{*}$
(3.8$\times$10$^{-3}$). Stellar age estimates indicate either an early
pre-main-sequence star or a middle/late-stage main-sequence
star.  This source falls in a region of color-color space that may be
explained by free-free emission or a hot dust component.  
Free-free emission is not sufficient to explain the large excess
at [24], suggesting the presence of circumstellar dust. 
This star exhibits both \ion{O}{1}
$\lambda$ 8446 and $\lambda$ 7772 \AA\ in emission without \ion{Fe}{2} or
Paschen lines in emission. We classify this source as a transition
disk and adopt the younger stellar age of log age (yr)=5.88.

\textbf{HD 118094 (G307.9784-00.7148)-B8V(N)}- \citet{Levenhagen:2004}
derived the fundamental parameters for this star, which they list as a
Be star. They found \textit{v} sin \textit{i}=239$\pm$30 km s$^{-1}$,
T$_{eff}$=13,500$\pm$550~K, log \textit{g}=3.90$\pm$0.15, log
age(yr)=8.03$\pm$0.04, log\ \lo=2.59$\pm$0.08, and \
\mo=4.2$\pm$0.2. Our derived values of projected rotational velocity,
240$\pm$20 km s$^{-1}$, and temperature, 13,220$\pm$520~K, are in
agreement with their derived values.  We determine this star to be
either a young pre-main-sequence object or a late-stage main-sequence
star, log age(yr)=6.00--6.23 or 7.66--8.03. Even using their derived
parameters and the \citet{Siess:2000} models we can not rule out the
possibility of a $\simeq$1 Myr star. The H$\alpha$ emission profile is
highly asymmetric and double-peaked. This star exhibits \ion{O}{1}
$\lambda$ 7772 \AA\ in emission without \ion{Fe}{2} or Paschen lines
in emission. Based on our stellar parameters, spectral features,
T$_{disk}$=557~K, $L_{IR}/L_{*}$=9.0$\times$10$^{-4}$, and the
possibility of a few Myr age, which we adopt, we classify this excess
owing to a transitional disk.

\textbf{HD 122620 (G311.6185+00.2469)-B8/9IV/V}- \citet{Westin:1985}
derived M$_{V}$=-0.613, distance=965 pc, T$_{eff}$=11,190~K, and log
age(yr)=8.275. Our derived effective temperature of 10,540$\pm$255 K
is lower. Our age estimates are log age(yr)=6.08--6.18 and
8.16--8.18. The older age is consistent with the literature but
inconsistent with the nature of the excess. The free-free component of
the [24] excess is one of the smallest of the sample. The
circumstellar disk parameters (T$_{disk}$=306~K and
$L_{IR}/L_{*}$=1.2$\times$10$^{-3}$) are consistent with a
warm massive debris disk. The weak double-peaked H$\alpha$ emission
profile is inconsistent with a debris disk nature for this source. We
classify this source as a transitional circumstellar disk and adopt
the younger stellar age range.

\textbf{HD 126578 (G314.3136-00.6977)-A1IV}- This star exhibits the
weakest H$\alpha$ emission profile. The circumstellar parameters
(T$_{disk}$=328~K and $L_{IR}/L_{*}$=9.6$\times$10$^{-4}$)
indicate a debris disk, but the weak H$\alpha$ emission precludes such
classification. \ion{O}{1} $\lambda$ 7772 \AA\ is seen in absorption
with an EW greater than the best-fit stellar model would indicate. 
This star falls in a region of color-color space where
excesses due to free-free emission or hot dust are permissible. 
This star may be at the end of its transitional disk phase.

\subsection{Probable Be stars (Type I Sources)}

\textbf{HD 165854 (G011.2691+00.4208)-B9V}- We find that the excess
for this star can be explained by free-free emission which supports
the standing classification of this star as a classical Be star. The only
emission we find is a weak double-peaked H$\alpha$
profile. \citet{Halbedel:1996} found the rotational velocity to be
250$\pm$10 km s$^{-1}$, and \citet{Yudin:2001} found it to be
242$\pm$10 km s$^{-1}$. These values are slightly greater than our
derived value of 220$\pm$15 km s$^{-1}$.

\textbf{HD 172030 (G027.0268+00.7224)-A0V}- This star exhibits a
double-peaked H$\alpha$ emission profile and \ion{Fe}{2} emission. The
spectrum for this star also shows the \ion{Ca}{2} IR triplet,
\ion{O}{1} $\lambda$ 8446 \AA, \ion{O}{1} $\lambda$ 7772 \AA\ , and
the Paschen lines in emission. The excess for this source can be fully
explained by free-free emission. It is interesting to note that this
star has a low projected rotational velocity, 25$\pm$5 km s$^{-1}$,
which is not typical for classical Be stars.

\textbf{HD 183035 (G051.6491-00.1182)-A0V}- The core of the marginally
double-peaked H$\alpha$ profile is nearly filled in, and no other
emission lines are present.  This star does exhibit broad absorption
features, i.e., $\lambda$ $\sim$4455 \AA\ inconsistent with the
measured stellar temperature. This absorption may be an indication of
circumstellar gas. However, this star falls in a region of color-color
space explainable by free-free emission and has an excess consistent
with such an origin. We conclude that this is a classical Be star.

\textbf{(G299.1677-00.3922)-B8}- This star exhibits 
double-peaked H$\alpha$ emission, Paschen emission, and
\ion{O}{1} $\lambda$ 8446 \AA\ emission. This type of
spectrum is typical for classical Be stars \citep{Andrillat:1990}.

\textbf{HD 107609 (G299.7090-00.9704)-B8/9IV}- 
This star exhibits a large fraction of the excess at [24] as free-free
emission. However, this source does not exhibit any 
lines other than H$\alpha$ in emission. 
Nevertheless, it is likely that this is a classical Be star.

\textbf{HD 121195 (G310.5420+00.4120)-B8IV(N)}- \citet{Westin:1985}
derived M$_{V}$=-2.953, distance=1,945 pc, and T$_{eff}$=12,940~K for
this star. Their temperature is in agreement with our derived
effective temperature of 12,430$\pm$510~K. We determined log
age(yr)=$\simeq$5.30 utilizing the \citet{Westin:1985} absolute visual
magnitude.  This extreme youth is probably not realistic given that,
at such an age, this star would be embedded and undergo substantial
extinction, which is not observed.  The H$\alpha$ emission profile is
weak and double-peaked, while \ion{O}{1} $\lambda$ 7772 \AA\ is found
in absorption. Since the excess can be explained fully by free-free
emission we classify this source as a Be star, but the optical
spectrum is similar to some transition disks.

\subsection{Other Circumstellar Disks and Infrared Excesses}

\textbf{HD 163296 (G007.2369+01.4894)-A1V}- This $MSX$ star has been
studied numerous times for both stellar parameters and circumstellar
disk evolution. This Herbig Ae (HAe) star is only 122$^{+17}_{-13}$ pc
distant. Rotational velocities for this source are consistent in the
literature, \citet{Finkenzeller:1985} derived a rotational velocity of
120$^{+20}_{-30}$ km s$^{-1}$, \citet{Halbedel:1996} derived a
rotational velocity of 120 km s$^{-1}$, \citet{Mora:2001} derived a
velocity of 130$\pm$6 km s$^{-1}$, all consistent with our calculated
value of 145$\pm$15 km s$^{-1}$. Stellar fundamental parameters were
calculated to be T$_{eff}$=9,200$\pm$270~K, M$_{Bol}$=0.48$\pm$0.10
from \citet{Cidale:2001}, and T$_{eff}$=9,475K, M$_{*}$=2.4\ \mo, and
5 Myr old from \citet{Mannings:1997}.  The age of this star varies in
the literature but is 4--6 Myr with H$\alpha$ equivalent width varying
between -12 -- -17 \AA \ \citep{Kessler-Silacci:2006}.
\citet{Pogodin:1994} found this star to exhibit rapid H$\alpha$
variability. Our derived T$_{eff}$ of 10,350$\pm$230~K is hotter than
literature values.  We find that the H$\alpha$ emission line is
stronger than the values reported in the literature
(EW(H$\alpha$)=-26.15$\pm$0.15 \AA). The \ion{Mg}{2} $\lambda$ 4481
\AA\ line is known to vary, and this may explain the difference
between our measured effective temperature and those in literature
\citep{Catala:1989}. Spectral types for this star vary between A0--A7,
confirming a range of stellar temperatures for this star
(\citealt{Finkenzeller:1985}; \citealt{Kessler-Silacci:2006}). Our
measured parameters are consistent with the pre-main-sequence nature
of this star.

\textbf{HD 167905 (G008.3752-03.6697)-F3V}- We classify this $MSX$
star as a T-Tauri object based on large $L_{IR}/L_{*}$
(2$\times$10$^{-1}$), weak H$\alpha$ emission with a P-Cygni profile,
low effective stellar temperature (T$_{eff}$=7260~K), and low
rotational velocity.

\textbf{HD 161643 (G009.4973+05.2441)-A7V}- We do not have enough data
to propose an evolutionary status for this $MSX$ star. This star is a
binary, and this may explain its slightly asymmetric H$\alpha$
absorption profile \citep{Fabricius:2000}. The secondary star is $\simeq$3
magnitudes fainter at $V$ in the $Hipparcos$ catalog.  Longer
wavelength confirmation of the excess is needed. Confirmation of an
excess having small $L_{IR}/L_{*}$ would indicate a debris disk.
This star does not exhibit any emission lines.

\textbf{HD 171149 (G025.6122+01.3020)-B9V}- \citet{Abt:1995} derived a
rotational velocity for this $MSX$ star of 280 km s$^{-1}$, while
\citet{Royer:2002b} found the rotational velocity to be $>$301 km
s$^{-1}$. Our derived value of 290$\pm$20 km s$^{-1}$ is consistent
with both prior measurements. We find H$\alpha$ in absorption. We also
find this source to exhibit broad absorption at $\sim$4455 \AA\ that
may be due to circumstellar material. Additional longer wavelength
observations are needed to classify this star. The H$\alpha$ profile
needs to be studied for variability and emission. Longer wavelength
infrared/radio measurements are needed to confirm the excess. With
additional optical and infrared/radio measurements it will be possible
to distinguish between an excess due to a classical Be star, debris
disk, or false excess.

\textbf{HD 182293 (G054.5163+02.4470)-K1IV}- Although listed
as a K1IV in \citet{Wright:2003}, our spectrum (not shown) indicates
a spectral type of II-III. We omit further
analysis of this source due to its evolved nature.

\textbf{HD 58647 (G229.4514+01.0143)-B9V}- There is a disparity
between measured projected rotational velocities for this star.
\citet{Grady:1996} and \citet{Manoj:2002} found it to be 280 km
s$^{-1}$, however, \citet{Mora:2001} found the projected rotational
velocity to be 118$\pm$4 km s$^{-1}$. Historically this star has been
classified as a Herbig AeBe star, but in \citet{Manoj:2002} they
conclude that this star is a classical Be star based on the low
fractional infrared luminosity. \citet{Monnier:2005} studied the
near-IR and disk properties of this source. They measured a stellar
luminosity of 295$\pm$50 \ \lo, and found a very low fractional
infrared luminosity when compared to other Herbig AeBe
stars. \citet{Vink:2002} found that H$\alpha$ line polarization for
this star does not agree with that of classical Be stars but exhibits
similar properties to Herbig AeBe stars in the (Q,U) plane. They also
measured the H$\alpha$ equivalent width to be -8.6\ \AA \ which is
consistent with our measurement of -7.82$\pm$0.45\ \AA \ assuming they
have similar uncertainties. We find this source to exhibit the
\ion{Ca}{2} IR triplet, \ion{O}{1} $\lambda$ 8446 \AA\, and
\ion{Fe}{2} in emission.  This source also shows strong \ion{O}{1}
$\lambda$ 7772 \AA\ , and $\sim$4455 \AA\ absorption. Based upon
the spectral properties of this source we conclude that HD 58647 is a
Herbig AeBe star. Given the low fractional infrared luminosity, this
star may be a more evolved form of Herbig AeBe star.

\textbf{HD 72106 (G257.6236+00.2459)-A0IV}- This is a visual double
star with a separation of 0.8 arcseconds \citep{Viera:2003}. We have a
spectrum of only the primary star, and no emission lines are
found. \citet{Viera:2003} conclude that the secondary of this system
exhibits the infrared excess, the H$\alpha$ emission, and is an evolved
HAeBe star. The mid-IR spectrum of this source exhibits similar
features of solar system comets also indicating the evolved nature of
a Herbig AeBe star \citep{Schutz:2005}. \citet{Wade:2005} measured the
magnetic field of both the primary and secondary of this system and
found the primary to exhibit a strong magnetic field. They argue that
the primary may be a progenitor to the Ap/Bp stars.

\textbf{HD 73461 (G265.5536-03.9951)-A5V\&A6/8V}- This is a known wide
visual binary \citep{Brosche:1989}. We have spectra of both the
primary and secondary. Neither source exhibits any emission lines. The
primary exhibits stronger \ion{O}{1} $\lambda$ 7772 \AA\ absorption
that may indicate circumstellar gas.  Longer wavelength confirmation
of the excess is necessary to determine the nature of the
circumstellar disk. Assuming the binary system is
coeval, the stellar parameters constrain the primary to 2.2$\pm$0.2\
\mo\ and secondary to 2.0$\pm$0.2\ \mo\ with an age of 6.81--6.83 Log
yr.

\textbf{HD 74534 (G269.5873-05.8882)-G0V}- \citet{Cutispoto:2002} detect
\ion{Ca}{2} $\lambda$ K emission. We do not detect this in our spectrum.
\citet{Cutispoto:2003} derived a mass of 2\ \mo, an age of $<$ 2 Gyr,
and a projected rotational velocity of 17 km s$^{-1}$. They find this
star to be an evolved giant. We also find this star to be an evolved
giant based on our spectrum. Our derived projected rotational velocity
of 10$\pm$5 km s$^{-1}$ is the lowest determined for our sample. 
We do not derive additional
parameters for this source due to its evolved nature.

\textbf{HD 135354 (G321.7868+00.4102)-B8VN}- \citet{Kozok:1985}
derived distances, absolute visual magnitudes, and reddenings for Be
stars, including this star. The absolute visual magnitude is -1.18 at
a distance of 1.01 kpc. Utilizing these parameters, we determined a
stellar age range of log age(yr)=5.83--6.12 and 8.01--8.13.  This star
exhibits H$\alpha$, \ion{O}{1} $\lambda$ 8446 \AA, \ion{O}{1}
$\lambda$ 7772 \AA, Paschen, and \ion{Fe}{2} emission.
\citet{Andrillat:1990} found that classical Be stars exhibiting
\ion{O}{1} emission also show Paschen emission. \citet{Jaschek:1993}
found that classical Be stars exhibit both \ion{O}{1} and \ion{Fe}{2}
emission.  This star exhibits both of these characteristics. The
fractional infrared luminosity and disk temperature
($L_{IR}/L_{*}$=2.1$\times$10$^{-3}$ and T$_{disk}$=556~K) are
consistent with transition disks but the measured H$\alpha$ EW is the
second strongest of the sample.  Given that the H$\alpha$ flux is
insufficient to explain the excess at [24] and the EW of H$\alpha$ is
consistent with Herbig AeBe stars we classify this star as a Herbig
AeBe star.

\textbf{HD 151228 (G339.4392-00.7791)-A0/1 IV/V}- The excess
at [8.0] is not confirmed for this star at [24], making it
a false mid-IR excess.

\textbf{HD 150625 (G340.0517+00.6687)-B8}- \citet{Kozok:1985} derived
an absolute visual magnitude of -1.87 and a distance of 0.83 kpc.
Using these parameters we constrain stellar ages to log
age(yr)=5.62--5.85 and 7.81--7.90. This star exhibits a strong
H$\alpha$ emission profile, \ion{Fe}{2}, Paschen, and \ion{O}{1}
$\lambda$ 8446 and 7772 \AA\ emission lines. The measured H$\alpha$
flux is insufficient to explain the excess at [24]. This is the other
Type II source that we classify as a Herbig AeBe star based on the
emission line spectrum.  The fractional infrared luminosity and disk
temperature ($L_{IR}/L_{*}$=1.5$\times$10$^{-2}$ and
T$_{disk}$=823~K) for this star is the largest in the GLIMPSE sample
and falls between Herbig AeBe stars and debris disks. This may be an
evolved Herbig AeBe star.

\textbf{HD 152404 (G347.3777+04.2010)-F5V}- This star is a
spectroscopic binary in the Upper Centarus-Lupus with an orbital
period of 13.6 days \citep{Manset:2005}. This star is $\simeq$ 17 Myr
old and classified as F5V \citep{Chen:2005}. This star has a strong
excess ratio given its age and may be a long lived primordial disk
\citep{Siegler:2007}.  Our projected rotational velocity for the two
stars is 15$\pm$5 km s$^{-1}$ consistent with 18.5$\pm$1.0 km s$^{-1}$
from \citep{Alencar:2003}. Due to the nature of this binary source we
do not derive effective temperature since our method of deriving
temperature does not extend to spectroscopic binaries of similar
mass. The H$\alpha$ profile indicates an unusual double peaked inverse
P-Cygni profile.  This may indicate circumstellar accretion.

\section{Sample Comparisons}

Given the similarity of the H$\alpha$ profiles for both stars
classified as candidate transition disks systems and classical Be
stars, we make a comparison of the spectral features for our sample
with a group of well-studied classical Be stars from
\citet{Andrillat:1990} and \citet{Jaschek:1993}. We use the following
conclusions from \citet{Andrillat:1990} for classical Be stars: 1)
\ion{Ca}{2} emission is infrequent for spectral types later than B6
and nearly all stars that exhibit \ion{Ca}{2} in emission exhibit
Paschen emission, 2) Paschen lines are in emission when \ion{O}{1} is
in emission, 3) When $IRAS$ 12 $\mu$m is in excess greater than 0.55
magnitudes of the predicted measurement then \ion{Ca}{2} is in
emission. In order to make the comparison with the last criterion we
used the [8.0] measurement in excess greater than 0.55 magnitudes.
Table~\ref{comp} lists the 10 stars we are able to classify as either
consistent or inconsistent with a classical Be star classification
based on the three criteria we mentioned above. In all 10 instances
the classifications we derive using the criteria of
\citet{Andrillat:1990} are consistent with our classifications.

In order to validate our classification scheme we also applied the
classifications in Figure~\ref{flow} to the classical Be stars in
\citet{Andrillat:1990}. In order to make the comparison over all
spectral features classified, we supplemented the spectra from
\citet{Andrillat:1990} with spectra from \citet{Jaschek:1993} to gain
the \ion{O}{1} $\lambda$ 7772 \AA\ feature. There were only 19 stars
in both \citet{Andrillat:1990} and \citet{Jaschek:1993}.  Using our
classification criteria we were able to classify only one of the 19
stars as a classical Be star based on spectra alone.  The remaining 18
stars required some knowledge of the SED. We fit Kurucz stellar models
to 2MASS photometry for these 18 stars in a manner similar to our
GLIMPSE stars in order to estimate the amount of extraphotospheric
flux emanating from these sources. Since these 18 stars are all bright
($\sim$V$\leq$ 5) and 2MASS photometry for such stars have large
uncertainties, we chose not to determine the free-free emission from
H$\alpha$ flux because of the larger normalization uncertainties
associated with the $K$ band. In order to determine the nature of the
excess emission we determined the slope of the excess component,
$\beta$, where F$_{\nu}$ $\propto$ $\nu^{\beta}$ after an estimate of
the stellar flux was removed. The slope of free-free emission in the
infrared varies from $\beta$ $\sim$ -0.1--2.0 depending on the optical
depth of the gas \citep{Wright:1974}. By contrast, sources with dust
components would have slopes $\beta$ $<$ -0.1. In order to determine
the slope of the infrared excess, we used flux measurements at 8.7
$\mu$m from \citet{Gehrz:1974} and the $IRAS$ Point Source Catalog at
25 $\mu$m measurements. Only 10 of the 18 stars had photometry at both
8.7 and 25 $\mu$m. We found that eight out of the 10 stars exhibit
positive slopes consistent with optically thick free-free
emission. $\eta$ Tau and 48 Per exhibited negative slopes.  Only 48
Per exhibits a slope less than -0.1 at the 3$\sigma$ level.  Given the
length of time between the two infrared measurements ($>$ 10 years)
the large negative slope may be a consequence of the star's known
variability (e.g., \citealt{Koen:2002}).  Assuming that slopes
consistent with free-free emission would cause the stars to be
classified as Type I sources (Be stars), 10 of 11 known classical Be
stars would be classified as Be stars using our classification
methods. By comparison, all three of our candidate debris disk systems
exhibit slopes less than -0.1 at a 3$\sigma$ or greater level. None of
the six classical Be stars within our sample exhibit slopes less than
-0.1 at a 3$\sigma$ level, meaning that their excesses are consistent with
optically thin or thick free-free emission.  Neither of the two
candidate Herbig AeBe stars exhibit a slope less than -0.1 at the
3$\sigma$ level. Four of the eight candidate transition disk
systems exhibit a slope less than -0.1 at greater than a 3$\sigma$,
level indicative of dust. These four stars are G014.4239-00.7657,
G036.8722-00.4112, G311.6185+00.2469, and G314.3136-00.6977.  The
explanation for the four remaining candidate transition disk systems
not exhibiting a $\beta$ $<$ -0.1 at a 3$\sigma$ level may be
variability. An additional scenario for the shallower slopes
is a narrow ring of single temperature hot dust, something analogous
to $\zeta$ Leporis \citep{Chen:2001}. Another possible
explanation for the excess for these four stars may be optically thick
free-free emission. In order to be conservative, we added a possible
Be classification to the remaining four candidate transition disk
systems, G047.3677+00.6199, G300.0992-00.0627, G305.4232-00.8229, and
G307.9784-00.7148 in Table~\ref{tsum}.

\section{Conclusion}

In an effort to identify the nature of the infrared excesses, we
obtained high-resolution optical spectroscopy for 31 luminosity class
V or IV stars of spectral type B8 or later exhibiting 8 $\mu$m mid-IR
excesses in the GLIMPSE and $MSX$ catalogs. We measured stellar
effective temperatures for 28 stars and projected rotational
velocities and H$\alpha$ equivalent widths for 30 stars. The projected
rotational velocities of the 20 GLIMPSE stars lie between 25 and 310
km s$^{-1}$ with a mean and dispersion of 209$\pm$76 km s$^{-1}$
consistent with the mixed nature of the sample.  We used derived
stellar parameters and distances to estimate stellar ages ranging from
$\simeq$1 Myr to 100 Myr for 16 stars (9 of the 20 GLIMPSE
stars). Twenty of the 31 stars (16 of the 20 GLIMPSE stars) exhibit
H$\alpha$ emission, consistent with characteristics of
pre-main-sequence stars (i.e., Herbig AeBe stars) or classical Be
stars.  H$\alpha$ equivalent widths range between -39.38 \AA\
(characteristic of classical Be stars or Herbig AeBe stars) and 9.41
\AA\ (main-sequence B and A stars), suggesting a variety of origins in
a sample of objects spanning a range of stellar evolutionary states.

For 16 of the 20 GLIMPSE stars showing H$\alpha$ emission, we use the
H$\alpha$ flux to estimate the expected contribution of (free-free and
free-bound) emission to the mid-IR excesses. In six out of the 16
cases, the expected IR excess can fully account for the observed
infrared excess (Type I sources). We did not confirm the infrared
excess at 24 $\mu$m in one case. In the remaining nine cases (Type II
sources) the expected contribution from free-free emission cannot
account for all of the observed IR excess, implying that either 1) the
circumstellar excess is time-variable and the non-contemporaneous
nature of H$\alpha$ and mid-IR data render this estimate invalid, or
2) there is a dust component to the IR excess. The lack of evidence
for H$\alpha$ variability for the majority of our sources, combined
with the fact that in no instance does the H$\alpha$ \textit{over
predict} the IR excess component, makes the first scenario
unlikely. We conclude that the second scenario, the presence of a
circumstellar dust component, is most probable. Combining correlations
of spectral features for sources with known characteristics we find
that the nine Type II sources with IR excesses require a dust
component and are likely to be young main-sequence or near
main-sequence objects harboring warm (300--800 K) circumstellar dust.
The nominal fractional infrared luminosities for these nine sources
after removal of the free-free component is
$\simeq$10$^{-2}$--10$^{-3}$. For the six of these nine sources with
age estimates, there is a correlation between estimated age and
H$\alpha$ EW consistent with a progression of disk clearing with
time. Such characteristics make these stars candidates for being
``transition disk'' or ``evolved-HAeBe'' systems containing rare
dissipating primordial gas and dust disks.  One of the four systems
lacking H$\alpha$ emission has a very large $K$-[24] color, $>$ 4
magnitudes. Such a large color difference implies that the excess is
optically thick at [24]. We conclude that this source is also a
transition disk. The remaining three systems lacking H$\alpha$
emission have disk temperatures of 300--400 K and fractional infrared
luminosities of 10$^{-3}$.  These characteristics make them candidates
for dusty debris disk systems.

We show that $K$-8, 8-24 color-color diagrams are a useful tool to
distinguish cool (T$_{d}$ $<$ 300~K) dusty mid-IR excess objects from
hot (T$_{d}$ $>$ 500~K) dusty or free-free mid-IR excesses.  Such
diagrams reveal that additional information beyond near/mid-IR colors
is necessary to distinguish hot dust from free-free excesses. This
diagram also reinforces the idea that some stars cataloged as
classical Be stars may in fact be late-stage pre-main-sequence stars
containing circumstellar dust.

\acknowledgments

We would like to thank Mary Putnam and Sally Oey for their assistance
in obtaining the Magellan observations. We would like to thank Karen
Kinemuchi for her helpful discussions. We kindly thank M.R. Meade,
B.L. Babler, R. Indebetouw, B. A. Whitney, C. Watson, and
E. Churchwell for their use of the GLIMPSE data reduction
pipeline. B.U. acknowledges support from a NASA Graduate Student
Researchers Program fellowship, grant NNX06AI28H.  This research has
made use of the SIMBAD database, operated at CDS, Strasbourg,
France. This publication makes use of data products from the Two
Micron All Sky Survey, which is a joint project of the University of
Massachusetts and the Infrared Processing and Analysis
Center/California Institute of Technology, funded by the National
Aeronautics and Space Administration and the National Science
Foundation.


\begin{thebibliography}{0}
\expandafter\ifx\csname natexlab\endcsname\relax\def\natexlab#1{#1}\fi

\end{thebibliography}


\begin{thebibliography}{113}
\expandafter\ifx\csname natexlab\endcsname\relax\def\natexlab#1{#1}\fi

\bibitem[{{Abt} {et~al.}(2002){Abt}, {Levato}, \& {Grosso}}]{Abt:2002}
{Abt}, H.~A., {Levato}, H., \& {Grosso}, M. 2002, \apj, 573, 359

\bibitem[{{Abt} \& {Morrell}(1995)}]{Abt:1995}
{Abt}, H.~A., \& {Morrell}, N.~I. 1995, \apjs, 99, 135

\bibitem[{{Alencar} {et~al.}(2003){Alencar}, {Melo}, {Dullemond}, {Andersen},
  {Batalha}, {Vaz}, \& {Mathieu}}]{Alencar:2003}
{Alencar}, S.~H.~P., {Melo}, C.~H.~F., {Dullemond}, C.~P., {Andersen}, J.,
  {Batalha}, C., {Vaz}, L.~P.~R., \& {Mathieu}, R.~D. 2003, \aap, 409, 1037

\bibitem[{{Andrillat} {et~al.}(1990){Andrillat}, {Jaschek}, \&
  {Jaschek}}]{Andrillat:1990}
{Andrillat}, A., {Jaschek}, M., \& {Jaschek}, C. 1990, \aaps, 84, 11

\bibitem[{{Ashok} {et~al.}(1984){Ashok}, {Bhatt}, {Kulkarni}, \&
  {Joshi}}]{Ashok:1984}
{Ashok}, N.~M., {Bhatt}, H.~C., {Kulkarni}, P.~V., \& {Joshi}, S.~C. 1984,
  \mnras, 211, 471

\bibitem[{{Aumann} {et~al.}(1984){Aumann}, {Beichman}, {Gillett}, {de Jong},
  {Houck}, {Low}, {Neugebauer}, {Walker}, \& {Wesselius}}]{Aumann:1984}
{Aumann}, H.~H., {Beichman}, C.~A., {Gillett}, F.~C., {de Jong}, T., {Houck},
  J.~R., {Low}, F.~J., {Neugebauer}, G., {Walker}, R.~G., \& {Wesselius}, P.~R.
  1984, \apjl, 278, L23

\bibitem[{{Aumann} \& {Probst}(1991)}]{Aumann:1991}
{Aumann}, H.~H., \& {Probst}, R.~G. 1991, \apj, 368, 264

\bibitem[{{Backman} \& {Paresce}(1993)}]{Backman:1993}
{Backman}, D.~E., \& {Paresce}, F. 1993, in Protostars and Planets III, ed.
  E.~H. {Levy} \& J.~I. {Lunine}, 1253--1304

\bibitem[{{Balona}(2000)}]{Balona:2000}
{Balona}, L.~A. 2000, in Astronomical Society of the Pacific Conference Series,
  Vol. 214, IAU Colloq. 175: The Be Phenomenon in Early-Type Stars, ed. M.~A.
  {Smith}, H.~F. {Henrichs}, \& J.~{Fabregat}, 1--58381

\bibitem[{{Benjamin} {et~al.}(2003){Benjamin}, {Churchwell}, {Babler}, {Bania},
  {Clemens}, {Cohen}, {Dickey}, {Indebetouw}, {Jackson}, {Kobulnicky},
  {Lazarian}, {Marston}, {Mathis}, {Meade}, {Seager}, {Stolovy}, {Watson},
  {Whitney}, {Wolff}, \& {Wolfire}}]{Benjamin:2003}
{Benjamin}, R.~A., {Churchwell}, E., {Babler}, B.~L., {Bania}, T.~M.,
  {Clemens}, D.~P., {Cohen}, M., {Dickey}, J.~M., {Indebetouw}, R., {Jackson},
  J.~M., {Kobulnicky}, H.~A., {Lazarian}, A., {Marston}, A.~P., {Mathis},
  J.~S., {Meade}, M.~R., {Seager}, S., {Stolovy}, S.~R., {Watson}, C.,
  {Whitney}, B.~A., {Wolff}, M.~J., \& {Wolfire}, M.~G. 2003, \pasp, 115, 953

\bibitem[{{Bernstein} {et~al.}(2003){Bernstein}, {Shectman}, {Gunnels},
  {Mochnacki}, \& {Athey}}]{Bernstein:2003}
{Bernstein}, R., {Shectman}, S.~A., {Gunnels}, S.~M., {Mochnacki}, S., \&
  {Athey}, A.~E. 2003, in Presented at the Society of Photo-Optical
  Instrumentation Engineers (SPIE) Conference, Vol. 4841, Instrument Design and
  Performance for Optical/Infrared Ground-based Telescopes. Edited by Iye,
  Masanori; Moorwood, Alan F. M. Proceedings of the SPIE, Volume 4841, pp.
  1694-1704 (2003)., ed. M.~{Iye} \& A.~F.~M. {Moorwood}, 1694--1704

\bibitem[{{Bjorkman} {et~al.}(1998){Bjorkman}, {Miroshnichenko}, {Bjorkman},
  {Meade}, {Babler}, {Code}, {Anderson}, {Fox}, {Johnson}, {Weitenbeck},
  {Zellner}, \& {Lupie}}]{Bjorkman:1998}
{Bjorkman}, K.~S., {Miroshnichenko}, A.~S., {Bjorkman}, J.~E., {Meade}, M.~R.,
  {Babler}, B.~L., {Code}, A.~D., {Anderson}, C.~M., {Fox}, G.~K., {Johnson},
  J.~J., {Weitenbeck}, A.~J., {Zellner}, N.~E.~B., \& {Lupie}, O.~L. 1998,
  \apj, 509, 904

\bibitem[{{Boehm} \& {Catala}(1995)}]{Bohm:1995}
{Boehm}, T., \& {Catala}, C. 1995, \aap, 301, 155

\bibitem[{{Bowen}(1947)}]{Bowen:1947}
{Bowen}, I.~S. 1947, \pasp, 59, 196

\bibitem[{{Brocklehurst}(1971)}]{Brocklehurst:1971}
{Brocklehurst}, M. 1971, \mnras, 153, 471

\bibitem[{{Brosche} \& {Sinachopoulos}(1989)}]{Brosche:1989}
{Brosche}, P., \& {Sinachopoulos}, D. 1989, Bulletin d'Information du Centre de
  Donnees Stellaires, 37, 109

\bibitem[{{Calvet} {et~al.}(2005){Calvet}, {D'Alessio}, {Watson},
  {Franco-Hern{\'a}ndez}, {Furlan}, {Green}, {Sutter}, {Forrest}, {Hartmann},
  {Uchida}, {Keller}, {Sargent}, {Najita}, {Herter}, {Barry}, \&
  {Hall}}]{Calvet:2005}
{Calvet}, N., {D'Alessio}, P., {Watson}, D.~M., {Franco-Hern{\'a}ndez}, R.,
  {Furlan}, E., {Green}, J., {Sutter}, P.~M., {Forrest}, W.~J., {Hartmann}, L.,
  {Uchida}, K.~I., {Keller}, L.~D., {Sargent}, B., {Najita}, J., {Herter},
  T.~L., {Barry}, D.~J., \& {Hall}, P. 2005, \apjl, 630, L185

\bibitem[{{Carroll}(1933)}]{Carroll:1933}
{Carroll}, J.~A. 1933, \mnras, 93, 478

\bibitem[{{Catala} {et~al.}(1986){Catala}, {Czarny}, {Felenbok}, \&
  {Praderie}}]{Catala:1986}
{Catala}, C., {Czarny}, J., {Felenbok}, P., \& {Praderie}, F. 1986, \aap, 154,
  103

\bibitem[{{Catala} {et~al.}(1989){Catala}, {Simon}, {Praderie}, {Talavera},
  {The}, \& {Tjin A Djie}}]{Catala:1989}
{Catala}, C., {Simon}, T., {Praderie}, F., {Talavera}, A., {The}, P.~S., \&
  {Tjin A Djie}, H.~R.~E. 1989, \aap, 221, 273

\bibitem[{{Chen} \& {Jura}(2001)}]{Chen:2001}
{Chen}, C.~H., \& {Jura}, M. 2001, \apjl, 560, L171

\bibitem[{{Chen} {et~al.}(2005){Chen}, {Jura}, {Gordon}, \&
  {Blaylock}}]{Chen:2005}
{Chen}, C.~H., {Jura}, M., {Gordon}, K.~D., \& {Blaylock}, M. 2005, \apj, 623,
  493

\bibitem[{{Chen} {et~al.}(2006){Chen}, {Sargent}, {Bohac}, {Kim},
  {Leibensperger}, {Jura}, {Najita}, {Forrest}, {Watson}, {Sloan}, \&
  {Keller}}]{Chen:2006}
{Chen}, C.~H., {Sargent}, B.~A., {Bohac}, C., {Kim}, K.~H., {Leibensperger},
  E., {Jura}, M., {Najita}, J., {Forrest}, W.~J., {Watson}, D.~M., {Sloan},
  G.~C., \& {Keller}, L.~D. 2006, \apjs, 166, 351

\bibitem[{{Cidale} {et~al.}(2001){Cidale}, {Zorec}, \&
  {Tringaniello}}]{Cidale:2001}
{Cidale}, L., {Zorec}, J., \& {Tringaniello}, L. 2001, \aap, 368, 160

\bibitem[{{Cox}(2000)}]{Cox:2000}
{Cox}, A.~N. 2000, {Allen's astrophysical quantities} (Allen's astrophysical
  quantities, 4th ed.~Publisher: New York: AIP Press; Springer, 2000.~Editedy
  by Arthur N.~Cox.~ ISBN: 0387987460)

\bibitem[{{Cutispoto} {et~al.}(2002){Cutispoto}, {Pastori}, {Pasquini}, {de
  Medeiros}, {Tagliaferri}, \& {Andersen}}]{Cutispoto:2002}
{Cutispoto}, G., {Pastori}, L., {Pasquini}, L., {de Medeiros}, J.~R.,
  {Tagliaferri}, G., \& {Andersen}, J. 2002, \aap, 384, 491

\bibitem[{{Cutispoto} {et~al.}(2003){Cutispoto}, {Tagliaferri}, {de Medeiros},
  {Pastori}, {Pasquini}, \& {Andersen}}]{Cutispoto:2003}
{Cutispoto}, G., {Tagliaferri}, G., {de Medeiros}, J.~R., {Pastori}, L.,
  {Pasquini}, L., \& {Andersen}, J. 2003, \aap, 397, 987

\bibitem[{{Dachs} {et~al.}(1988){Dachs}, {Kiehling}, \& {Engels}}]{Dachs:1988}
{Dachs}, J., {Kiehling}, R., \& {Engels}, D. 1988, \aap, 194, 167

\bibitem[{{Deroo} {et~al.}(2007){Deroo}, {Acke}, {Verhoelst}, {Dominik},
  {Tatulli}, \& {van Winckel}}]{Deroo:2007}
{Deroo}, P., {Acke}, B., {Verhoelst}, T., {Dominik}, C., {Tatulli}, E., \&
  {van Winckel}, H. 2007, \aap, 474, 45

\bibitem[{{Egan} {et~al.}(2003){Egan}, {Price}, {Kraemer}, {Mizuno}, {Carey},
  {Wright}, {Engelke}, {Cohen}, \& {Gugliotti}}]{Egan:2003}
{Egan}, M.~P., {Price}, S.~D., {Kraemer}, K.~E., {Mizuno}, D.~R., {Carey},
  S.~J., {Wright}, C.~O., {Engelke}, C.~W., {Cohen}, M., \& {Gugliotti}, M.~G.
  2003, VizieR Online Data Catalog, 5114, 0

\bibitem[{{Fabricius} \& {Makarov}(2000)}]{Fabricius:2000}
{Fabricius}, C., \& {Makarov}, V.~V. 2000, VizieR Online Data Catalog, 335,
  60141

\bibitem[{{Finkenzeller}(1985)}]{Finkenzeller:1985}
{Finkenzeller}, U. 1985, \aap, 151, 340

\bibitem[{{Finkenzeller} \& {Mundt}(1984)}]{Finkenzeller:1984}
{Finkenzeller}, U., \& {Mundt}, R. 1984, \aaps, 55, 109

\bibitem[{{Furlan} {et~al.}(2007){Furlan}, {Sargent}, {Calvet}, {Forrest},
  {D'Alessio}, {Hartmann}, {Watson}, {Green}, {Najita}, \&
  {Chen}}]{Furlan:2007}
{Furlan}, E., {Sargent}, B., {Calvet}, N., {Forrest}, W.~J., {D'Alessio}, P.,
  {Hartmann}, L., {Watson}, D.~M., {Green}, J.~D., {Najita}, J., \& {Chen},
  C.~H. 2007, \apj, 664, 1176

\bibitem[{{Gehrz} {et~al.}(1974){Gehrz}, {Hackwell}, \& {Jones}}]{Gehrz:1974}
{Gehrz}, R.~D., {Hackwell}, J.~A., \& {Jones}, T.~W. 1974, \apj, 191, 675

\bibitem[{{Grady} {et~al.}(1996){Grady}, {Perez}, {Talavera}, {Bjorkman}, {de
  Winter}, {The}, {Molster}, {van den Ancker}, {Sitko}, {Morrison}, {Beaver},
  {McCollum}, \& {Castelaz}}]{Grady:1996}
{Grady}, C.~A., {Perez}, M.~R., {Talavera}, A., {Bjorkman}, K.~S., {de Winter},
  D., {The}, P.-S., {Molster}, F.~J., {van den Ancker}, M.~E., {Sitko}, M.~L.,
  {Morrison}, N.~D., {Beaver}, M.~L., {McCollum}, B., \& {Castelaz}, M.~W.
  1996, \aaps, 120, 157

\bibitem[{{Gray}(1992)}]{Gray:1992}
{Gray}, D.~F. 1992, {The Observation and Analysis of Stellar Photospheres} (The
  Observation and Analysis of Stellar Photospheres, by David F.~Gray,
  pp.~470.~ISBN 0521408687.~Cambridge, UK: Cambridge University Press, June
  1992.)

\bibitem[{{Gursky}(1972)}]{Gursky:1972}
{Gursky}, H. 1972, \apjl, 175, L141

\bibitem[{{Halbedel}(1996)}]{Halbedel:1996}
{Halbedel}, E.~M. 1996, \pasp, 108, 833

\bibitem[{{Hamann} \& {Persson}(1992)}]{Hamann:1992}
{Hamann}, F., \& {Persson}, S.~E. 1992, \apjs, 82, 285

\bibitem[{{Hamdy} {et~al.}(1991){Hamdy}, {Elazm}, \& {Saad}}]{Hamdy:1991}
{Hamdy}, M.~A., {Elazm}, M.~S.~A., \& {Saad}, S.~M. 1991, \apss, 186, 161

\bibitem[{{Hanuschik} {et~al.}(1996){Hanuschik}, {Hummel}, {Sutorius},
  {Dietle}, \& {Thimm}}]{Hanuschik:1996}
{Hanuschik}, R.~W., {Hummel}, W., {Sutorius}, E., {Dietle}, O., \& {Thimm}, G.
  1996, \aaps, 116, 309

\bibitem[{{Hartmann} {et~al.}(1993){Hartmann}, {Kenyon}, \&
  {Calvet}}]{Hartmann:1993}
{Hartmann}, L., {Kenyon}, S.~J., \& {Calvet}, N. 1993, \apj, 407, 219

\bibitem[{{Hartmann} {et~al.}(2005){Hartmann}, {Megeath}, {Allen}, {Luhman},
  {Calvet}, {D'Alessio}, {Franco-Hernandez}, \& {Fazio}}]{Hartmann:2005}
{Hartmann}, L., {Megeath}, S.~T., {Allen}, L., {Luhman}, K., {Calvet}, N.,
  {D'Alessio}, P., {Franco-Hernandez}, R., \& {Fazio}, G. 2005, \apj, 629, 881

\bibitem[{{Herbig}(1960)}]{Herbig:1960}
{Herbig}, G.~H. 1960, \apjs, 4, 337

\bibitem[{{Hern{\'a}ndez} {et~al.}(2006){Hern{\'a}ndez}, {Brice{\~n}o},
  {Calvet}, {Hartmann}, {Muzerolle}, \& {Quintero}}]{Hernandez:2006}
{Hern{\'a}ndez}, J., {Brice{\~n}o}, C., {Calvet}, N., {Hartmann}, L.,
  {Muzerolle}, J., \& {Quintero}, A. 2006, \apj, 652, 472

\bibitem[{{Hern{\'a}ndez} {et~al.}(2004){Hern{\'a}ndez}, {Calvet},
  {Brice{\~n}o}, {Hartmann}, \& {Berlind}}]{Hernandez:2004}
{Hern{\'a}ndez}, J., {Calvet}, N., {Brice{\~n}o}, C., {Hartmann}, L., \&
  {Berlind}, P. 2004, \aj, 127, 1682

\bibitem[{{Hernandez} {et~al.}(2007){Hernandez}, {Hartmann}, {Megeath},
  {Gutermuth}, {Muzerolle}, {Calvet}, {Vivas}, {Briceno}, {Allen}, {Stauffer},
  {Young}, \& {Fazio}}]{Hernandez:2007}
{Hernandez}, J., {Hartmann}, L., {Megeath}, T., {Gutermuth}, R., {Muzerolle},
  J., {Calvet}, N., {Vivas}, A.~K., {Briceno}, C., {Allen}, L., {Stauffer}, J.,
  {Young}, E., \& {Fazio}, G. 2007, ArXiv Astrophysics e-prints

\bibitem[{{Hillenbrand} {et~al.}(2008){Hillenbrand}, {Carpenter}, {Kim},
  {Meyer}, {Backman}, {Moro-Martin}, {Hollenbach}, {Hines}, {Pascucci}, \&
  {Bouwman}}]{Hillenbrand:2008}
{Hillenbrand}, L.~A., {Carpenter}, J.~M., {Kim}, J.~S., {Meyer}, M.~R.,
  {Backman}, D.~E., {Moro-Martin}, A., {Hollenbach}, D.~J., {Hines}, D.~C.,
  {Pascucci}, I., \& {Bouwman}, J. 2008, ArXiv e-prints, 801

\bibitem[{{Hillenbrand} {et~al.}(1992){Hillenbrand}, {Strom}, {Vrba}, \&
  {Keene}}]{Hillenbrand:1992}
{Hillenbrand}, L.~A., {Strom}, S.~E., {Vrba}, F.~J., \& {Keene}, J. 1992, \apj,
  397, 613

\bibitem[{{Israel} \& {Kennicutt}(1980)}]{Israel:1980}
{Israel}, F.~P., \& {Kennicutt}, R.~C. 1980, \aplett, 21, 1

\bibitem[{{Jaschek} \& {Jaschek}(1992)}]{Jaschek:1992}
{Jaschek}, C., \& {Jaschek}, M. 1992, \aaps, 95, 535

\bibitem[{{Jaschek} {et~al.}(1988){Jaschek}, {Jaschek}, {Egret}, \&
  {Andrillat}}]{Andrillat:1988}
{Jaschek}, C., {Jaschek}, M., {Egret}, D., \& {Andrillat}, Y. 1988, \aap, 192,
  285

\bibitem[{{Jaschek} \& {Egret}(1982)}]{Jaschek:1982}
{Jaschek}, M., \& {Egret}, D. 1982, in IAU Symposium, Vol.~98, Be Stars, ed.
  M.~{Jaschek} \& H.-G. {Groth}, 261

\bibitem[{{Jaschek} {et~al.}(1993){Jaschek}, {Jaschek}, \&
  {Andrillat}}]{Jaschek:1993}
{Jaschek}, M., {Jaschek}, C., \& {Andrillat}, Y. 1993, \aaps, 97, 781

\bibitem[{{Jaschek} {et~al.}(1980){Jaschek}, {Jaschek}, {Hubert-Delplace}, \&
  {Hubert}}]{Jaschek:1980}
{Jaschek}, M., {Jaschek}, C., {Hubert-Delplace}, A.-M., \& {Hubert}, H. 1980,
  \aaps, 42, 103

\bibitem[{{Jenniskens} \& {Desert}(1994)}]{Jenniskens:1994}
{Jenniskens}, P., \& {Desert}, F.-X. 1994, \aaps, 106, 39

\bibitem[{{Jura}(1999)}]{Jura:1999}
{Jura}, M. 1999, \apj, 515, 706

\bibitem[{{Karzas} \& {Latter}(1961)}]{Karzas:1961}
{Karzas}, W.~J., \& {Latter}, R. 1961, \apjs, 6, 167

\bibitem[{{Kessler-Silacci} {et~al.}(2006){Kessler-Silacci}, {Augereau},
  {Dullemond}, {Geers}, {Lahuis}, {Evans}, {van Dishoeck}, {Blake}, {Boogert},
  {Brown}, {J{\o}rgensen}, {Knez}, \& {Pontoppidan}}]{Kessler-Silacci:2006}
{Kessler-Silacci}, J., {Augereau}, J.-C., {Dullemond}, C.~P., {Geers}, V.,
  {Lahuis}, F., {Evans}, II, N.~J., {van Dishoeck}, E.~F., {Blake}, G.~A.,
  {Boogert}, A.~C.~A., {Brown}, J., {J{\o}rgensen}, J.~K., {Knez}, C., \&
  {Pontoppidan}, K.~M. 2006, \apj, 639, 275

\bibitem[{{Koen} \& {Eyer}(2002)}]{Koen:2002}
{Koen}, C., \& {Eyer}, L. 2002, \mnras, 331, 45

\bibitem[{{Kozok}(1985)}]{Kozok:1985}
{Kozok}, J.~R. 1985, \aaps, 62, 7

\bibitem[{{Kurucz}(1993)}]{Kurucz:1993}
{Kurucz}, R.~L. 1993, {SYNTHE spectrum synthesis programs and line data}
  (Kurucz CD-ROM, Cambridge, MA: Smithsonian Astrophysical Observatory, |c1993,
  December 4, 1993)

\bibitem[{{Lagrange} {et~al.}(2000){Lagrange}, {Backman}, \&
  {Artymowicz}}]{Lagrange:2000}
{Lagrange}, A.-M., {Backman}, D.~E., \& {Artymowicz}, P. 2000, Protostars and
  Planets IV, 639

\bibitem[{{Levenhagen} \& {Leister}(2004)}]{Levenhagen:2004}
{Levenhagen}, R.~S., \& {Leister}, N.~V. 2004, \aj, 127, 1176

\bibitem[{{Mannings} \& {Sargent}(1997)}]{Mannings:1997}
{Mannings}, V., \& {Sargent}, A.~I. 1997, \apj, 490, 792

\bibitem[{{Manoj} {et~al.}(2006){Manoj}, {Bhatt}, {Maheswar}, \&
  {Muneer}}]{Manoj:2006}
{Manoj}, P., {Bhatt}, H.~C., {Maheswar}, G., \& {Muneer}, S. 2006, \apj, 653,
  657

\bibitem[{{Manoj} {et~al.}(2002){Manoj}, {Maheswar}, \& {Bhatt}}]{Manoj:2002}
{Manoj}, P., {Maheswar}, G., \& {Bhatt}, H.~C. 2002, \mnras, 334, 419

\bibitem[{{Manset} {et~al.}(2005){Manset}, {Bastien}, \&
  {Bertout}}]{Manset:2005}
{Manset}, N., {Bastien}, P., \& {Bertout}, C. 2005, \aj, 129, 480

\bibitem[{{McDavid}(2001)}]{McDavid:2001}
{McDavid}, D. 2001, \apj, 553, 1027

\bibitem[{{McSwain} \& {Gies}(2005)}]{McSwain:2005}
{McSwain}, M.~V., \& {Gies}, D.~R. 2005, \apjs, 161, 118

\bibitem[{{Meyer} {et~al.}(2004){Meyer}, {Hillenbrand}, {Backman}, {Beckwith},
  {Bouwman}, {Brooke}, {Carpenter}, {Cohen}, {Gorti}, {Henning}, {Hines},
  {Hollenbach}, {Kim}, {Lunine}, {Malhotra}, {Mamajek}, {Metchev},
  {Moro-Martin}, {Morris}, {Najita}, {Padgett}, {Rodmann}, {Silverstone},
  {Soderblom}, {Stauffer}, {Stobie}, {Strom}, {Watson}, {Weidenschilling},
  {Wolf}, {Young}, {Engelbracht}, {Gordon}, {Misselt}, {Morrison}, {Muzerolle},
  \& {Su}}]{Meyer:2004}
{Meyer}, M.~R., {Hillenbrand}, L.~A., {Backman}, D.~E., {Beckwith}, S.~V.~W.,
  {Bouwman}, J., {Brooke}, T.~Y., {Carpenter}, J.~M., {Cohen}, M., {Gorti}, U.,
  {Henning}, T., {Hines}, D.~C., {Hollenbach}, D., {Kim}, J.~S., {Lunine}, J.,
  {Malhotra}, R., {Mamajek}, E.~E., {Metchev}, S., {Moro-Martin}, A., {Morris},
  P., {Najita}, J., {Padgett}, D.~L., {Rodmann}, J., {Silverstone}, M.~D.,
  {Soderblom}, D.~R., {Stauffer}, J.~R., {Stobie}, E.~B., {Strom}, S.~E.,
  {Watson}, D.~M., {Weidenschilling}, S.~J., {Wolf}, S., {Young}, E.,
  {Engelbracht}, C.~W., {Gordon}, K.~D., {Misselt}, K., {Morrison}, J.,
  {Muzerolle}, J., \& {Su}, K. 2004, \apjs, 154, 422

\bibitem[{{Monet}(1998)}]{Monet:1998}
{Monet}, D.~G. 1998, in Bulletin of the American Astronomical Society, Vol.~30,
  Bulletin of the American Astronomical Society, 1427--+

\bibitem[{{Monet} {et~al.}(2003){Monet}, {Levine}, {Canzian}, {Ables}, {Bird},
  {Dahn}, {Guetter}, {Harris}, {Henden}, {Leggett}, {Levison}, {Luginbuhl},
  {Martini}, {Monet}, {Munn}, {Pier}, {Rhodes}, {Riepe}, {Sell}, {Stone},
  {Vrba}, {Walker}, {Westerhout}, {Brucato}, {Reid}, {Schoening}, {Hartley},
  {Read}, \& {Tritton}}]{Monet:2003}
{Monet}, D.~G., {Levine}, S.~E., {Canzian}, B., {Ables}, H.~D., {Bird}, A.~R.,
  {Dahn}, C.~C., {Guetter}, H.~H., {Harris}, H.~C., {Henden}, A.~A., {Leggett},
  S.~K., {Levison}, H.~F., {Luginbuhl}, C.~B., {Martini}, J., {Monet},
  A.~K.~B., {Munn}, J.~A., {Pier}, J.~R., {Rhodes}, A.~R., {Riepe}, B., {Sell},
  S., {Stone}, R.~C., {Vrba}, F.~J., {Walker}, R.~L., {Westerhout}, G.,
  {Brucato}, R.~J., {Reid}, I.~N., {Schoening}, W., {Hartley}, M., {Read},
  M.~A., \& {Tritton}, S.~B. 2003, \aj, 125, 984

\bibitem[{{Monnier} {et~al.}(2005){Monnier}, {Millan-Gabet}, {Billmeier},
  {Akeson}, {Wallace}, {Berger}, {Calvet}, {D'Alessio}, {Danchi}, {Hartmann},
  {Hillenbrand}, {Kuchner}, {Rajagopal}, {Traub}, {Tuthill}, {Boden}, {Booth},
  {Colavita}, {Gathright}, {Hrynevych}, {Le Mignant}, {Ligon}, {Neyman},
  {Swain}, {Thompson}, {Vasisht}, {Wizinowich}, {Beichman}, {Beletic},
  {Creech-Eakman}, {Koresko}, {Sargent}, {Shao}, \& {van Belle}}]{Monnier:2005}
{Monnier}, J.~D., {Millan-Gabet}, R., {Billmeier}, R., {Akeson}, R.~L.,
  {Wallace}, D., {Berger}, J.-P., {Calvet}, N., {D'Alessio}, P., {Danchi},
  W.~C., {Hartmann}, L., {Hillenbrand}, L.~A., {Kuchner}, M., {Rajagopal}, J.,
  {Traub}, W.~A., {Tuthill}, P.~G., {Boden}, A., {Booth}, A., {Colavita}, M.,
  {Gathright}, J., {Hrynevych}, M., {Le Mignant}, D., {Ligon}, R., {Neyman},
  C., {Swain}, M., {Thompson}, R., {Vasisht}, G., {Wizinowich}, P., {Beichman},
  C., {Beletic}, J., {Creech-Eakman}, M., {Koresko}, C., {Sargent}, A., {Shao},
  M., \& {van Belle}, G. 2005, \apj, 624, 832

\bibitem[{{Mora} {et~al.}(2001){Mora}, {Mer{\'{\i}}n}, {Solano}, {Montesinos},
  {de Winter}, {Eiroa}, {Ferlet}, {Grady}, {Davies}, {Miranda}, {Oudmaijer},
  {Palacios}, {Quirrenbach}, {Harris}, {Rauer}, {Cameron}, {Deeg},
  {Garz{\'o}n}, {Penny}, {Schneider}, {Tsapras}, \& {Wesselius}}]{Mora:2001}
{Mora}, A., {Mer{\'{\i}}n}, B., {Solano}, E., {Montesinos}, B., {de Winter},
  D., {Eiroa}, C., {Ferlet}, R., {Grady}, C.~A., {Davies}, J.~K., {Miranda},
  L.~F., {Oudmaijer}, R.~D., {Palacios}, J., {Quirrenbach}, A., {Harris},
  A.~W., {Rauer}, H., {Cameron}, A., {Deeg}, H.~J., {Garz{\'o}n}, F., {Penny},
  A., {Schneider}, J., {Tsapras}, Y., \& {Wesselius}, P.~R. 2001, \aap, 378,
  116

\bibitem[{{Munari} {et~al.}(2005){Munari}, {Sordo}, {Castelli}, \&
  {Zwitter}}]{Munari:2005}
{Munari}, U., {Sordo}, R., {Castelli}, F., \& {Zwitter}, T. 2005, \aap, 442,
  1127

\bibitem[{{Pogodin}(1994)}]{Pogodin:1994}
{Pogodin}, M.~A. 1994, \aap, 282, 141

\bibitem[{{Porter} \& {Rivinius}(2003)}]{Porter:2003}
{Porter}, J.~M., \& {Rivinius}, T. 2003, \pasp, 115, 1153

\bibitem[{{Rhee} {et~al.}(2007){Rhee}, {Song}, \& {Zuckerman}}]{Rhee:2007}
{Rhee}, J.~H., {Song}, I., \& {Zuckerman}, B. 2007, \apj, 671, 616

\bibitem[{{Royer}(2005)}]{Royer:2005}
{Royer}, F. 2005, Memorie della Societa Astronomica Italiana Supplement, 8, 124

\bibitem[{{Royer} {et~al.}(2002{\natexlab{a}}){Royer}, {Gerbaldi},
  {Faraggiana}, \& {G{\'o}mez}}]{Royer:2002a}
{Royer}, F., {Gerbaldi}, M., {Faraggiana}, R., \& {G{\'o}mez}, A.~E.
  2002{\natexlab{a}}, \aap, 381, 105

\bibitem[{{Royer} {et~al.}(2002{\natexlab{b}}){Royer}, {Grenier}, {Baylac},
  {G{\'o}mez}, \& {Zorec}}]{Royer:2002b}
{Royer}, F., {Grenier}, S., {Baylac}, M.-O., {G{\'o}mez}, A.~E., \& {Zorec}, J.
  2002{\natexlab{b}}, \aap, 393, 897

\bibitem[{{Sch{\"u}tz} {et~al.}(2005){Sch{\"u}tz}, {Meeus}, \&
  {Sterzik}}]{Schutz:2005}
{Sch{\"u}tz}, O., {Meeus}, G., \& {Sterzik}, M.~F. 2005, \aap, 431, 165

\bibitem[{{Seidensticker}(1989)}]{Seidensticker:1989}
{Seidensticker}, K.~J. 1989, \aaps, 79, 61

\bibitem[{{Shu} {et~al.}(1987){Shu}, {Adams}, \& {Lizano}}]{Shu:1987}
{Shu}, F.~H., {Adams}, F.~C., \& {Lizano}, S. 1987, \araa, 25, 23

\bibitem[{{Sicilia-Aguilar} {et~al.}(2006{\natexlab{a}}){Sicilia-Aguilar},
  {Hartmann}, {Calvet}, {Megeath}, {Muzerolle}, {Allen}, {D'Alessio},
  {Mer{\'{\i}}n}, {Stauffer}, {Young}, \& {Lada}}]{Sicilia-Aguilar:2006a}
{Sicilia-Aguilar}, A., {Hartmann}, L., {Calvet}, N., {Megeath}, S.~T.,
  {Muzerolle}, J., {Allen}, L., {D'Alessio}, P., {Mer{\'{\i}}n}, B.,
  {Stauffer}, J., {Young}, E., \& {Lada}, C. 2006{\natexlab{a}}, \apj, 638, 897

\bibitem[{{Sicilia-Aguilar} {et~al.}(2006{\natexlab{b}}){Sicilia-Aguilar},
  {Hartmann}, {F{\"u}r{\'e}sz}, {Henning}, {Dullemond}, \&
  {Brandner}}]{Sicilia-Aguilar:2006b}
{Sicilia-Aguilar}, A., {Hartmann}, L.~W., {F{\"u}r{\'e}sz}, G., {Henning}, T.,
  {Dullemond}, C., \& {Brandner}, W. 2006{\natexlab{b}}, \aj, 132, 2135

\bibitem[{{Siegler} {et~al.}(2007){Siegler}, {Muzerolle}, {Young}, {Rieke},
  {Mamajek}, {Trilling}, {Gorlova}, \& {Su}}]{Siegler:2007}
{Siegler}, N., {Muzerolle}, J., {Young}, E.~T., {Rieke}, G.~H., {Mamajek},
  E.~E., {Trilling}, D.~E., {Gorlova}, N., \& {Su}, K.~Y.~L. 2007, \apj, 654,
  580

\bibitem[{{Siess} {et~al.}(2000){Siess}, {Dufour}, \& {Forestini}}]{Siess:2000}
{Siess}, L., {Dufour}, E., \& {Forestini}, M. 2000, \aap, 358, 593

\bibitem[{{Slettebak} {et~al.}(1975){Slettebak}, {Collins}, {Parkinson},
  {Boyce}, \& {White}}]{Slettebak:1975}
{Slettebak}, A., {Collins}, II, G.~W., {Parkinson}, T.~D., {Boyce}, P.~B., \&
  {White}, N.~M. 1975, \apjs, 29, 137

\bibitem[{{Song} {et~al.}(2001){Song}, {Caillault}, {Barrado y Navascu{\'e}s},
  \& {Stauffer}}]{Song:2001}
{Song}, I., {Caillault}, J.-P., {Barrado y Navascu{\'e}s}, D., \& {Stauffer},
  J.~R. 2001, \apj, 546, 352

\bibitem[{{Spangler} {et~al.}(2001){Spangler}, {Sargent}, {Silverstone},
  {Becklin}, \& {Zuckerman}}]{Spangler:2001}
{Spangler}, C., {Sargent}, A.~I., {Silverstone}, M.~D., {Becklin}, E.~E., \&
  {Zuckerman}, B. 2001, \apj, 555, 932

\bibitem[{{Strom} {et~al.}(1989){Strom}, {Strom}, {Edwards}, {Cabrit}, \&
  {Skrutskie}}]{Strom:1989}
{Strom}, K.~M., {Strom}, S.~E., {Edwards}, S., {Cabrit}, S., \& {Skrutskie},
  M.~F. 1989, \aj, 97, 1451

\bibitem[{{Strom}(1972)}]{Strom:1972}
{Strom}, S.~E. 1972, \pasp, 84, 745

\bibitem[{{Struve}(1931)}]{Struve:1931}
{Struve}, O. 1931, \apj, 73, 94

\bibitem[{{The} {et~al.}(1994){The}, {de Winter}, \& {Perez}}]{The:1994}
{The}, P.~S., {de Winter}, D., \& {Perez}, M.~R. 1994, \aaps, 104, 315

\bibitem[{{Tucker}(1975)}]{Tucker:1975}
{Tucker}, W. 1975, {Radiation processes in astrophysics} (Cambridge, Mass., MIT
  Press, 1975.~320 p.)

\bibitem[{{Uzpen} {et~al.}(2007){Uzpen}, {Kobulnicky}, {Monson}, {Pierce},
  {Clemens}, {Backman}, {Meade}, {Babler}, {Indebetouw}, {Whitney}, {Watson},
  {Wolfire}, {Benjamin}, {Bracker}, {Bania}, {Cohen}, {Cyganowski}, {Devine},
  {Heitsch}, {Jackson}, {Mathis}, {Mercer}, {Povich}, {Rho}, {Robitaille},
  {Sewilo}, {Stolovy}, {Watson}, {Wolff}, \& {Churchwell}}]{Uzpen:2007}
{Uzpen}, B., {Kobulnicky}, H.~A., {Monson}, A.~J., {Pierce}, M.~J., {Clemens},
  D.~P., {Backman}, D.~E., {Meade}, M.~R., {Babler}, B.~L., {Indebetouw}, R.,
  {Whitney}, B.~A., {Watson}, C., {Wolfire}, M.~G., {Benjamin}, R.~A.,
  {Bracker}, S., {Bania}, T.~M., {Cohen}, M., {Cyganowski}, C.~J., {Devine},
  K.~E., {Heitsch}, F., {Jackson}, J.~M., {Mathis}, J.~S., {Mercer}, E.~P.,
  {Povich}, M.~S., {Rho}, J., {Robitaille}, T.~P., {Sewilo}, M., {Stolovy},
  S.~R., {Watson}, D.~F., {Wolff}, M.~J., \& {Churchwell}, E. 2007, \apj, 658,
  1264

\bibitem[{{Uzpen} {et~al.}(2005){Uzpen}, {Kobulnicky}, {Olsen}, {Clemens},
  {Laurance}, {Meade}, {Babler}, {Indebetouw}, {Whitney}, {Watson}, {Wolfire},
  {Wolff}, {Benjamin}, {Bania}, {Cohen}, {Devine}, {Dickey}, {Heitsch},
  {Jackson}, {Marston}, {Mathis}, {Mercer}, {Stauffer}, {Stolovy}, {Backman},
  \& {Churchwell}}]{Uzpen:2005}
{Uzpen}, B., {Kobulnicky}, H.~A., {Olsen}, K.~A.~G., {Clemens}, D.~P.,
  {Laurance}, T.~L., {Meade}, M.~R., {Babler}, B.~L., {Indebetouw}, R.,
  {Whitney}, B.~A., {Watson}, C., {Wolfire}, M.~G., {Wolff}, M.~J., {Benjamin},
  R.~A., {Bania}, T.~M., {Cohen}, M., {Devine}, K.~E., {Dickey}, J.~M.,
  {Heitsch}, F., {Jackson}, J.~M., {Marston}, A.~P., {Mathis}, J.~S., {Mercer},
  E.~P., {Stauffer}, J.~R., {Stolovy}, S.~R., {Backman}, D.~E., \&
  {Churchwell}, E. 2005, \apj, 629, 512

\bibitem[{{Verhoelst} {et~al.}(2007){Verhoelst}, {van Aarle}, \&
  {Acke}}]{Verhoelst:2007}
{Verhoelst}, T., {van Aarle}, E., \& {Acke}, B. 2007, \aap, 470, 21

\bibitem[{{Viallefond} {et~al.}(1983){Viallefond}, {Donas}, \&
  {Goss}}]{Viallefond:1983}
{Viallefond}, F., {Donas}, J., \& {Goss}, W.~M. 1983, \aap, 119, 185

\bibitem[{{Viallefond} \& {Goss}(1986)}]{Viallefond:1986a}
{Viallefond}, F., \& {Goss}, W.~M. 1986, \aap, 154, 357

\bibitem[{{Viallefond} {et~al.}(1986){Viallefond}, {Goss}, {van der Hulst}, \&
  {Crane}}]{Viallefond:1986b}
{Viallefond}, F., {Goss}, W.~M., {van der Hulst}, J.~M., \& {Crane}, P.~C.
  1986, \aaps, 64, 237

\bibitem[{{Vieira} {et~al.}(2003){Vieira}, {Corradi}, {Alencar}, {Mendes},
  {Torres}, {Quast}, {Guimar{\~a}es}, \& {da Silva}}]{Viera:2003}
{Vieira}, S.~L.~A., {Corradi}, W.~J.~B., {Alencar}, S.~H.~P., {Mendes},
  L.~T.~S., {Torres}, C.~A.~O., {Quast}, G.~R., {Guimar{\~a}es}, M.~M., \& {da
  Silva}, L. 2003, \aj, 126, 2971

\bibitem[{{Vink} {et~al.}(2002){Vink}, {Drew}, {Harries}, \&
  {Oudmaijer}}]{Vink:2002}
{Vink}, J.~S., {Drew}, J.~E., {Harries}, T.~J., \& {Oudmaijer}, R.~D. 2002,
  \mnras, 337, 356

\bibitem[{{Voges} {et~al.}(2000){Voges}, {Aschenbach}, {Boller}, {Brauninger},
  {Briel}, {Burkert}, {Dennerl}, {Englhauser}, {Gruber}, {Haberl}, {Hartner},
  {Hasinger}, {Pfeffermann}, {Pietsch}, {Predehl}, {Schmitt}, {Trumper}, \&
  {Zimmermann}}]{Voges:2000}
{Voges}, W., {Aschenbach}, B., {Boller}, T., {Brauninger}, H., {Briel}, U.,
  {Burkert}, W., {Dennerl}, K., {Englhauser}, J., {Gruber}, R., {Haberl}, F.,
  {Hartner}, G., {Hasinger}, G., {Pfeffermann}, E., {Pietsch}, W., {Predehl},
  P., {Schmitt}, J., {Trumper}, J., \& {Zimmermann}, U. 2000, VizieR Online
  Data Catalog, 9029, 0

\bibitem[{{Wade} {et~al.}(2005){Wade}, {Drouin}, {Bagnulo}, {Landstreet},
  {Mason}, {Silvester}, {Alecian}, {B{\"o}hm}, {Bouret}, {Catala}, \&
  {Donati}}]{Wade:2005}
{Wade}, G.~A., {Drouin}, D., {Bagnulo}, S., {Landstreet}, J.~D., {Mason}, E.,
  {Silvester}, J., {Alecian}, E., {B{\"o}hm}, T., {Bouret}, J.-C., {Catala},
  C., \& {Donati}, J.-F. 2005, \aap, 442, L31

\bibitem[{{Walker} {et~al.}(1989){Walker}, {Volk}, {Wainscoat}, {Schwartz}, \&
  {Cohen}}]{Walker:1989}
{Walker}, H.~J., {Volk}, K., {Wainscoat}, R.~J., {Schwartz}, D.~E., \& {Cohen},
  M. 1989, \aj, 98, 2163

\bibitem[{{Waters} \& {Waelkens}(1998)}]{Waters:1998}
{Waters}, L.~B.~F.~M., \& {Waelkens}, C. 1998, \araa, 36, 233

\bibitem[{{Westin}(1985)}]{Westin:1985}
{Westin}, T.~N.~G. 1985, \aaps, 60, 99

\bibitem[{{Wright} {et~al.}(2003){Wright}, {Egan}, {Kraemer}, \&
  {Price}}]{Wright:2003}
{Wright}, C.~O., {Egan}, M.~P., {Kraemer}, K.~E., \& {Price}, S.~D. 2003, \aj,
  125, 359

\bibitem[{{Wright} \& {Barlow}(1974)}]{Wright:1974}
{Wright}, A.~E., \& {Barlow}, M.~F. 1974, \mnras, 170, 41

\bibitem[{{Yudin}(2001)}]{Yudin:2001}
{Yudin}, R.~V. 2001, \aap, 368, 912

\bibitem[{{Zhang} {et~al.}(2004){Zhang}, {Chen}, \& {He}}]{Zhang:2004}
{Zhang}, P., {Chen}, P.~S., \& {He}, J.~H. 2004, New Astronomy, 9, 509

\bibitem[{{Zhang} {et~al.}(2005){Zhang}, {Chen}, \& {Yang}}]{Zhang:2005}
{Zhang}, P., {Chen}, P.~S., \& {Yang}, H.~T. 2005, New Astronomy, 10, 325

\bibitem[{{Zuckerman}(2001)}]{Zuckerman:2001}
{Zuckerman}, B. 2001, \araa, 39, 549

\end{thebibliography}

\clearpage

\begin{deluxetable}{lcccccc}
\tabletypesize{\tiny}
\setlength{\tabcolsep}{0.01in}
\tablewidth{0pc}
\tablecaption{Projected Rotational Velocities}
\tablehead{
\colhead{Target} &
\colhead{HD} &
\colhead{Spec.} &
\colhead{\textit{v} sin \textit{i}} &
\colhead{Lines Used} & 
\colhead{\textit{v} sin \textit{i} (Literature)} &
\colhead{References} \\
\colhead{} &
\colhead{} &
\colhead{} &
\colhead({km s$^{-1}$)} &
\colhead{} &
\colhead{(km s$^{-1}$)} & 
\colhead{} }
\startdata 
G007.2369+01.4894 & 163296 & A1V & 145$\pm$15 & M & & \\
G008.3752-03.6697 & 167905 & F3V & 35$\pm$5 & F & & \\
G009.4973+05.2441 & 161643 & A7V & 140$\pm$15 & M & & \\
G011.2691+00.4208 & 165854 & B9V & 220$\pm$15 & H,M & 250$\pm$10, 242$\pm$10 & (1), (2) \\
G014.4239--00.7657 & 165854 & B8 & 165$\pm$10 & H,M & & \\
G025.6122+01.3020 & 171149 & B9V & 290$\pm$20 & H,M & 280, 301 & (3), (4)  \\
G027.0268+00.7224 & 172030 & A0V & 25$\pm$5 & M & & \\
G036.8722--00.4112 & ... & B8V & 295$\pm$20 & H,M & & \\
G047.3677+00.6199 & 180398 & B8Ve & 310$\pm$25 & H,M & & \\
G047.4523+00.5132 & 231033 & A0V & 150$\pm$10 & H,M & & \\
G051.6491--00.1182 & 183035 & A0V & 270$\pm$20 & H,M & & \\
G229.4514+01.0145 & 58647 & B9V & 295$\pm$20 & H,M & 118, 280 & (5), (6) \\
G257.6236+00.2459 & 72106 & A0IV & 240$\pm$25 & M & & \\
G265.5536-03.9951-north(A) & 73461 & A5V & 40$\pm$5 & F & & \\
G265.5536-03.9951-south(B) & ... & A6/8V & 25$\pm$5 & F & & \\
G269.5873-05.8882 & 74534 & G0V & 10$\pm$5 & F & 17 & (7) \\
G299.1677-00.3922 & ... & B8 & 130$\pm$10 & H,M & & \\
G299.7090-00.9704 & 107609 & B8/9IV & 175$\pm$15 & H,M & & \\
G300.0992-00.0627 & ... & A1 & 255$\pm$20 & H,M & & \\
G305.4232-00.8229 & 114757 & B6/8V(E) & 230$\pm$20 & H,M & & \\
G307.9784-00.7148 & 118094 & B8V(N) & 240$\pm$20 & H,M & & \\
G310.5420+00.4120 & 121195 & B8IV(N) & 270$\pm$20 & H,M & & \\
G311.0099+00.4156 & 121808 & A3IV & 235$\pm$25 & M & & \\
G311.6185+00.2469 & 122620 & B8/9 IV/V & 265$\pm$20 & H,M & & \\
G314.3136-00.6977 & 126578 & A1IV & 205$\pm$20 & M & & \\
G321.7868+00.4102 & 135354 & B8VN & 230$\pm$20 & H,M & & \\
G339.4392--00.7791 & 151228 & A0/1 IV/V & 40$\pm$5 & M & & \\
G339.7415--00.1904 & 151017 & A0 & 240$\pm$15 & H,M & & \\
G340.0517+00.6687 & 150625 & B8 & 230$\pm$15 & H,M & & \\
G347.3777+04.2010 & 152404 & F5V & 15$\pm$5 & F & & \\
\hline
 & 22252 & B8V & 255$\pm$20 & M & 249 & (8) \\
 & 48757 & A0V & 30$\pm$5 & M & 27 & (8) \\
 & 50853 & A0V & 205$\pm$20 & M & 195 & (8)  \\
 & 181454 & B8V & 80$\pm$10 & H,M & 75 & (9) \\
 & 100673 & B9Ve & 150$\pm$15 & M & 160,125 & (8), (9) \\
 & 97277 & A2IV & 55$\pm$5 & M & 45 & (9) \\
 & 83953 & B6V & 285$\pm$20 & H,M & 260 & (9) \\
 & 95370 & A3IV & 95$\pm$10 & M & 86, 90 & (8), (9) \\
 & 83754 & B5V & 190$\pm$20 & H,M & 150 & (9) \\
 & 98718 & B5Vn & 325$\pm$35 & H & 280 & (9) \\
 & 90994 & B6V & 105$\pm$10 & H,M & 80 & (9) \\
 & 78045 & A3V & 50$\pm$ 5 & M & 34, 25 & (8), (9) \\
 & 83446 & A5V & 135$\pm$15 & M & 133, 125 & (8), (9) \\
\enddata
\tablecomments{ See \citet{Uzpen:2007} for further information about
GLIMPSE or MSX Catalog ID.  Line profiles used to derive projected
rotational velocities are M for Mg \textsc{II} $\lambda$ 4481, H for H
\textsc{I} $\lambda$ 4471 $\lambda$ and F for Fe \textsc{I} $\lambda$
4476. (1) \citet{Halbedel:1996}; (2) \citet{Yudin:2001}; (3)
\citet{Abt:1995}; (4) \citet{Royer:2002b}; (5) \citet{Mora:2001}; (6)
\citet{Grady:1996}; (7) \citet{Cutispoto:2003}; (8)
\citet{Royer:2002a}; (9) \citet{Slettebak:1975}}
\label{trot}
\end{deluxetable}

\clearpage

\begin{deluxetable}{lcc}
\tabletypesize{\tiny}
\setlength{\tabcolsep}{0.01in}
\tablewidth{0pc}
\tablecaption{Effective Temperatures}
\tablehead{
\colhead{ID} &
\colhead{Spec.} &
\colhead{T$_{eff}$} \\
\colhead{} &
\colhead{} &
\colhead{(K)} \\
\colhead{(1)} &
\colhead{(2)} &
\colhead{(3)} }
\startdata 
G007.2369+01.4894 & A1V & 10350$\pm$230 \\
G008.3752-03.6697 & F3V & 7260$\pm$160 \\
G009.4973+05.2441 & A7V & 8320$\pm$130 \\
G011.2691+00.4208 & B9V & 12350$\pm$520 \\
G014.4239--00.7657 & B8 & 16460$\pm$640 \\
G025.6122+01.3020 & B9V & 12520$\pm$510 \\
G027.0268+00.7224 & A0V & 10510$\pm$300 \\
G036.8722--00.4112 & B8V & 11740$\pm$260 \\
G047.3677+00.6199 & B8Ve & 15700$\pm$600 \\
G047.4523+00.5132 & A0V & 12260$\pm$510 \\
G051.6491--00.1182 & A0V & 13100$\pm$530 \\
G229.4514+01.0145 & B9V & 10220$\pm$390 \\
G257.6236+00.2459 & A0IV & 11530$\pm$300 \\
G265.5536-03.9951-A & A5V & 7040$\pm$170 \\
G265.5536-03.9951-B & A6/8V & 7660$\pm$140 \\
G269.5873-05.8882$^{a}$ & G0III & ... \\
G299.1677-00.3922 & B8 & 16170$\pm$730 \\
G299.7090-00.9704 & B8/9IV & 12520$\pm$505 \\
G300.0992-00.0627 & A1 & 13510$\pm$510 \\
G305.4232-00.8229 & B6/8V(E) & 13520$\pm$510 \\ 
G307.9784-00.7148 & B8V (N) & 13220$\pm$520 \\
G310.5420+00.4120 & B8IV(N) & 12430$\pm$510 \\
G311.0099+00.4156 & A3IV & 9800$\pm$130 \\
G311.6185+00.2469 & B8/9 IV/V & 10540$\pm$255 \\
G314.3136-00.6977 & A1IV & 9870$\pm$130 \\
G321.7868+00.4102 & B8V N & 12340$\pm$505 \\
G339.4392--00.7791 & A0/1 IV/V & 8800$\pm$150 \\
G339.7415--00.1904 & A0 & 12220$\pm$510 \\
G340.0517+00.6687 & B8 & 14080$\pm$520 \\
G347.3777+04.2010$^{b}$ & F5V & ... \\
\enddata
\tablecomments{ (a) This
star is a giant. See \S 8 for further details.
(b) This star is a spectroscopic binary. Equivalent
width measurements vary too greatly to determine
temperature. See \S 8 for further details.}
\label{ttemp}
\end{deluxetable}

\clearpage

\begin{deluxetable}{lcccccccc}
\tabletypesize{\tiny}
\rotate
\tablewidth{0pc}
\tablecaption{H$\alpha$ Equivalent Width Measurements}
\tablehead{
\colhead{ID} & 
\colhead{Spec. Type} & 
\colhead{Date} & 
\colhead{EW(H$\alpha$)} & 
\colhead{EW(H$\alpha_{abs}$)} & 
\colhead{EW(H$\alpha_{corr}$)} & 
\colhead{R$_{0}$ Flux} & 
\colhead{H$\alpha$ Flux} &
\colhead{H$\alpha$ Profile} \\
\colhead{} & 
\colhead{} & 
\colhead{} & 
\colhead{(\AA)} & 
\colhead{(\AA)} &
\colhead{(\AA)} &
\colhead{(erg cm$^{-2}$ s$^{-1}$ \AA$^{-1}$)} & 
\colhead{(erg cm$^{-2}$ s$^{-1}$) } &
\colhead{} \\ 
\colhead{(1)} &
\colhead{(2)} &
\colhead{(3)} &
\colhead{(4)} &
\colhead{(5)} &
\colhead{(6)} &
\colhead{(7)} &
\colhead{(8)} &
\colhead{(9)} }
\startdata
G007.2369+01.4894 & A1V & 20060928 & -26.15$\pm$0.15 & ...  & ... & 1.42$\times$10$^{-11}$ & ... & S \\     
G008.3752-03.6697 & F3V & 20060928 &  -0.17$\pm$0.08 & ...  & ... & 2.12$\times$10$^{-12}$ & ... & P \\
G009.4973+05.2441 & A7V & 20060928 &   5.11$\pm$0.02 & ...  & ... & 2.52$\times$10$^{-14}$ & ... & A \\
G011.2691+00.4208 & B9V & 20060927 &  -4.24$\pm$0.16 & 9.62 & -13.86$\pm$0.16 & 1.30$\times$10$^{-12}$ & 1.80$\times$10$^{-11}$ & D \\
G014.4239-00.7657 & B8  & 20060927 &   4.45$\pm$0.03 & ...  &   0.00 & 1.40$\times$10$^{-13}$ & ... & A \\
G025.6122+03.3020 & B9V & 20060928 &   6.42$\pm$0.09 & ...  & ... & 5.99$\times$10$^{-12}$ & ... & A \\
G027.0268+00.7224 & A0V & 20070407 &  -1.97$\pm$0.43 & 11.89 & -13.86$\pm$0.43 & 1.74$\times$10$^{-12}$ & 2.41$\times$10$^{-11}$ & D \\
G036.8722-00.4112 & B8V & 20060927 &  -2.39$\pm$0.15 & 12.68 & -15.07$\pm$0.15 & 1.68$\times$10$^{-12}$ & 2.53$\times$10$^{-11}$ & D \\
G047.3677+00.6199 & B9Ve & 20060927 &-14.08$\pm$0.24 & 8.39 &  -22.47$\pm$0.24 & 1.79$\times$10$^{-12}$ & 4.02$\times$10$^{-11}$ & D \\
G047.4523+00.5132 & A0V & 20060927 &   6.49$\pm$0.02 & ...  &   0.00 & 5.12$\times$10$^{-13}$ & ... & A \\
G051.6491-00.1182 & A0V & 20060928 &   5.76$\pm$0.15 & 8.65 & -2.89$\pm$0.15 & 1.30$\times$10$^{-12}$ & 3.76$\times$10$^{-12}$ & D \\
G229.4514+01.0145 & B9V & 20060926 &  -7.82$\pm$0.45 & ...  & ... & 8.66$\times$10$^{-12}$ & ... & D \\
G257.6236+00.2459 & A0IV & 20060926 &  9.41$\pm$0.06 & ...  & ... & 1.03$\times$10$^{-12}$ & ... & A \\
G265.5536-03.9951-A & A5V & 20060927 & 3.77$\pm$0.06 & ... & ... & 2.36$\times$10$^{-12}$ & ... & A \\
G265.5536-03.9951-B & A6/8V & 20060927 & 4.93$\pm$0.11 & ...& ... & 7.01$\times$10$^{-13}$ & ... & A \\
G269.5873-05.8882 & G0III & 20060926 & 2.12$\pm$0.03 & ... & ... & 1.04$\times$10$^{-12}$ & ... & A \\
G299.1677-00.3922 & B8 & 20070407 & -4.37$\pm$0.12 & 9.62 & -13.99$\pm$0.12 & 5.51$\times$10$^{-13}$ & 7.71$\times$10$^{-12}$ & D \\
G299.7090-00.9704 & B8/9IV & 20070407 & -1.99$\pm$0.16 & 8.13 & -10.12$\pm$0.16 & 5.36$\times$10$^{-13}$ & 5.42$\times$10$^{-12}$ & D \\
G300.0992-00.0627 & A1 & 20070407 & -13.89$\pm$0.32 & 7.76 & -21.65$\pm$0.32 & 2.41$\times$10$^{-13}$ & 5.22$\times$10$^{-12}$ & D \\
G305.4232-00.8229 & B6/8V(E) & 20070407 & -14.22$\pm$0.42 & 7.76 & -21.98$\pm$0.42 & 7.34$\times$10$^{-13}$ & 1.61$\times$10$^{-11}$ & D \\
G307.9784-00.7148 & B8V(N) & 20070407 & -4.63$\pm$0.69 & 8.64 & -13.27$\pm$0.69 & 1.62$\times$10$^{-12}$ & 2.15$\times$10$^{-11}$ & D \\
G310.5420+00.4120 & B8IV(N) & 20070407 & -0.09$\pm$0.09 & 9.88 & -9.97$\pm$0.09 & 6.88$\times$10$^{-13}$ & 6.86$\times$10$^{-12}$ & D \\
G311.0099+00.4156 & A3IV & 20070407 & 5.21$\pm$0.03 & ... & 0.00 & 1.67$\times$10$^{-12}$ & ... & A \\
G311.6185+00.2469 & B8/9IV/V & 20070407 & 1.96$\pm$0.23 & 12.86 & -10.90$\pm$0.23 & 3.12$\times$10$^{-13}$ & 3.40$\times$10$^{-12}$ & D \\
G314.3136-00.6977 & A1IV & 20070407 & 4.20$\pm$0.29 & 16.83 & -12.63$\pm$0.29 & 9.41$\times$10$^{-13}$ & 1.19$\times$10$^{-11}$ & D \\
G321.7686+00.4102 & B8V N & 20070407 & -16.93$\pm$0.53 & 9.88 & -26.81$\pm$0.53 & 5.88$\times$10$^{-13}$ & 1.58$\times$10$^{-11}$ & D \\
G339.4392-00.7791 & A0/1IV/V & 20060927 & 3.28$\pm$0.76 & 14.48 & -11.20$\pm$0.76 & 8.58$\times$10$^{-13}$ & 9.61$\times$10$^{-12}$ & D \\
G339.7415-00.1904 & A0 & 20060927 & 5.50$\pm$0.03 & ... & 0.00  & 4.15$\times$10$^{-13}$ & ... & A \\
G340.0517+00.6687 & B8 & 20060927 & -31.62$\pm$0.08 & 7.76 & -39.38$\pm$0.03 & 1.17$\times$10$^{-12}$ & 4.61$\times$10$^{-11}$ & D \\
G347.3777+04.2010 & B8 & 20060927 & -0.02$\pm$0.10 & ... & ... & 2.08$\times$10$^{-12}$ & ... & IP \\
\enddata
\tablecomments{(5) H$\alpha$ equivalent width of the nearest stellar
model. This equivalent width was then used to determine
EW(H$\alpha_{corr}$) and subsequently calculate the H$\alpha$
flux. (7) The reddening corrected R flux. (8) H$\alpha$ flux was determined only for GLIMPSE mid-IR excess
stars with 24 $\mu$m measurements and H$\alpha$ emission. (9) D:
double-peaked profile, A: absorption only profile, S: single emission
only profile, P: P-Cygni profile, IP: inverse P-Cygni profile. }
\label{tha}
\end{deluxetable}

\begin{deluxetable}{lcccccc}
\tabletypesize{\tiny}
\setlength{\tabcolsep}{0.01in}
\tablewidth{0pc}
\tablecaption{Spectral Features}
\tablehead{
\colhead{ID} &
\colhead{H$\alpha$ Emission} &
\colhead{Paschen Emission} &
\colhead{Ca \textsc{II} Emission} &
\colhead{O \textsc{I} 8446 Emission} &
\colhead{O \textsc{I} 7772 Absorption} &
\colhead{Fe \textsc{II} Emission} \\
\colhead{} &
\colhead{} &
\colhead{} &
\colhead{} &
\colhead{} &
\colhead{} &
\colhead{} \\
\colhead{(1)} &
\colhead{(2)} &
\colhead{(3)} &
\colhead{(4)} &
\colhead{(5)} &
\colhead{(6)} &
\colhead{(7)} }
\startdata 
G007.2369+01.4894 & y & n & y & y & s & n \\
G008.3752-03.6697 & y & n & n & n & s & n \\
G009.4973+05.2441 & n & n & n & n & n & n \\
G011.2691+00.4208 & y & n & n & n & n & n \\
G014.4239-00.7657 & n & n & n & n & s & n \\
G025.6122+01.3020 & n & n & n & n & n & n \\
G027.0268+00.7224 & y & y & y & y & e & y \\
G036.8722-00.4112 & y & n & n & n & n & n \\
G047.3677+00.6199 & y & n & n & y & n & n \\
G047.4523+00.5132 & n & n & n & n & s & n \\
G051.6491-00.1182 & y & n & n & n & n & n \\
G229.4514+01.0145 & y & n & y & y & s & y \\
G257.6236+00.2459 & n & n & n & n & n & n \\
G265.5536-03.9951-A & n & n & n & n & s & n \\
G265.5536-03.9951-B & n & n & n & n & n & n \\
G269.5873-05.8882$^{a}$ & n & n & n & n & N/A & n \\
G299.1677-00.3922 & y & y & n & y & e$?$ & n \\
G299.7090-00.9704 & y & n & n & n & n & n \\
G300.0992-00.0627 & y & n & n & y & e & n \\
G305.4232-00.8229 & y & n & n & y & e & n \\ 
G307.9784-00.7148 & y & n & n & y & n & n \\
G310.5420+00.4120 & y & n & n & n & s & n \\
G311.0099+00.4156 & n & n & n & n & s & n \\
G311.6185+00.2469 & y & n & n & n & n & n \\
G314.3136-00.6977 & y & n & n & n & s & n \\
G321.7868+00.4102 & y & y & n & y & e & y \\
G339.4392-00.7791 & y & n & n & n & n & n \\
G339.7415-00.1904 & n & n & n & n & n & n \\
G340.0517+00.6687 & y & y & n & y & e & y \\
G347.3777+04.2010$^{b}$ & y & n & n & n & N/A & n \\
\enddata
\tablecomments{(2) Stars that exhibit H$\alpha$ in emission are
denoted by (y). (3) Stars with Paschen lines in emission are denoted
by (y). (4) Stars that exhibit Ca \textsc{II} IR triplet emission are
denoted by (y). (5) Stars that exhibit O \textsc{I} $\lambda$ 8446
\AA\ in emission are noted by (y) (6) Stars that are shown to have O
\textsc{I} $\lambda$ 7772 \AA\ in absorption stronger than stellar
models predict are denoted by (s), those that exhibit it in emission
(e), and those that exhibit absorption are denoted by
(n). G269.5873-05.8882 and G347.3777+04.2010 are listed as N/A because
we do not determine their stellar temperature and cannot make a
comparison to stellar models. (7) Stars that exhibit Fe \textsc{II}
emission are denoted by (y). (a) Spectral features indicate that this
star is a giant. See \S 8 for further details.  (b) This star is a
spectroscopic binary. See \S 8 for further details.}
\label{tspec}
\end{deluxetable}

\clearpage

\begin{deluxetable}{lcccc}
\tabletypesize{\tiny}
\setlength{\tabcolsep}{0.01in}
\tablewidth{0pc}
\tablecaption{Estimated Stellar Ages}
\tablehead{
\colhead{ID} &
\colhead{Distance} &
\colhead{Adopted M$_{Bol}$} &
\colhead{Age Derived} &
\colhead{Age in Literature} \\
\colhead{} &
\colhead{(pc)} &
\colhead{[mag]} &
\colhead{(log yr)} &
\colhead{(log yr)} \\
\colhead{(1)} &
\colhead{(2)} &
\colhead{(3)} &
\colhead{(4)} & 
\colhead{(5)} }
\startdata 
G007.2369+01.4894 & 122$^{+17}_{-13}$ & -1.15$^{+1.34}_{-0.34}$ & 6.13-6.63, 8.04-8.27 & 6.70$^{a}$ \\
G009.4973+05.2441 & 137$^{+32}_{-22}$ & 1.27$^{+0.40}_{-0.45}$ & 6.65-6.95, 8.42-8.74 & ... \\
G025.6122+01.3020 & 114$^{+12}_{-9}$ & -0.10$^{+0.34}_{-0.36}$ & 7.09-8.09 & ... \\
G036.8722--00.4112 & 424$^{+423}_{-142}$ & -1.53$^{+3.40}_{-1.68}$ & 6.11-6.32, 8.10-8.20 & ... \\
G047.3677+00.6199 & 353$^{+171}_{-86}$ & -2.05$^{+0.78}_{-1.06}$ & 6.87-7.72 & ... \\
G047.4523+00.5132 & 909$^{+91}_{-755}$ & -2.00$^{+4.01}_{-0.37}$ & 6.11-6.37, 7.58-8.19 & ... \\
G051.6491--00.1182 & 345$^{+173}_{-87}$ & -0.98$^{+0.79}_{-1.03}$ & 6.13-8.05 & ... \\
G229.4514+01.0145 & 277$^{+79}_{-50}$ & -2.25$^{+0.59}_{-0.50}$ & 6.09-6.43, 8.37-8.43 &  \\
G257.6236+00.2459 & 288$^{+202}_{-84}$ & -0.13$^{+0.84}_{-1.26}$ & 6.26-6.55, 7.33-8.24 &  \\
G265.5536--03.9951-A$^{b}$ & 139$^{+39}_{-22}$ & 0.70$^{+0.42}_{-0.49}$ & 6.81-6.83 & ... \\
G265.5536--03.9951-B$^{b}$ & 139$^{+39}_{-22}$ & 1.85$^{+0.42}_{-0.50}$ & 6.81-6.83 & ... \\
G305.4232--00.8229$^{c}$ & 1009$^{+244}_{-188}$ & -3.05$\pm$0.45 & 5.88, 7.87-7.95 &  \\ 
G307.9784--00.7148 & 758$^{+2272}_{-326}$ & -3.04$^{+1.38}_{-2.10}$ & 6.00-6.23, 7.66-8.03 & 8.03$\pm$0.04$^{d}$ \\
G311.6185+00.2469$^{e}$ & 965 & -1.13 $\pm$0.50 & 6.08-6.18, 8.16-8.18 &  8.275$^{f}$ \\
G321.7868+00.4102$^{g}$ & 1010 & -2.09$\pm$0.51 & 5.83-6.12, 8.01-8.13 & ... \\
G340.0517+00.6687$^{g}$ & 830 & -3.14$\pm$0.14 & 5.62-5.85, 7.81-7.90 & ... \\
G347.3777+04.2010 & ... & ... & ... & 7.23$^{h}$ \\
\enddata
\tablecomments{ (a) \citep{Mannings:1997}; (b) Assuming both stars 
to be coeval only a 2.2 and 2.0\ \mo \ binary is possible
given temperature measurements. (c) Utilizing
\citet{Seidensticker:1989} extinction and absolute visual magnitude
measurements (d) \citep{Levenhagen:2004} (e) Utilizing
\citet{Westin:1985} absolute visual magnitude.  (f)
\citet{Westin:1985}; (g) Utilizing \citet{Kozok:1985}; (h)
\citet{Chen:2005} }
\label{tage}
\end{deluxetable}
\clearpage

\begin{deluxetable}{cccccccc}
\tabletypesize{\tiny}
\setlength{\tabcolsep}{0.01in}
\tablewidth{0pc}
\tablecaption{Free-Free Contribution to the 24 $\mu$m Excess}
\tablehead{
\colhead{ID} &
\colhead{[24] Flux} &
\colhead{[24] error$^{a}$} &
\colhead{Excess flux at [24]} &
\colhead{Free-free component} &
\colhead{$\%$ free-free } &
\colhead{$\sigma$ excess} &
\colhead{Excess Type} \\
\colhead{} &
\colhead{(mJy)} &
\colhead{(mJy)} &
\colhead{(mJy)} &
\colhead{(mJy)} &
\colhead{} & 
\colhead{} &
\colhead{} \\
\colhead{(1)} &
\colhead{(2)} &
\colhead{(3)} &
\colhead{(4)} &
\colhead{(5)} &
\colhead{(6)} &
\colhead{(7)} &
\colhead{(8)} }
\startdata 
G011.2691+00.4208 & 18.4 & 2.7 & 13.1 & 6.0 & 46$\pm$21$\%$ & 2.6 & I \\
G014.4239--00.7657 & 71.9 & 13.7 & 71.2 & ... & ... & 5.2 & III \\
G027.0268+00.7224 & 17.7 & 2.1 & 8.7 & 8.5 & 98$\pm$24$\%$ & 0 & I \\
G036.8722--00.4112 & 21.7 & 2.2 & 16.3 & 7.4 & 45$\pm$13$\%$ & 4.0 & II \\
G047.3677+00.6199 & 44.6 & 4.5 & 38.9 & 12.1 & 31$\pm$12$\%$ & 6.0 & II \\
G047.4523+00.5132 & 8.5 & 1.1 & 6.3 & ... & ... & 5.5 & III \\
G051.6491--00.1182 & 17.8 & 2.8 & 12.7 & 6.1 & 48$\pm$22$\%$ & 2.4 & I \\
G299.1677--00.3922 & 7.0 & 0.8 & 5.4 & 5.6 & 104$\pm$15$\%$ & 0 & I \\
G299.7090--00.9704 & 5.1 & 0.5 & 3.1 & 1.9 & 61$\pm$16$\%$ & 2.4 & I \\
G300.0992--00.0627 & 7.3 & 0.8 & 6.1 & 1.7 & 28$\pm$13$\%$ & 5.5 & II \\
G305.4232--00.8229 & 16.7 & 1.8 & 14.2 & 5.1 & 36$\pm$13$\%$ & 5.1 & II \\
G307.9784--00.7148 & 21.5 & 2.2 & 16.1 & 7.0 & 43$\pm$14$\%$ & 4.1 & II \\
G310.5420+00.4120 & 5.6 & 0.7 & 3.3 & 2.3 & 70$\pm$21$\%$ & 1.4 & I \\
G311.0099+00.4156 & 149.4 & 17.9 & 140.9 & ... & ... & 7.9 & III \\
G311.6185+00.2469 & 8.6 & 1.5 & 7.4 & 1.2 & 16$\pm$20$\%$ & 4.1 & II \\
G314.3136--00.6977 & 22.8 & 2.3 & 17.8 & 4.0 & 22$\pm$13$\%$ & 6.0 & II \\
G321.7868+00.4102 & 12.6 & 1.3 & 10.7 & 5.0 & 47$\pm$12$\%$ & 4.4 & II \\
G339.4392--00.7791$^{b,c}$ & 7.7 & 1.5 & ... & ... & ... & ... & ... \\
G339.7415--00.1904$^{b}$ & 5.4 & 0.8 & 4.4 & ... & ... & 5.5 & III \\
G340.0517+00.6687$^{b}$ & 31.6 & 3.4 & 26.9 & 14.4 & 54$\pm$14$\%$ & 3.7 & II \\
\enddata
\tablecomments{(7) Determined by removing the free-free component from
the [24] excess and dividing by the [24] error with 10$\%$ calibration
uncertainty error added in quadrature with measured error. (a) [24]
flux measurement error with 10$\%$ calibration uncertainty error added
in quadrature with measured error. (b) found to have no real excess
at [24]. (c) New measurement, not in \citep{Uzpen:2007}.}
\label{tff}
\end{deluxetable}

\clearpage

\begin{deluxetable}{lcccccccc}
\tabletypesize{\tiny}
\setlength{\tabcolsep}{0.01in}
\tablewidth{0pc}
\tablecaption{Stellar and Circumstellar Disk Parameters}
\tablehead{
\colhead{ID} &
\colhead{$K$-8} &
\colhead{8-24} &
\colhead{Disk Temperature} &
\colhead{$\frac{L_{IR}}{L_{*}}$} & 
\colhead{EW(H$\alpha_{corr}$)} &
\colhead{T$_{eff}$} &
\colhead{\textit{v} sin \textit{i}} &
\colhead{Evolutionary Status} \\
\colhead{} &
\colhead{[mag]} &
\colhead{[mag]} &
\colhead{(K)} &
\colhead{} &
\colhead{(\AA)} &
\colhead{(K)} &
\colhead{(km s$^{-1}$)} &
\colhead{} \\
\colhead{(1)} &
\colhead{(2)} &
\colhead{(3)} &
\colhead{(4)} &
\colhead{(5)} &
\colhead{(6)} &
\colhead{(7)} &
\colhead{(8)} &
\colhead{(9)} }
\startdata 
G007.2369+01.4894 & ... & ... & 493$^{+4}_{-5}$ & 0.14$^{+0.002}_{-0.002}$ & -26.15$\pm$0.15$^{a}$ & 10350$\pm$230 & 145$\pm$15 & H \\
G008.3752--03.6697 & ... & ... & 566$^{+7}_{-8}$ & 0.20$^{+0.004}_{-0.004}$ & -0.17$\pm$0.08$^{a}$ & 7260$\pm$160 & 35$\pm$5 & TT \\
G009.4973+05.2441 & ... & ... & ... & ... & 5.11$\pm$0.02$^{a}$ & 8320$\pm$130 & 140$\pm$15 & $?$ \\
G011.2691+00.4208 & 0.34 & 1.01 & ... & ... & -13.86$\pm$0.16 & 12350$\pm$520 & 220$\pm$15 & Be \\
G014.4239--00.7657 & 0.47 & 4.78 &  191$^{+5}_{-4}$ & 0.013$^{+0.001}_{-0.001}$ & 4.45$\pm$0.03 & 16900$\pm$800 & 165$\pm$10 & T \\
G025.6122+01.3020 & ... & ... & ... & ... & 6.42$\pm$0.09 & 12520$\pm$510 & 290$\pm$20 & $?$ \\
G027.0268+00.7224 & 0.34 & 0.56 & ... & ... & -13.89$\pm$0.43 & 10510$\pm$300 & 25$\pm$5 & Be \\
G036.8722--00.4112 & 0.27 & 1.26 & 427$^{+16}_{-17}$ & 0.00072$^{+0.00007}_{-0.00006}$ & -15.07$\pm$0.15 & 11740$\pm$260 & 295$\pm$20 & T \\
G047.3677+00.6199 & 0.93 & 1.24 & 783$^{+19}_{-24}$ & 0.0028$^{+0.0001}_{-0.0001}$ & -22.47$\pm$0.24 & 15700$\pm$600 & 310$\pm$25 & T/Be \\
G047.4523+00.5132 & 0.29 & 1.25 & 355$^{+8}_{-8}$ & 0.0012$^{+0.0001}_{-0.0001}$ & 6.49$\pm$0.02 & 12260$\pm$510 & 150$\pm$10 & D \\
G051.6491--00.1182 & 0.19 & 1.12 & ... & ... & -2.89$\pm$0.15 & 13100$\pm$530 & 270$\pm$20 & Be \\
G229.4514+01.0145 & ... & ... & 579$^{+4}_{-5}$ & 0.058$^{+0.001}_{-0.001}$ & -7.82$\pm$0.45$^{a}$ & 10220$\pm$390 & 295$\pm$20 & H \\
G257.6236+00.2459 & ... & ... & 404$^{+9}_{-10}$ & 0.12$^{+0.002}_{-0.002}$ & 9.41$\pm$0.06$^{a}$ & 11530$\pm$300 & 240$\pm$25 & H \\
G265.5536--03.9951-A & ... & ... & ... & ... & 4.93$\pm$0.11$^{a}$ & 7660$\pm$140 & 25$\pm$5 & $?$ \\
G265.5536--03.9951-B & ... & ... & ... & ... & 3.77$\pm$0.06$^{a}$ & 7040$\pm$170 & 40$\pm$5 & $?$ \\
G269.5873--05.8882 & ... & ... & ... & ... & 2.12$\pm$0.03$^{a}$ & ... & 10$\pm$5 & $?$ \\
G299.1677--00.3922 & 0.35 & 1.18 & ... & ... & -13.99$\pm$0.12 & 16450$\pm$1030 & 130$\pm$10 & Be \\
G299.7090--00.9704 & 0.18 & 0.81 & ... & ... & -10.12$\pm$0.16 & 12520$\pm$505 & 175$\pm$15 & Be \\
G300.0992--00.0627 & 0.56 & 1.36 & 526$^{+41}_{-32}$ & 0.0021$^{+0.0002}_{-0.0002}$ & -21.65$\pm$0.32 & 13510$\pm$510 & 255$\pm$20 & T/Be \\
G305.4232--00.8229 & 0.79 & 1.27 & 675$^{+37}_{-33}$ & 0.0038$^{+0.0003}_{-0.0002}$ & -21.98$\pm$0.42 & 13520$\pm$510 & 230$\pm$20 & T/Be\\ 
G307.9784--00.7148 & 0.40 & 1.06 & 557$^{+22}_{-16}$ & 0.00090$^{+0.00007}_{-0.00008}$ & -13.27$\pm$0.69 & 13220$\pm$520 & 240$\pm$20 & T/Be \\
G310.5420+00.4120 & 0.25 & 0.70 & ... & ... & -9.97$\pm$0.09 & 12430$\pm$510 & 270$\pm$20 & Be \\
G311.0099+00.4156 & 0.37 & 2.79 & 315$^{+4}_{-3}$ & 0.0027$^{+0.0001}_{-0.0003}$ & 5.21$\pm$0.03 & 9800$\pm$130 & 235$\pm$25 & D \\
G311.6185+00.2469 & 0.28 & 1.82 &  306$^{+13}_{-6}$ & 0.0012$^{+0.0001}_{-0.0001}$ & -10.90$\pm$0.23 & 10540$\pm$255 & 265$\pm$20 & T \\
G314.3136--00.6977 & 0.22 & 1.45 & 328$^{+11}_{-9}$ & 0.00096$^{+0.00007}_{-0.00008}$ & -12.63$\pm$0.29 & 9870$\pm$130 & 205$\pm$20 & T \\
G321.7868+00.4102 & 0.77 & 1.25 & 556$^{+18}_{-12}$ & 0.0021$^{+0.0002}_{-0.0001}$ & -26.81$\pm$0.53 & 12340$\pm$505 & 230$\pm$20 & H \\
G339.4392--00.7791$^{b}$ & ... & ... & ... & ... & -11.20$\pm$0.76 & 8800$\pm$150 & 40$\pm$5 & ... \\
G339.7415--00.1904 & 0.27 & 1.93 & 311$^{+6}_{-6}$ & 0.0010$^{+0.0001}_{-0.0001}$ & 5.50$\pm$0.03 & 12220$\pm$510 & 240$\pm$15 & D \\
G340.0517+00.6687 & 1.11 & 1.04 & 823$^{+20}_{-20}$ & 0.015$^{+0.001}_{-0.001}$ & -39.38$\pm$0.08 & 14080$\pm$520 & 230$\pm$15 & H \\
G347.3777+04.2010 & ... & ... & 487$^{+9}_{-11}$ & 0.13$^{+0.02}_{-0.03}$ & -0.02$\pm$0.10$^{a}$ & ... & 15$\pm$5 & TT/T \\
\enddata
\tablecomments{ (9) H: Herbig AeBe star, TT: T-Tauri
  star, T: Transitional disk, D: Debris disk, Be: Excess due to
  free-free emission and therefore a classical Be-type star, $?$:
  additional measurements are needed to classify the nature of the IR
  excess.  (a) This value is EW(H$\alpha$), uncorrected for underlying
  stellar absorption.  (b) The [8.0] excess from \citet{Uzpen:2007} is
  not confirmed at [24].}
\label{tsum}
\end{deluxetable}

\clearpage

\begin{deluxetable}{lcccc}
\tabletypesize{\tiny}
\setlength{\tabcolsep}{0.01in}
\tablewidth{0pc}
\tablecaption{Classification Based on \citet{Andrillat:1990} Spectral Features}
\tablehead{
\colhead{ID} &
\colhead{Criterion 1} &
\colhead{Criterion 2} &
\colhead{Criterion 3} &
\colhead{Classical Be} \\ 
\colhead{(1)} &
\colhead{(2)} &
\colhead{(3)} &
\colhead{(4)} &
\colhead{(5)} }
\startdata 
G007.2369+01.4894 & n & n & y & n \\
G027.0268+00.7224 & y & y & N/A & y \\
G047.3677+00.6199 & N/A & N/A & n & n \\
G229.4514+01.0145 & n & n & y & n \\
G299.1677-00.3922 & N/A & y & N/A & y \\
G300.0992-00.0627 & N/A & n & n & n \\
G305.4232-00.8229 & N/A & n & n & n \\ 
G307.9784-00.7148 & N/A & n & n & n \\
G321.7868+00.4102 & n & y & n & n \\
G340.0517+00.6687 & n & y & n & n \\
\enddata
\tablecomments{(2) Criterion 1: Stars with Ca \textsc{II} in emission
and Paschen in emission are denoted by y.  Stars with Ca \textsc{II}
in emission without Paschen emission are denoted by n. Stars without
Ca \textsc{II} emission are denoted by N/A.  (3) Criterion 2: Stars
with O \textsc{I} in emission and Paschen in emission are denoted by
y. Stars with O \textsc{I} in emission and Paschen in absorption are
denoted by n. Stars without O \textsc{I} in emission are denoted by
N/A.  (4) Criterion 3: Stars with [8.0] excess $>$ 0.55 magnitudes
with Ca \textsc{II} in emission are denoted by y.  Stars with [8.0]
excess $>$ 0.55 magnitudes without Ca \textsc{II} are denoted by
n. Stars without [8.0] excess $>$ 0.55 magnitudes are denoted by
N/A. (5) Stars with criterion 1, 2, and/or 3 with y only are
consistent with classical Be stars.}
\label{comp}
\end{deluxetable}

\clearpage

\clearpage

\clearpage

\clearpage

\clearpage

\clearpage

\clearpage

\clearpage

\clearpage

\clearpage

\begin{figure}
    \includegraphics[width=4.0in,angle=90]{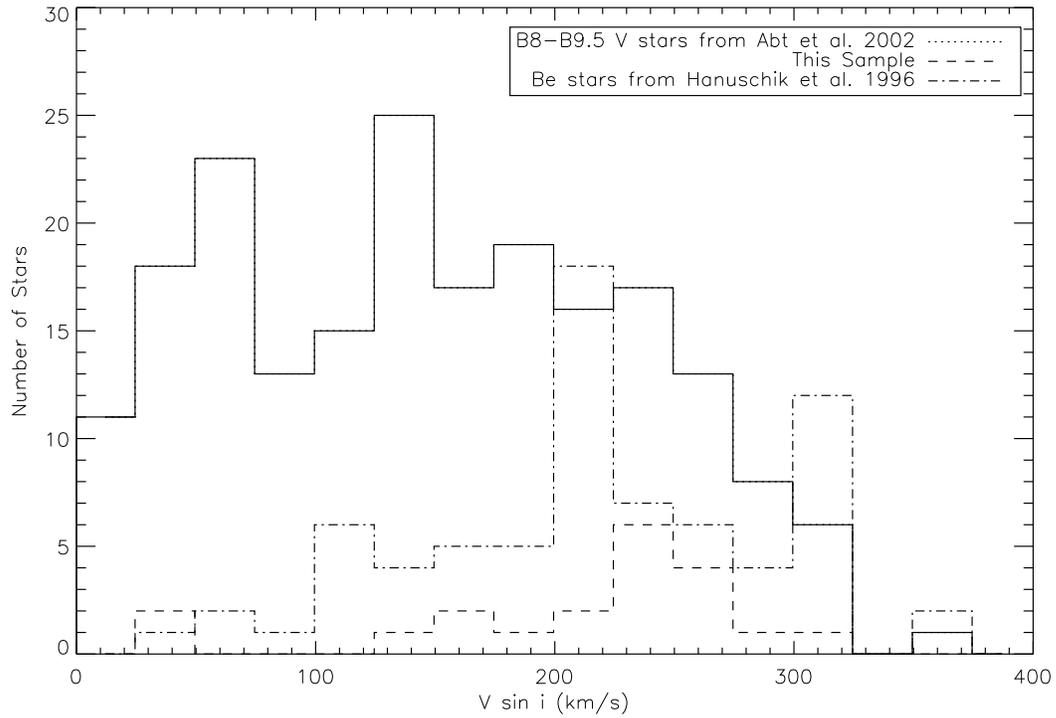}
    \caption{A comparison of projected rotational velocities of
B8--9.5 V stars from \citet{Abt:2002}, classical Be stars from
\citet{Hanuschik:1996}, and our GLIMPSE sample. Utilizing a KS test we
find that there is no significant probability, $<$ 1\% for the
\citet{Abt:2002} sample and 8\% for the \citet{Hanuschik:1996} sample,
that this sample may be derived from either population.}
    \label{comp}
\end{figure}

\clearpage

\begin{figure}
    \includegraphics[width=5.0in]{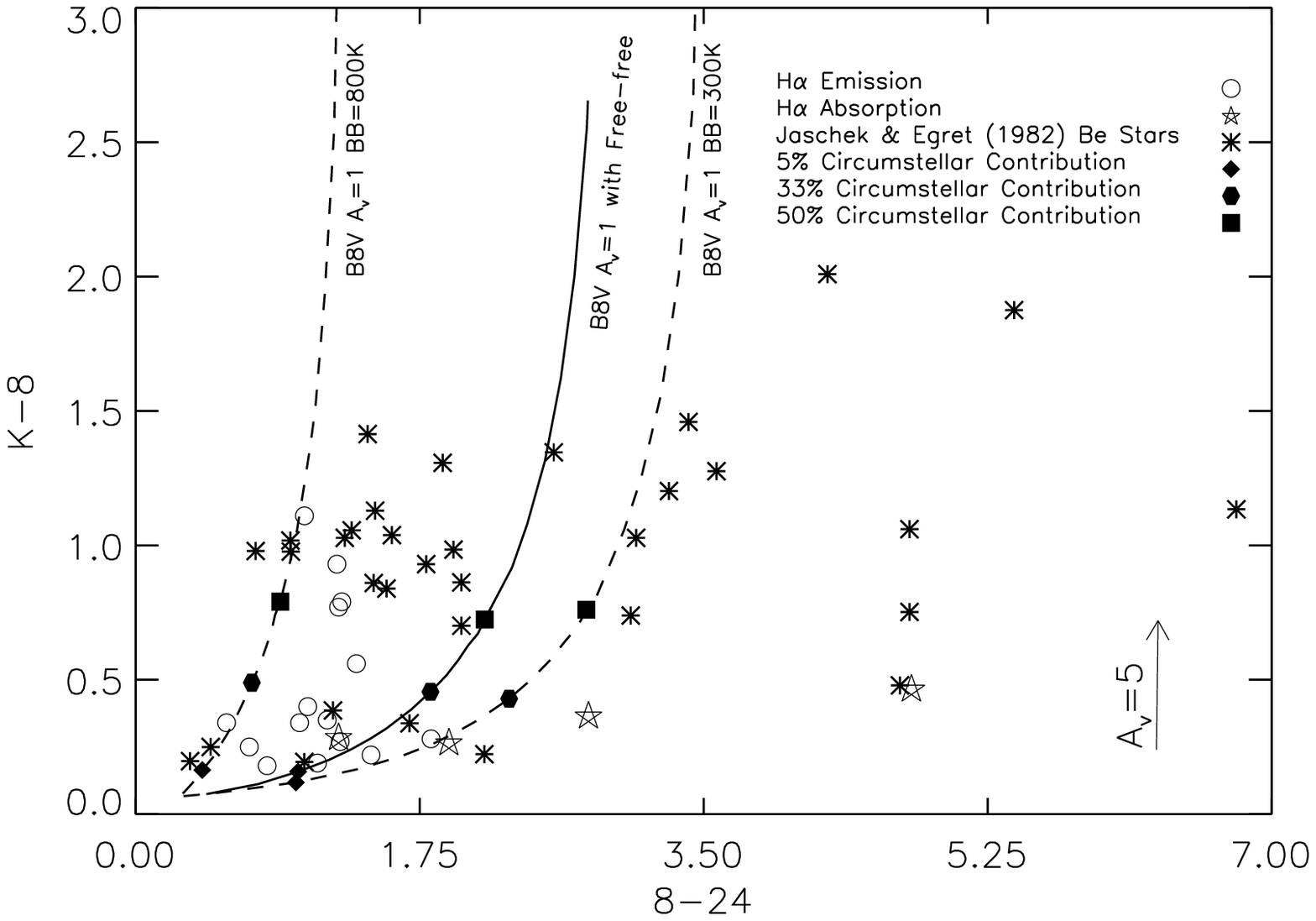}
    \caption{A $K$-8 $\mu$m vs. 8-24 $\mu$m color-color plot of our
    objects ($open$ $symbols$) and classical Be stars ($asterisks$)
    from \citet{Jaschek:1982} with $MSX$ A band and IRAS 25 $\mu$m
    (from \citealt{Zhang:2004,Zhang:2005}) were utilized as 8 $\mu$m
    and 24 $\mu$m measurements respectively. The arrow shows the
    reddening vector for A$_{V}$=5.0. The solid curve denotes a B8V
    main-sequence with A$_{V}$=1.0 and increasing contribution of
    free-free flux at 8 $\mu$m. The solid symbols denote 5\%, 33\%,
    and 50\% circumstellar contributions to the total flux at 8
    $\mu$m. The dashed curves denote B8V main-sequence stars with
    A$_{V}$=1.0 and increasing contributions from 800~K and 300~K
    blackbody circumstellar components, respectively.  This plot shows
    that excess emission owing to free-free processes can be
    distinguished from cool dust ($<$ 300~K).  Systems containing the
    former lie to the left of the solid line while the systems with
    the latter lie to the right. In this color-color space, free-free
    and hot dust excesses are indistinguishable.}
    \label{color}
\end{figure}

\clearpage

\begin{figure}
    \includegraphics[width=5.0in]{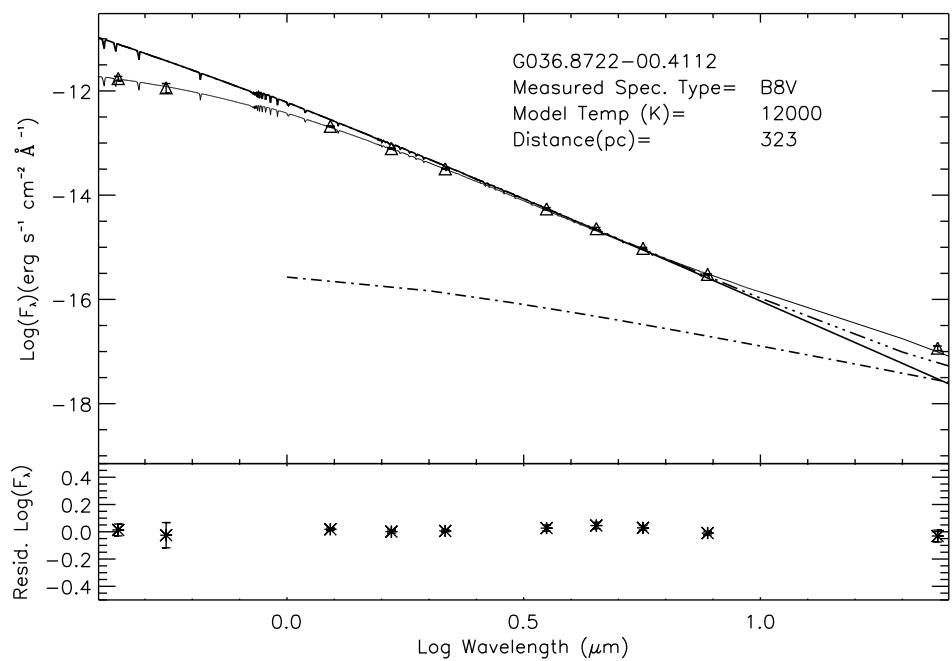}
    \caption{Spectral energy distribution of one of our target stars,
    G036.8722-00.4112. The thick solid line is the distance-normalized
    stellar model. The thin solid line is the stellar model with
    extinction and circumstellar blackbody component from
    \citet{Uzpen:2007}. The dash-dot curve is the modeled free-free
    component normalized using the H$\alpha$ flux. The
    dashed-dot-dot-dot curve is the sum of the stellar model and the
    free-free component. The free-free component is not sufficient to
    explain the measured excess at 24 $\mu$m. The residuals are for the
    stellar model with circumstellar blackbody component.}
    \label{G36}
\end{figure}

\clearpage

\begin{figure}
\hbox{
    \includegraphics[width=4.0in]{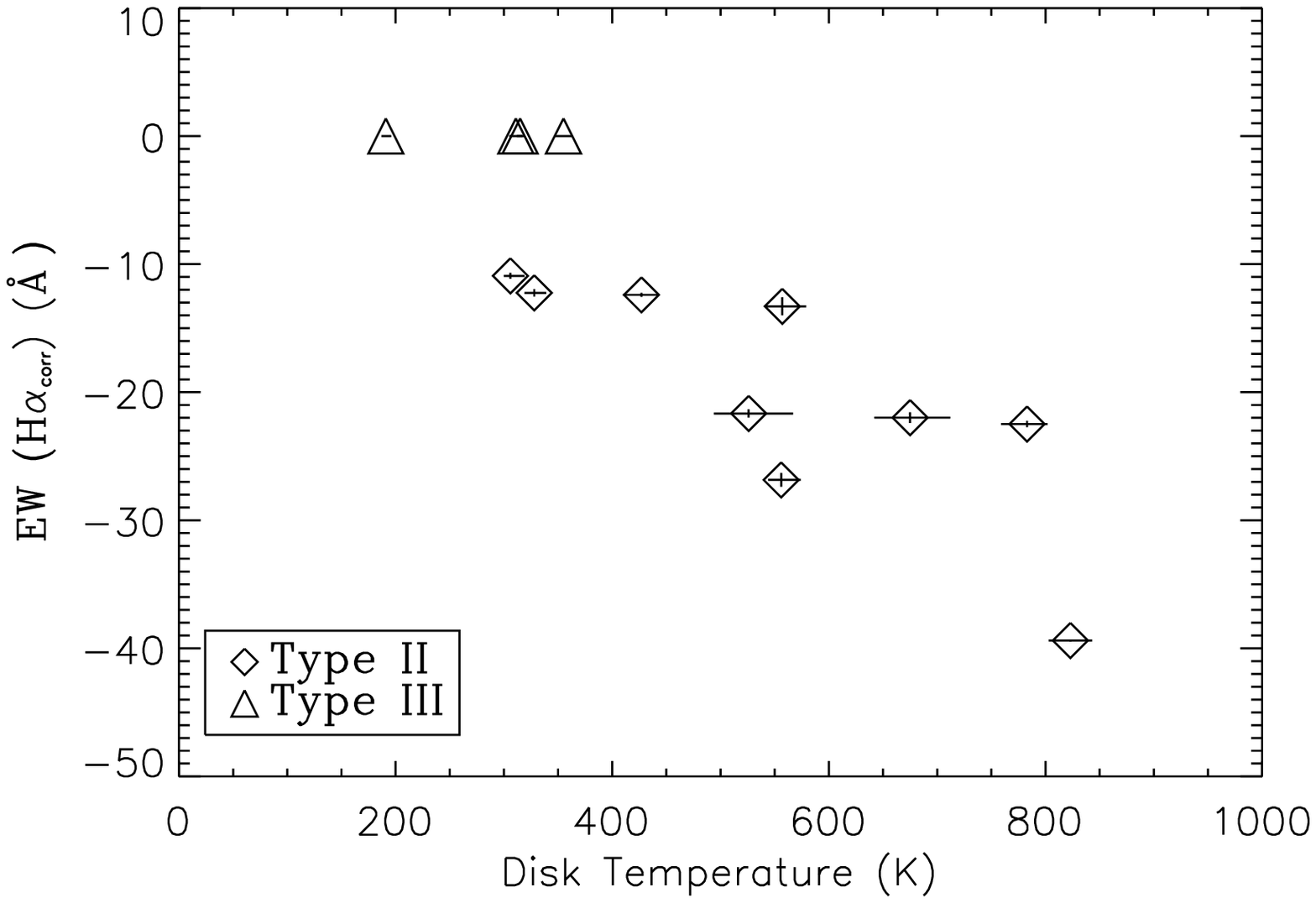}
}
\hbox{
    \includegraphics[width=4.0in]{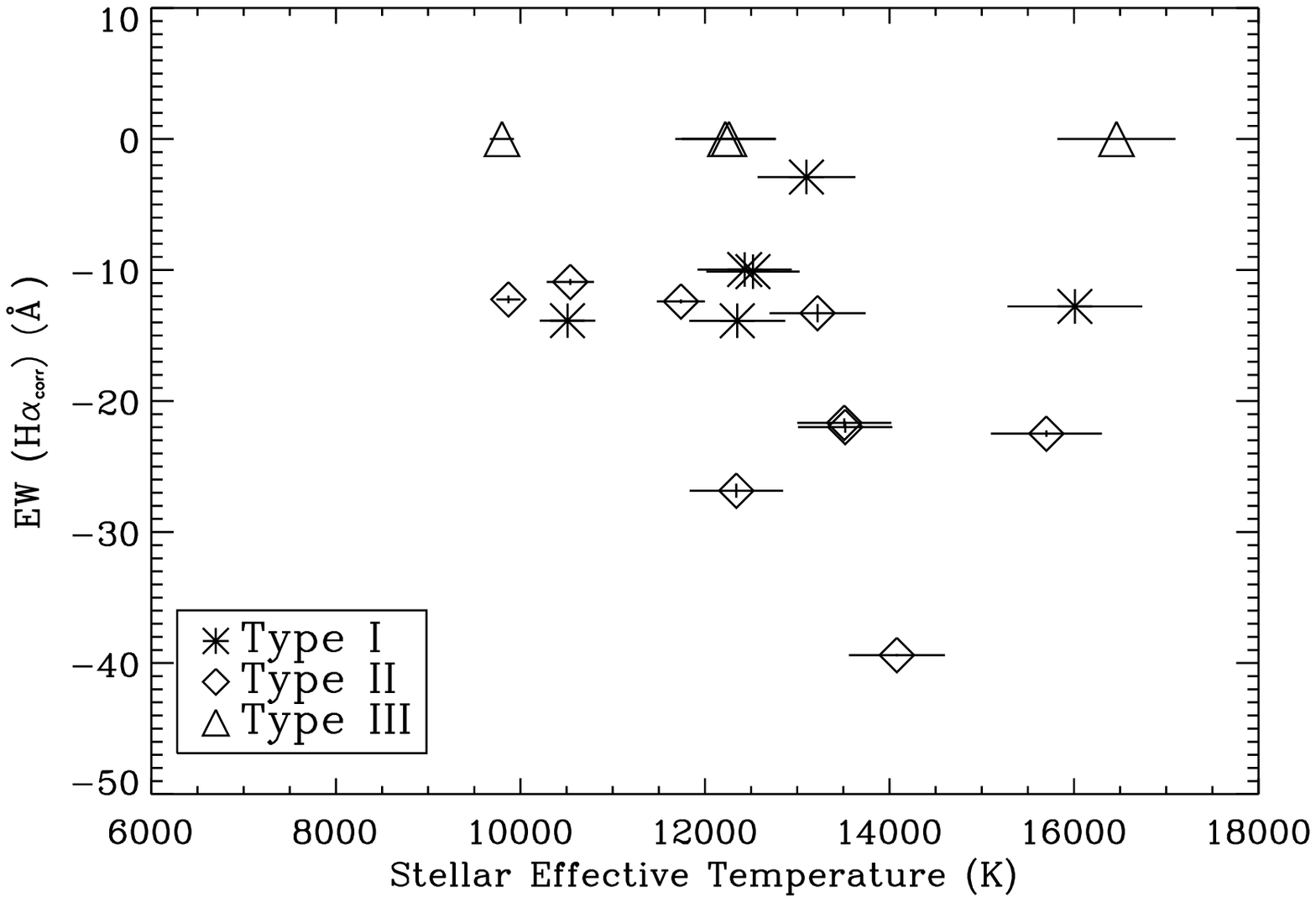}
}
 \caption{(a) EW(H$\alpha$$_{corr}$) vs. disk temperature. There
is a strong anti-correlation between these two parameters.  (b) H$\alpha$
equivalent width vs. stellar effective temperature. There is no
correlation between these two parameters for Type I, Type II, or Type III
sources.}
    \label{diskt}
\end{figure}

\clearpage

\begin{figure}
    \plotone{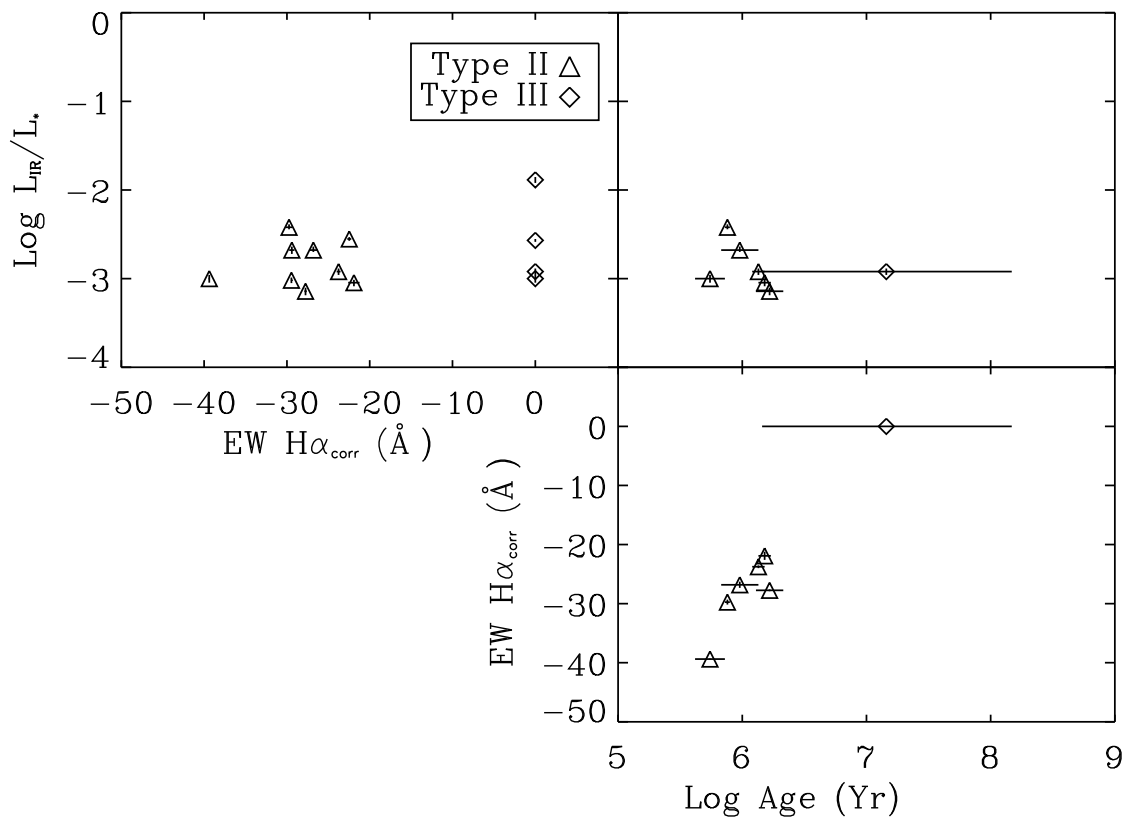}
    \caption{An inter-comparison of $\frac{L_{IR}}{L_{*}}$, age, and
H$\alpha$ equivalent width.  Using the younger age, when available,
age is correlated with H$\alpha$ equivalent width with a
correlation coefficient of 0.97 indicating $>$ 99\% probability that
the two parameters are correlated for these seven stars. Using the
older age when available, there is no correlation between age and
$\frac{L_{IR}}{L_{*}}$ or between H$\alpha$ equivalent width and
$\frac{L_{IR}}{L_{*}}$. }
    \label{3in1}
\end{figure}

\clearpage

\begin{figure}
    \includegraphics[width=5.0in]{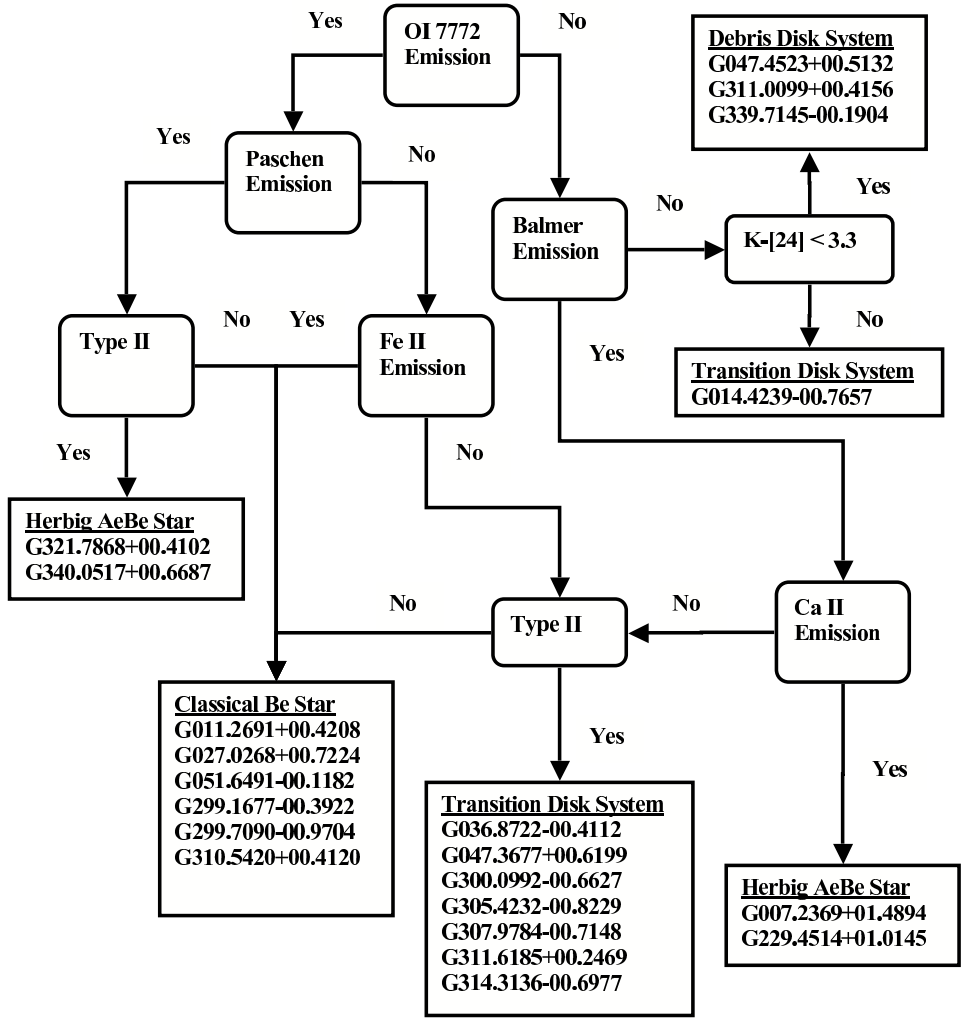}
    \caption{A flow chart showing how we determined the type of
    circumstellar disk system using echelle spectra and an estimate of
    the contribution from free-free emission for 19 GLIMPSE stars. Two
    MSX stars are characterized: G007.2369+01.4894, where extensive
    data exists in the literature, and G229.4514+01.0145, which
    exhibits spectral features similar to G007.2369+01.4894.}
    \label{flow}
\end{figure}

\end{document}